\RequirePackage{fix-cm}
\documentclass[twocolumn]{svjour3}          % twocolumn
\smartqed  % flush right qed marks, e.g. at end of proof

\usepackage{graphicx}
\usepackage{graphicx}
\usepackage{empheq}
\usepackage{subfigure}

\usepackage{tikz-cd}
\usepackage{tikz}

\usepackage[numbers]{natbib}
\usepackage[integrals]{wasysym}
\usepackage{changes}

\usetikzlibrary{decorations.markings}
\usetikzlibrary{shapes.geometric, arrows}
\usepackage{tikzscale}
\usepackage{wrapfig}
\usepackage{tcolorbox}
\usepackage{color}
\definecolor{blue1}{rgb}{0.0, 0.0, 1.0}
\definecolor{gray}{rgb}{0.9,0.9,0.9}
\definecolor{gray1}{rgb}{0.7,0.7,0.7}
\definecolor{gray2}{rgb}{0.8,0.8,0.8}
\definecolor{magenta}{rgb}{1.0, 0.0, 1.0}
\usepackage[title]{appendix}
\newcommand\backmatter{\appendix
\def\chaptermark##1{\markboth{%
\ifnum  \c@secnumdepth > \m@ne  \@chapapp\ \thechapter:  \fi  ##1}{%
\ifnum  \c@secnumdepth > \m@ne  \@chapapp\ \thechapter:  \fi  ##1}}%
\def\sectionmark##1{\relax}}

\newcommand{\ISI}[1]{\mathrm{ISI}}

\usepackage{hyperref}
\hypersetup{ colorlinks=true, urlcolor  = blue, linkcolor = blue,
citecolor = blue1,}

\usepackage{float}
\usepackage{amsmath,mathtools,amssymb,amsfonts}
\usepackage[scaled=.95]{helvet}

\definecolor{lime}{HTML}{A6CE39}
\DeclareRobustCommand{\orcidicon}{%
    \begin{tikzpicture}
    \draw[lime, fill=lime] (0,0) 
    circle [radius=0.16] 
    node[white] {{\fontfamily{qag}\selectfont \tiny ID}};
    \draw[white, fill=white] (-0.0625,0.095) 
    circle [radius=0.007];
    \end{tikzpicture}
    \hspace{-2mm}
}

\foreach \x in {A, ..., Z}{%
    \expandafter\xdef\csname orcid\x\endcsname{\noexpand\href{https://orcid.org/\csname orcidauthor\x\endcsname}{\noexpand\orcidicon}}
}
\newcommand{\orcid}[1]{\href{https://orcid.org/#1}{\textcolor[HTML]{A6CE39}{\orcidicon}}}

\newcommand{\uproman}[1]{\uppercase\expandafter{\romannumeral#1}}

\usepackage{ulem}
\journalname{Cognitive Neurodynamics: CODY-D-21-00122R2  }

\begin{document}

\title{Control of noise-induced coherent oscillations in three-neuron motifs}


\author{Florian B\"onsel\orcid{0000-0002-7193-9643}, Patrick Krauss\orcid{0000-0002-6611-7733}, Claus Metzner\orcid{0000-0002-5709-4306}, Marius E. Yamakou\orcid{0000-0002-2809-1739}}


\institute{Florian B\"onsel $\cdot$ Marius E. Yamakou \at Chair for Dynamics, Control and Numerics, Department of Data Science, Friedrich-Alexander-Universit\"at Erlangen-Nürnberg, Cauerstr. 11, 91058 Erlangen, Germany\\
Florian B\"onsel $\cdot$ Claus Metzner \at Biophysics Group, Friedrich-Alexander-Universit\"at Erlangen-Nürnberg, Henkestr. 91, 91052 Erlangen, Germany\\
Patrick Krauss \at Neuroscience Lab, University Hospital Erlangen, Friedrich-Alexander-Universit\"at Erlangen-Nürnberg,  Waldstr. 1, 91054 Erlangen, Germany\\
             \email{florian.boensel@gmail.com}\\  %\emph{Florian B\"onsel}\\
             \email{marius.yamakou@fau.de}           %  \\
            %of F. Author  %  if needed
}

\date{Received: 20th June 2021 / Accepted: 28th November 2021}

\maketitle
\begin{abstract}
The phenomenon of self-induced stochastic resonance (SISR) requires a nontrivial scaling limit between the deterministic and the stochastic timescales of an excitable system, leading to the emergence of coherent oscillations which are absent without noise. In this paper, we numerically investigate SISR and its control in single neurons and three-neuron motifs made up of the Morris-Lecar model. In single neurons, we compare the effects of electrical and chemical autapses on the degree of coherence of the oscillations due to SISR. In the motifs, we compare the effects of altering the synaptic time-delayed couplings and the topologies on the degree of SISR. Finally, we provide two enhancement strategies for a particularly poor degree of SISR in motifs with chemical synapses: $(\mathrm{i})$ we show that a poor SISR can be significantly enhanced by attaching an electrical or an excitatory chemical autapse on one of the neurons, and $(\mathrm{ii})$ we show that by multiplexing the motif with a poor SISR to another motif (with a high SISR in isolation), the degree of SISR in the former motif can be significantly enhanced. We show that the efficiency of these enhancement strategies depends on the topology of the motifs and the nature of synaptic time-delayed couplings mediating the multiplexing connections.

\end{abstract}

%%%%%%%%%%%%%%%%%%%%%%%%%%%%%%%%%%%%%%%%%%%%%%%%%%%%%%%%%%%%%%%%%%%%%%%%%%%%%%%%%%%%%%%%%%%%%%%%%%%%%%%%%%%%%%%%%%%%%%%%%
\section{Introduction}\label{sect1}
Noise is ubiquitous in biological systems and in particular in neural systems. Contrary to the intuitive perception of noise as deteriorating signal quality~\cite{mcdonnell2011benefits}, several studies have shown the constructive effects of noise on neural dynamics~\cite{longtin1993stochastic,patel2008stochastic,gang1993stochastic,gutkin2007transient}, perception and cognition~\cite{krauss2016stochastic,krauss2017adaptive,krauss2018cross,schilling2020intrinsic,SCHILLING2021139}. The most remarkable of these effects is the phenomenon of noise-induced resonance, in which an optimal amount of noise enhances the detection of weak oscillations and the coherence of these oscillations in a neural systems.  

There are several noise-induced resonance phenomena with different emergent conditions and mechanisms, and thus, may play different roles in information processing under different settings of the neural system, see, e.g.,~\cite{yamakou2017simple,yamakou2019control,deville2005two,zamani2020concomitance}. These include, amongst others, the well-known stochastic resonance (SR)~\cite{longtin1993stochastic,wiesenfeld1995stochastic,lindner2004effects,guo2017frequency,patel2008stochastic,benzi1981mechanism,gang1993stochastic},
coherence resonance (CR)~\cite{xu2019effects,gang1993stochastic,pikovsky1997coherence, PhysRevE.60.7270,gammaitoni1998stochastic,zhou2001array,neiman1997coherence,zhu2020phase}, inverse stochastic resonance (ISR)~\cite{gutkin2007transient,gutkin2009inhibition,uzuntarla2013dynamical,yamakou2017simple,yamakou2018weak}, recurrence resonance (RR)~\cite{krauss2019recurrence}, and  self-induced stochastic resonance (SISR) \cite{yamakou2017simple,freidlin2001stable,freidlin2001stochastic, muratov2005self,deville2005two,muratov2008noise,deville2007nontrivial,deville2007self,yamakou2018coherent,yamakou2019control}.  

In this work, we focus on SISR. Generically, SISR occurs when a multiple-timescale excitable dynamical system is driven by a noise of weak amplitude. During SISR (see, e.g.,  Fig.\,\ref{fig:time_series_phase_for_single_ML_SISR1}\textbf{(b)}), 
the escape timescale of trajectories from one attracting region (e.g., the left monotonically decreasing parts of the $S$-shaped nullcline in Fig.\,\ref{fig:time_series_phase_for_single_ML_SISR1}\textbf{(b)}) in phase space to another (e.g., the right monotonically decrease parts of the $S$-shape nullcline in Fig.\,\ref{fig:time_series_phase_for_single_ML_SISR1}\textbf{(b)}) is exponentially distributed, and the associated transition rate is governed by an activation energy (e.g., the energy barriers $\Delta U^l_i(w_e)$ and $\Delta U^r_i(w_{p,i})$ defined in Eq.\,\eqref{eqn:9}). If the excitable system (e.g., a neuron) is placed out-of-equilibrium, and its activation energy decreases monotonically as the neuron relaxes slowly to a stable quiescent state (stable fixed point), then at a 
specific instant during the relaxation, the timescale of escape due to noise and the timescale of relaxation match, and the neuron fires almost surely at this point. 
If this activation brings the neuron back out-of-equilibrium, the relaxation stage can start over again, and the scenario repeats itself indefinitely, 
leading to a coherent spiking activity which cannot occur without noise. SISR essentially depends on the interplay of three different timescales: the slow and fast timescales in the deterministic equation of the system, plus a third timescale characteristic to the noise. 

In 2005, Muratov et al.~\cite{muratov2005self} coined the term self-induced stochastic resonance after they discovered the mechanism behind this noise-induced resonance phenomena in a chemical model equation. After the 2005 paper, a series of papers on SISR in other models including neural systems \cite{deville2005two,muratov2008noise,yamakou2018coherent,yamakou2019control,yamakou2020optimal}, Brownian ratchets \cite{deville2007self}, cancer model \cite{shen2010self}, and even in bearing faults model \cite{zhang2021stochastic} were published, each showing how generic the mechanism of SISR in a slow-fast stochastic excitable system is, and how ubiquitous it is in physical, biological, and chemical systems.

All previous studies have investigated SISR in isolated oscillators, except in the case of neural systems, where only two studies have investigated SISR in networks of coupled neurons \cite{yamakou2019control,yamakou2020optimal}. It was shown in \cite{yamakou2019control} that, in contrast to SISR in a single isolated FitzHugh-Nagumo (FHN) neuron, the maximum noise amplitude at which SISR can occur in the network of coupled FHN neurons is not fixed (i.e., is controllable), especially in the regime of strong synaptic couplings and long time delays. And in \cite{yamakou2020optimal}, the performance of electrical and inhibitory chemical synapses in the enhancement of the degree of SISR in layer and multiplex networks of FHN neurons are compared. It was shown that for each isolated layer network, weaker electrical and chemical synaptic couplings are better enhancers of SISR. It was also shown that, regardless of the synaptic strengths, shorter electrical synaptic delays are better enhancers of SISR than shorter chemical synaptic delays, while longer chemical synaptic delays are better enhancers than longer electrical synaptic delays. Furthermore, it is found that electrical, inhibitory, or excitatory chemical multiplexing of the two layers having only electrical synapses at the intra-layer levels can each enhance SISR. Additionally, only excitatory chemical multiplexing of the two layers having only inhibitory chemical synapses at the intra-layer levels can enhance SISR. Furthermore, in \cite{yamakou2019control,yamakou2020optimal}, the enhancement of SISR is based on the configuration of the electrical and chemical synapses between the connected neurons within a layer network and between layers in a multiplex network. No studies have reported on the (in)efficiency of autapses --- self-feedback synapses --- on the enhancement of SISR in neurons. The current work aims at bridging this gap.

Moreover, in all previous studies of SISR in neural systems, including isolated neurons \cite{yamakou2017simple,deville2005two,muratov2008noise,yamakou2018coherent,yamakou2020levy,zhu2021stochastic} and neural networks \cite{yamakou2019control,yamakou2020optimal}, the mathematically simpler but biophysically less realistic FHN neuron model has been used. In this work and for the very first time, we study SISR and its control in a conductance-based neuron model, i.e., in the more realistic Morris-Lecar (ML) model ~\cite{morris1981voltage}. The mathematical structure of the ML neuron model --- low dimensional, existence of a (explicit) strong timescale separation between the dynamical variables (conditions required SISR), and to some degree, a tractable nonlinear vector field ---  makes it a perfect conductance-based model for the analysis of SISR. 
%Other conductance-based models are lacking in some or even all (e.g., the Hodgkin-Huxley model) of these desired mathematical structures.

In information processing, networks take different tasks of functionality~\cite{markram2012human,van2013wu}. Thus, a better understanding of their structure and connectivity should shed more light on the dynamics of the phenomena occurring on them~\cite{krauss2019weight}. It is well-known that large recurrent networks can be decomposed into smaller building blocks --- the so-called motifs~\cite{milo2002network}, whereby three-neuron motifs are the most basic motifs, which frequently appear in neural circuits and can be seen as basic computational units~\cite{krauss2019analysis}, each uniquely contributing to a large-scale neural behavior~\cite{li2008functions, song2005highly}. Thus, in this paper we focus on these basic computational units --- three-neuron motifs (3NMs).
Another important class of networks is the so-called multiplex network. It consists of two or several layer networks connected to each other, with each node in one layer connected only to its replica node in another layer~\cite{battiston2017structure}. This kind of inter-layer coupling can induce complex behaviors, namely: the emergence and suppression of chimera states~\cite{PhysRevE.94.052205, ghosh2016birth, ghosh2016emergence}, and the formation of synchronization patterns~\cite{sawicki2018synchronization,goremyko2017pattern}, including intra-layer synchronization effects~\cite{singh2015synchronization,singh2015synchronization}.
It has been shown that multiplexing of layer networks can be used to control the dynamics of one layer by tuning the parameters of another layer. For example, multiplexing of layer networks has been shown to be an efficient strategy for improving CR in one layer of a two-layer multiplex network by tuning the parameters of the other layer~\cite {semenova2018weak,yamakou2019control} network.
In particular, it was found that multiplexing can induce CR in a layer which does not exhibit this phenomenon in isolation. Moreover, it has been shown that the control of CR can be achieved even for weak multiplexing.
While these theoretical results are intriguing, it remains an open question to which extent they affect our understanding of the neural information processing underlying perception, cognition, and behavior in biological organisms. In the case of multiplexing, for example, a fundamental question is where the required point-to-point connections might be found in actual nervous systems. Possible candidates are the nerve fibres connecting the layers within cortical micro columns~\cite{kandel2000principles}, or also the commissural fibers of the corpus callosum, which are known to form point-to-point connections between homologous cortex areas in the two different  hemispheres~\cite{aboitiz1992individual,schuz1996basic}. Even if these anatomical structures can be interpreted as cases of multiplexing, it must however be assumed that heterogeneous multiplex conﬁgurations, i.e., those between diﬀerent types of motifs, are signiﬁcantly more likely than homogeneous conﬁgurations of identical motifs. We therefore consider in this work also the SISR phenomenon in systems of two structurally different, but mutually coupled motifs.

Due to the complexity of noise-induced resonance phenomena, most existing studies are forced to consider relatively small networks of idealized model neurons, and they are typically based on assumptions that make an extrapolation of the obtained results to larger neural networks quite difficult. For example, while many theoretical works consider electrical synapses, chemical synapses are by far the most common connections between neurons in the brain~\cite{pereda2014electrical}, and particularly in the mammalian central nervous system, their number exceeds that of electrical synapses by several orders of magnitude~\cite{kandel2000principles}. For this reason, we are considering both electrical and chemical synapses in this work. Moreover, we investigate the effect of so-called autapses~\cite{van1972autapses}. Autapses are synaptic contacts of a neuron’s axon onto its own dendrite and soma. In the neocortex, self-inhibiting autapses in GABAergic interneurons are abundant in number and play critical roles in regulating spike precision and network  activity~\cite{lubke1996frequency,yilmaz2016autapse, herrmann2004autapse, guo2016regulation,yin2018autapses,bacci2006enhancement}. Anatomical observations suggest that autapses  might be used as compensatory replacements for injured axons \cite{wang2017formation}, or to enhance persistent neural activity (that is supposed to be) elementary for short-term memory storage \cite{seung2000autapse}. Some research papers have shown that autapses can significantly influence the dynamics of single-neurons~\cite{wang2014effect,liu2018coherence} and neural networks, including synchronization~\cite{protachevicz2020influence,fan2018autapses}, SR~\cite{yang2017autapse}, CR~\cite{yilmaz2016autapse,song2018coherence,jia2021inhibitory}, and ISR~\cite{zhang2021autapse}. However, as we pointed out earlier, no study have investigated the effects of autapses on SISR and how these effects can be combined with the network topology to enhance the coherence of oscillations induced by SISR --- this is one objective of the current paper.

Furthermore, Fries~\cite{fries2005mechanism} suggested that coherence of neural activity is conducive to neural communication. For instance, it was demonstrated that coherence is advantageous for the signal transmission between spatially separated active brain regions~\cite{benchenane2010coherent}. This communication can be achieved simultaneously at different ranges of oscillation frequency~\cite{fries2015rhythms}, which would not be possible if they show incoherent behavior. Motivated by these studies, in this work, we focus on SISR in single ML neurons, 3NMs of time-delayed coupled neurons, and how it can be controlled through autapses~\cite{van1972autapses, lubke1996frequency} and multiplexing~\cite{battiston2017structure}. In this paper, we address the following four main questions:
\begin{itemize}
\item[(i)] How does the type of autaptic connections affect the degree of SISR in a single-ML neuron?
\item[(ii)] How does the type of synaptic connections and topology of a motif affect SISR?
\item[(iii)] Can a poor degree of SISR in a motif be enhanced by autapses?
\item[(iv)] Can a poor degree of SISR in a motif be enhanced by multiplexing?
\end{itemize}

The rest of the paper is organized as follows: In section~\ref{Section 2}, we introduce the model equations. In section~\ref{Section 3}, we represent the analytical conditions necessary for the occurrence of SISR in the model. In section~\ref{Section 4}, we represent the numerical methods used in our simulations. In section~\ref{Section 5}, we present and discuss the simulation results and in section~\ref{Section 6}, we have a summary with concluding remarks.
%%%%%%%%%%%%%%%%%%%%%%%%%%%%%%%%%%%%%%%%%%%%%%%%%%%%%%%%%%%%%%%%%%%%%%%%%%%%%%%%%%%%%%%%%%

\section{Model description}\label{Section 2}
Different neural models have been used to investigate several dynamical behaviors ranging from synchronization~\cite{wouapi2020various, wouapi2021complex, boaretto2021role, yu2021synchronization} to resonance~\cite{masoliver2017coherence, liu2018coherence, lu2020inverse, wang2021effects}. In this paper, we consider a network of ML neurons with type-II excitability and driven by Gaussian processes to investigate SISR. The network is described by the following coupled stochastic delayed differential equations:
\begin{subequations}\label{eqn:1}
\begin{empheq}[left=\empheqlbrace]{align}
\frac{dv_{p,i}}{dt} & = f(v_{p,i},w_{p,i}) + \sigma_{p,i} \frac{dW_{p,i}}{dt}\label{eqn:dv_single_neuron}\\
& +\kappa^a_{\mathrm{e}}f^a_e(v_{p,i})+\kappa^a_{c}f^a_{c}(v_{p,i})\label{eqn:dv_autapse}\\[0.5mm]
& + \kappa_{\mathrm{e}}f^G_e(v_{p,i},v_{p,j})+\kappa_{c}f^G_{c}(v_{p,i},v_{p,j})\label{eqn:dv_3NM}\\[0.5mm]
& + \kappa^m_{\mathrm{e}}f^m_e(v_{p,i}, v_{q,i}) +\kappa^m_{c}f^m_{c}(v_{p,i}, v_{q,i}),\label{eqn:dv_multiplexing}\\[0.5mm]
\frac{dw_{p,i}}{dt} & = \varepsilon g(v_{p,i},w_{p,i}).\label{eqn:dw_single_neuron}
\end{empheq}
\end{subequations}
Here, the membrane potential and the recovery current variables of neuron $i$ in the motif layer $p$ are given by $v_{p,i}\in \mathbb{R}$ and $w_{p,i}\in\mathbb{R}$, respectively. To avoid confusion, one may keep in mind that the first indices, i.e., $p,q\in \{1,2\}$ ($p\neq q$), denote the motif layer in which the neuron is located, while the second indices, i.e., $i,j \in \{1,2,3\}$ ($i\neq j$), denote the $i$th and $j$th neuron within a given layer: $p$ or $q$.

Eq.~\eqref{eqn:1} in the absence of Eqs.~\eqref{eqn:dv_autapse}, \eqref{eqn:dv_3NM}, and \eqref{eqn:dv_multiplexing}, represents a single isolated stochastic ML neuron without autapses, where the deterministic nonlinear vector fields $f(v_{p,i},w_{p,i})$ in Eq.~\eqref{eqn:dv_single_neuron} and $g(v_{p,i},w_{p,i})$ in Eq.~\eqref{eqn:dw_single_neuron} are, after dropping the indices, given by 
\begin{equation}\label{eqn:2}
\begin{split}
\left\{\begin{array}{lcl}
f(v,w)&=& \overline{g}_c m_{\infty}(v)(1-v) +  \overline{g}_l(v_l-v)\\
&+&\overline{g}_k w (v_k-v),\\[1.0mm]
g(v,w) &=&  \displaystyle{\mathrm{cosh}\left(\frac{v-v_3}{v_4}  \right)\left(w_{\infty}(v)- w\right)},
\end{array}\right.
\end{split}
\end{equation}
where the nonlinearities get in via $m_{\infty}(v)$ and $w_{\infty}(v)$, each given by
\begin{equation}\label{eq:3}
\begin{split}
\left\{\begin{array}{lcl}
m_{\infty}(v)= \displaystyle{\frac{1}{2}\left[1+ \mathrm{tanh}\left(\frac{v-v_1}{v_2}  \right)\right]},\\[4.0mm]
w_{\infty}(v)= \displaystyle{\frac{1}{2}\left[1+ \mathrm{tanh}\left(\frac{v-v_3}{v_4}  \right)\right]},
\end{array}\right.
\end{split}
\end{equation}
with $\overline{g}_c=1.0$, $\overline{g}_k=1.0$, and $\overline{g}_l=0.1$ representing the conductances; and $v_k=-2.0$, $v_1=0.0$, $v_2=0.36$, $v_3=-0.2$, and $v_4=0.52$ representing constant parameters \cite{liu2014bifurcation}. The excitability parameter $v_l$ is a codimension-one Hopf bifurcation parameter for the ML neuron. And $0<\varepsilon\ll1$ is a small positive parameter
that sets the timescale separation between the fast membrane potential and the slow recovery
current variables. 

$dW_{p,i}/dt$ are uncorrelated Gaussian white noises, that is the formal derivative of Brownian motion with $\langle dW_{p,i}(t),dW_{p,i}(t')\rangle_t =\delta(t-t')$ and variance (intensity) $\sigma_{p,i}$. For the sake of simplicity, we assume that the noise intensities $\sigma_{p,i}$ within a given layer $p$ are all the same, i.e., we choose $\sigma_{p,1}=\sigma_{p,2}=\sigma_{p,3}$. The parameters $\sigma_{p,i}$ and $\varepsilon$ are crucial for the occurrence of SISR.

\begin{figure*}%[ht]
\centering
\subfigure[Isolated neurons without and with autapses indicated by the feedback loops.]{\includegraphics[width=0.4750\textwidth]{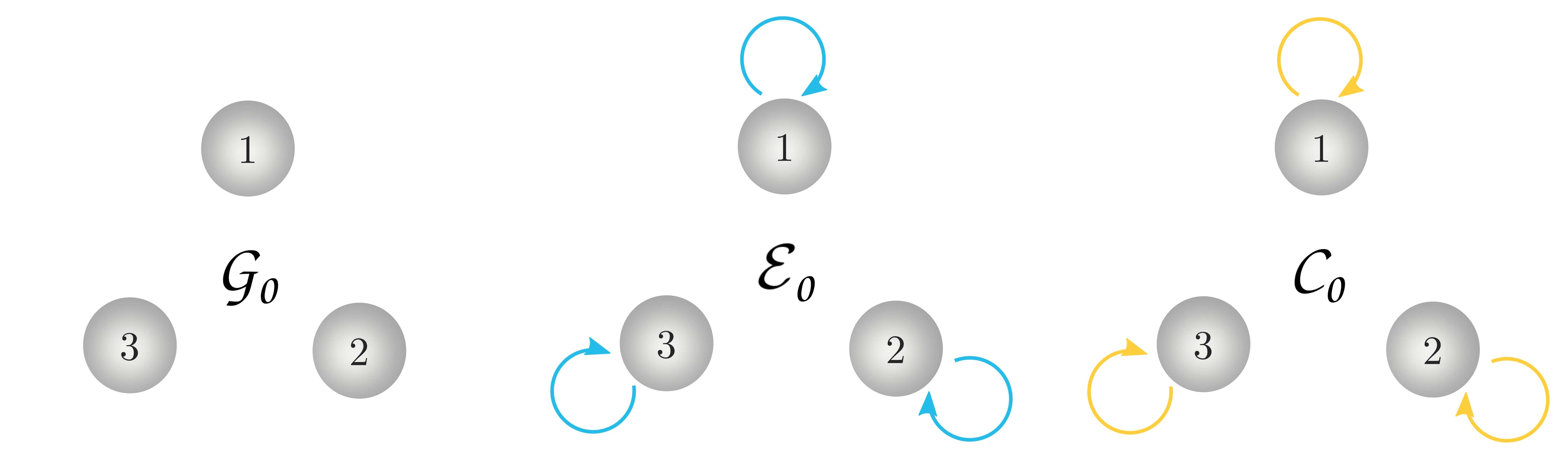}}
\hspace{2.0cm}
\subfigure[Bidirectional electrical synapses.]
{\includegraphics[width=0.2750\textwidth]{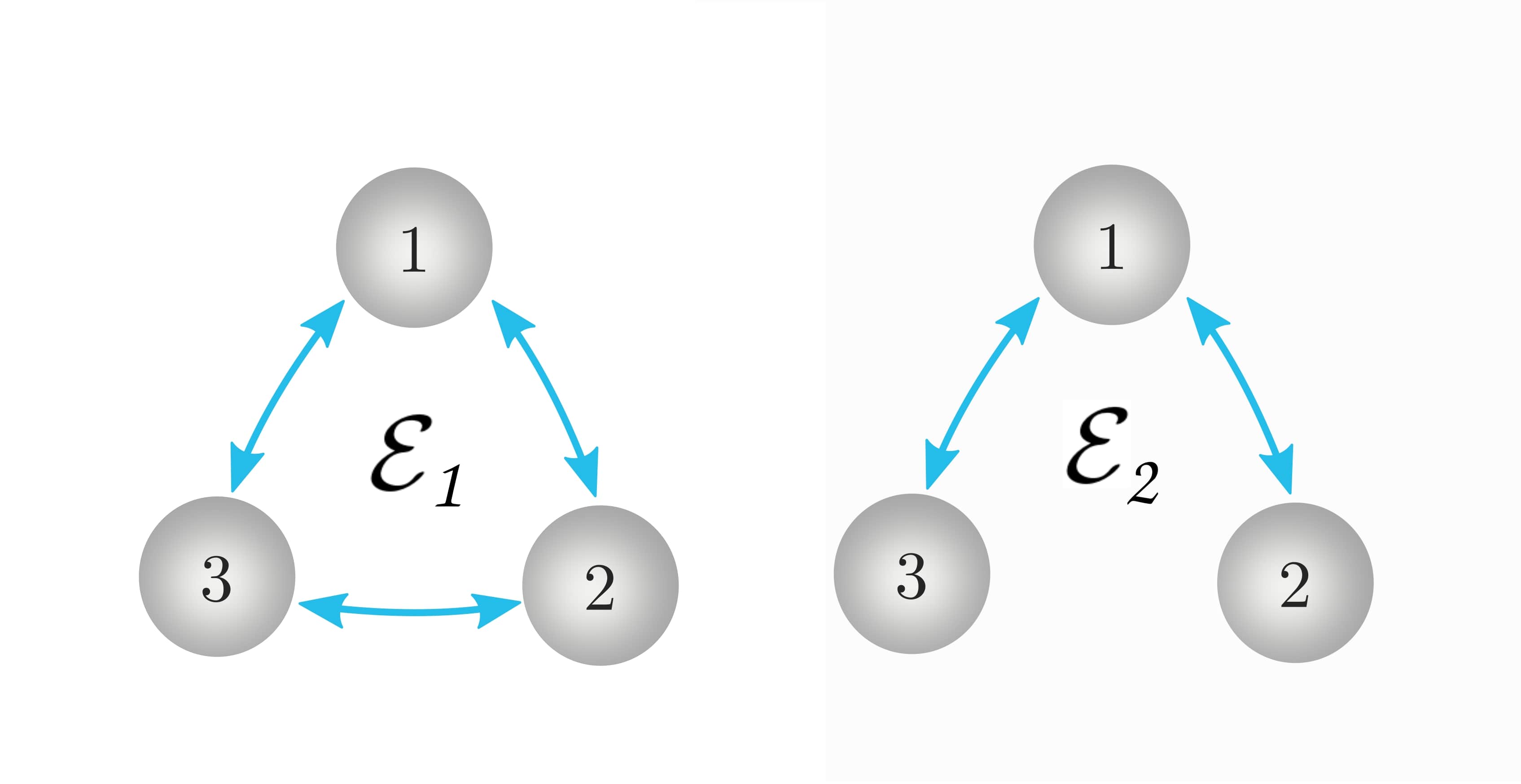}}\hspace{2.5cm}
\subfigure[Unidirectional chemical synapses.]
{\includegraphics[width=0.400\textwidth]{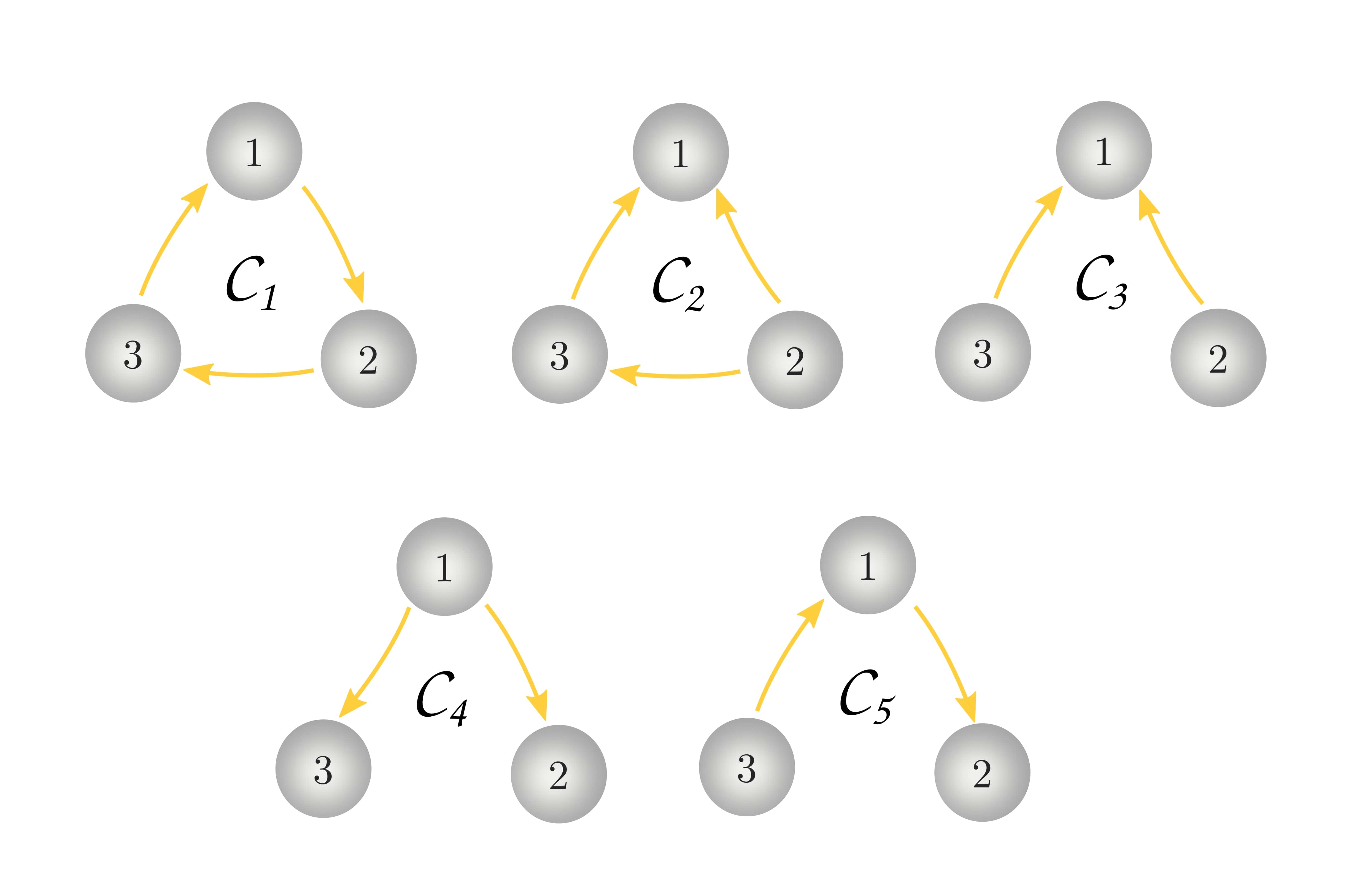}}\hspace{3.50cm}
\subfigure[Double unidirectional chemical synapses]{\raisebox{1.1cm}
{\includegraphics[width=0.2750\textwidth]{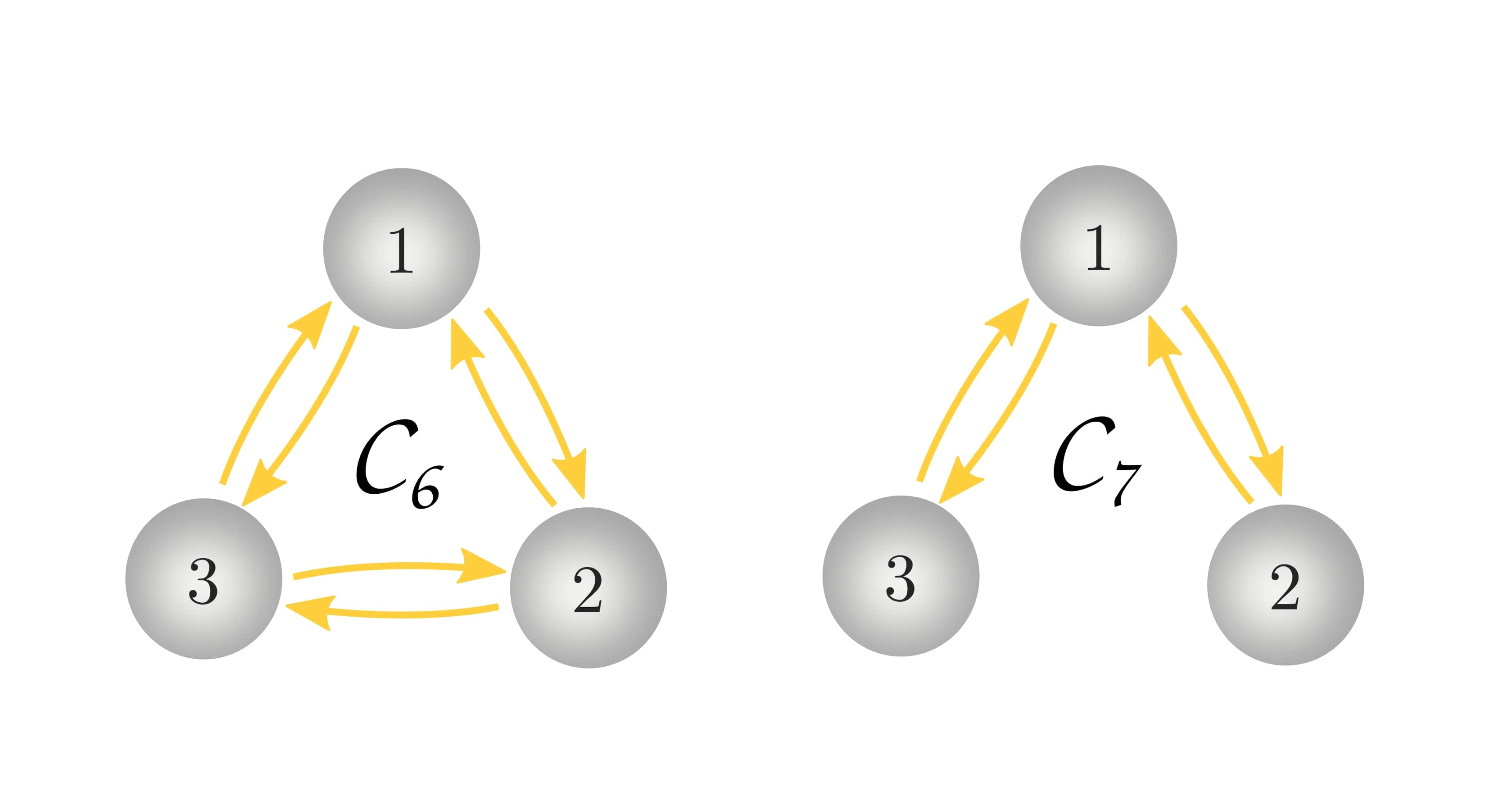}}}
\subfigure[Multiplex networks consisting of $\mathcal{C}_2-\mathcal{C}_2$ and $\mathcal{C}_2-\mathcal{C}_3$ motif layers with multiplexing connections mediated by bidirectional electrical synapses (in blue) or unidirectional inhibitory chemical synapses (in yellow) in each case. The green motif layers represent the ones with a high degree of SISR and the red ones represent the motifs with a poor degree of SISR.]{\raisebox{0.1cm}
{\includegraphics[width=0.7000\textwidth]{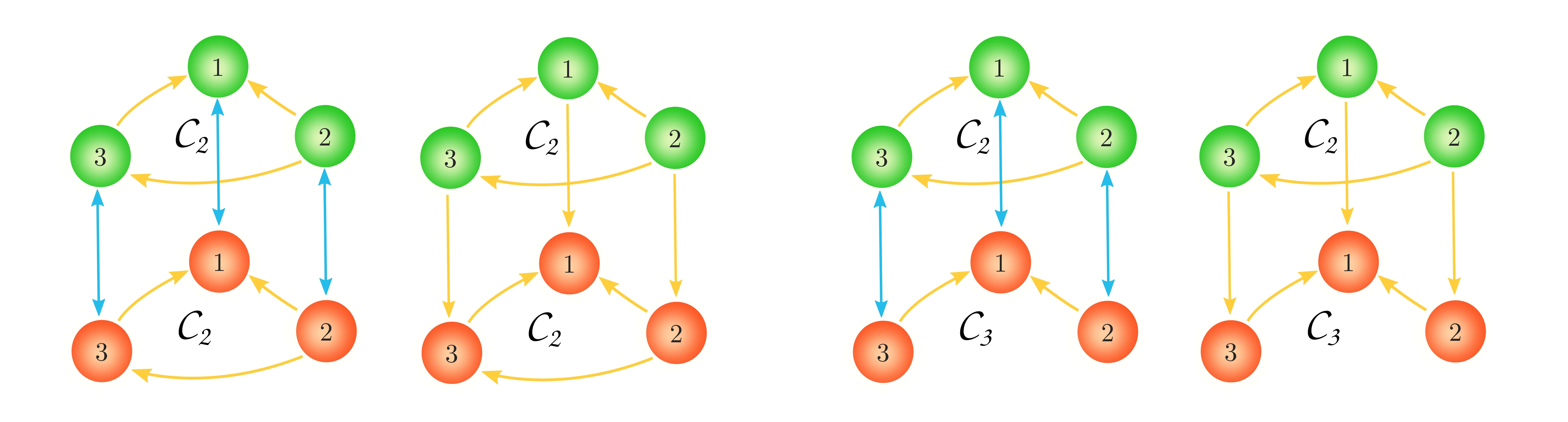}}}
\caption{Schematic representations of the network topologies investigated in this work, indicating the types and configurations of the synaptic connections.}
\label{fig:network_motifs_topologies}
\end{figure*}

Fig.\,\ref{fig:network_motifs_topologies} shows the specific motif layer networks and the multiplex network configurations, including isolated neurons, that will be considered in this work. In Eq.~\eqref{eqn:dv_autapse}, $\kappa^a_{\mathrm{e}}f^a_e(v_{p,i})$ and $\kappa^a_{c}f^a_{c}(v_{p,i})$ respectively represent the electrical and chemical autaptic terms of the $i$th neuron in the $p$th layer. Here, $\kappa^a_{\mathrm{e}}$ represents the strength of the electrical autapse and $\kappa^a_{c}$ the strength of the chemical autapse. Furthermore, $f^a_e(v_{p,i})$ and $f^a_c(v_{p,i})$ are respectively given by the well-known \cite{iqbal2017modeling, wang2006chaos, autaptic_regulation_ying, destexhe1994efficient, destexhe1998kinetic, greengard2001neurobiology} forms of the electrical and chemical autapses:
\begin{equation}\label{eqn:4}
\begin{split}
\left\{\begin{array}{lcl}
f^a_e(v_{p,i}) &=& \big(v_{p,i}(t-\tau^a_e) - v_{p,i}(t)\big),\\[2.0mm]
f^a_c(v_{p,i}) &=& \displaystyle{ \frac{\big(v_{p,i}(t)-v_{\mathrm{syn}}\big)}{1+ e^{-\lambda\big[v_{p,i}(t-\tau^a_c) -\theta_{\mathrm{syn}}\big]}}}.
\end{array}\right.
\end{split}
\end{equation}

In Eq.~\eqref{eqn:dv_3NM}, $\kappa_{\mathrm{e}}f^G_e(v_{p,i},v_{p,j})$ and  $\kappa_{c}f^G_{c}(v_{p,i},v_{p,j})$ respectively represent the electrical interaction (gap junctions) between the $i$th and $j$th neurons in layer $p$ and the chemical interaction from the $i$th to the $j$th neurons in the $p$th layer. Here, $\kappa_{\mathrm{e}}$ and $\kappa_{c}$ represents the strength of the electrical and chemical couplings, respectively. The terms 
$f^G_{\mathrm{e}}(v_{p,i},v_{p,j})$ and $f^G_{c}(v_{p,i},v_{p,j})$ are given by
\begin{equation}\label{eqn:5}
\begin{split}
\left\{\begin{array}{lcl}
f^G_e(v_{p,i},v_{p,j}) &=&\sum\limits_{j\neq i}\mathcal{G}_{ij}\big(v_{p,i}(t-\tau_e) - v_{p,j}(t)\big),\\[2.0mm]
f^G_c(v_{p,i},v_{p,j}) &=&\sum\limits_{j\neq i}\mathcal{G}_{ij}\displaystyle{ \frac{\big(v_{p,i}(t)-v_{\mathrm{syn}}\big)}{1+e^{-\lambda\big[v_{p,j}(t-\tau_c) -\theta_{\mathrm{syn}}\big]}}}.
\end{array}\right.
\end{split}
\end{equation}

In Eq.~\eqref{eqn:dv_multiplexing}, $\kappa^m_{\mathrm{e}}f^m_e(v_{p,i}, v_{q,i})$ and $\kappa^m_{c}f^m_{c}(v_{p,i},v_{q,i})$ respectively represent the electrical and chemical interactions (multiplexing) between the $i$th neuron in layer $p$ and the $i$th neuron in layer $q$. That is, in multiplex networks, connections exist only between replica neurons. Similarly, $\kappa^m_{\mathrm{e}}$ and $\kappa^m_{c}$ represent the strengths of the electrical and chemical multiplexing, respectively. Here, $f^m_e(v_{p,i}, v_{q,i})$ and $f^m_{c}(v_{p,i},v_{q,i})$ are given by
\begin{equation}\label{eqn:6a}
\begin{split}
\left\{\begin{array}{lcl}
f^m_e(v_{p,i},v_{q,i}) &=&\big(v_{q,i}(t-\tau^m_e) - v_{p,i}(t)\big),\\[2.0mm]
f^m_c(v_{p,i},v_{q,i}) &=&\displaystyle{ \frac{\big(v_{p,i}(t)-v_{\mathrm{syn}}\big)}{1+e^{-\lambda\big[v_{q,i}(t-\tau^m_c) -\theta_{\mathrm{syn}}\big]}}}.
\end{array}\right.
\end{split}
\end{equation}

It is worth noting from Fig.\,\ref{fig:network_motifs_topologies} and Eqs.~\eqref{eqn:5} and \eqref{eqn:6a} that the electrical synapses are always bidirectional (represented in Fig.\,\ref{fig:network_motifs_topologies}\textbf{(b)} by the blue links with double arrow going from the $i$th to the $j$th neuron and also back, i.e., from the $j$th to the $i$th neuron). On the other hand, chemical synapses can either be (i) a single unidirectional connection (i.e., for Eq.~\eqref{eqn:5}, it is represented in Fig.\,\ref{fig:network_motifs_topologies}\textbf{(c)} by the yellow links with a single arrow from the $i$th to the $j$th neuron; and for Eq.~\eqref{eqn:6a}, it is represented in Fig.\,\ref{fig:network_motifs_topologies}\textbf{(e)} by the single yellow links with a single arrow going from the $i$th neuron in the $p$th layer to the $i$th neuron in the $q$th layer) or 
(ii) a double unidirectional connection (i.e., for Eq.~\eqref{eqn:5}, it is represented in Fig.\,\ref{fig:network_motifs_topologies}\textbf{(d)} by two yellow links each representing a single unidirectional chemical connection, one going from the $i$th to the $j$th neuron and the other from the $j$th to the $i$th neuron.
This kind of reciprocal connection is a very common and universal design principle of biological nervous systems~\cite{ markram1997network,pitkanen2000reciprocal,song2005highly,zupanc2005reciprocal,perin2011synaptic, bastos2012canonical}.
For Eq.~\eqref{eqn:6a} (i.e., the coupling terms between two motif layers) we consider one chemical unidirectional connection from the $i$th neuron in the $p$th layer to the $i$th neuron in the $q$th layer and the other from the $i$th neuron in the $q$th layer to the $i$th neuron in the $p$th layer). The electrical and chemical autapses given by Eq.~\eqref{eqn:4} are self-loops on each neuron and are represented in Fig.\,\ref{fig:network_motifs_topologies}\textbf{(a)} by the blue and yellow loops, respectively.

In Eq.~\eqref{eqn:4}, $\tau^a_e$ and $\tau^a_c$ respectively represent the electrical and chemical autaptic time delays. In Eq.~\eqref{eqn:5}, $\tau_e$ and $\tau_c$ respectively represent the time delays involved in the electrical and chemical interactions between neurons within the same layer. While in Eq.~\eqref{eqn:6a}, $\tau^m_e$ and $\tau^m_c$ represent the electrical and chemical multiplexing time delays, respectively. Moreover, the parameter $\lambda$ (which is fixed at $\lambda=5.0$ in this work) determines the slope of the sigmoidal input-output function $\Gamma(v)=\displaystyle{ 1/(1+e^{-\lambda(v -\theta_{\mathrm{syn}})}})$, where $\theta_{\mathrm{syn}}$ represents the synaptic firing threshold (which is fixed at $\theta_{\mathrm{syn}}=0.0$ in this work). And $v_{\mathrm{syn}}$ represents the synaptic reversal potential. 

When $v_{\mathrm{syn}} < v_{p,i}$, the chemical interaction has a depolarizing effect which makes the synapse inhibitory, and when $v_{\mathrm{syn}} > v_{p,i}$, the chemical interaction has a hyper-polarizing effect, making the synapse excitatory. This means that we can choose a value for $v_{\mathrm{syn}}$ such that the inhibitory and excitatory nature of the chemical synapse is determined only by the sign in front of the chemical coupling strengths $\kappa^a_c$, $\kappa_c$, and $\kappa^m_c$. For the ML neuron model used in this study, the membrane potential variables are certainly bounded as: $-1.4<v_{p,i}(t)<2.5$ ($p = 1,2$; $i = 1, 2, 3$) for all time t. We fix $v_{\mathrm{syn}} = -1.5$ (maintained throughout our computations), a value with which the term $(v_{p,i}(t) - v_{\mathrm{syn}})$ is always positive. In this way, a positive sign in front of $\kappa^a_c$, $\kappa_c$ and $\kappa^m_c$ will always make these chemical synapses excitatory, represented by $\kappa^a_{\mathrm{c,exc}}$, $\kappa_{\mathrm{c,exc}}$, and $\kappa^m_{\mathrm{c,exc}}$, respectively. While a negative sign in front of $\kappa^a_c$, $\kappa_c$ and $\kappa^m_c$ will always make these chemical synapses inhibitory, represented $\kappa^a_{\mathrm{c,inh}}$, $\kappa_{\mathrm{c,inh}}$, and $\kappa^m_{\mathrm{c,inh}}$, respectively.

The matrix $\mathcal{G}$ in Eq.~\eqref{eqn:5} represents the adjacency matrix of a motif layer network. The entry $\mathcal{G}_{ij}$ is 1 if the $i$th neuron is connected to the $j$th neuron and 0 otherwise.

\section{Deterministic predisposition and necessary conditions for SISR}\label{Section 3}
A ML neuron with a unique and stable fixed point and in the complete absence of random perturbations (or even in the presence of a sub-threshold deterministic perturbation) cannot maintain a self-sustained oscillation (i.e., no limit cycle solution can emerge). One says in this case that the neuron is in the excitable regime \cite{izhikevich2000neural}. The predisposition state for the occurrence of SISR in an isolated neuron and in a network of neurons is precisely excitability. In an excitable state, choosing an initial condition in the basin of attraction of the unique and stable fixed point will result in at most one large non-monotonic excursion in the phase space after which the trajectory returns exponentially fast to this fixed point and stays there until a disturbance like a random perturbation is introduced in the neuron model.

\begin{figure}%[H]
    \centering
    \raisebox{0.02cm}{\subfigure[]{\includegraphics[width=0.248\textwidth]{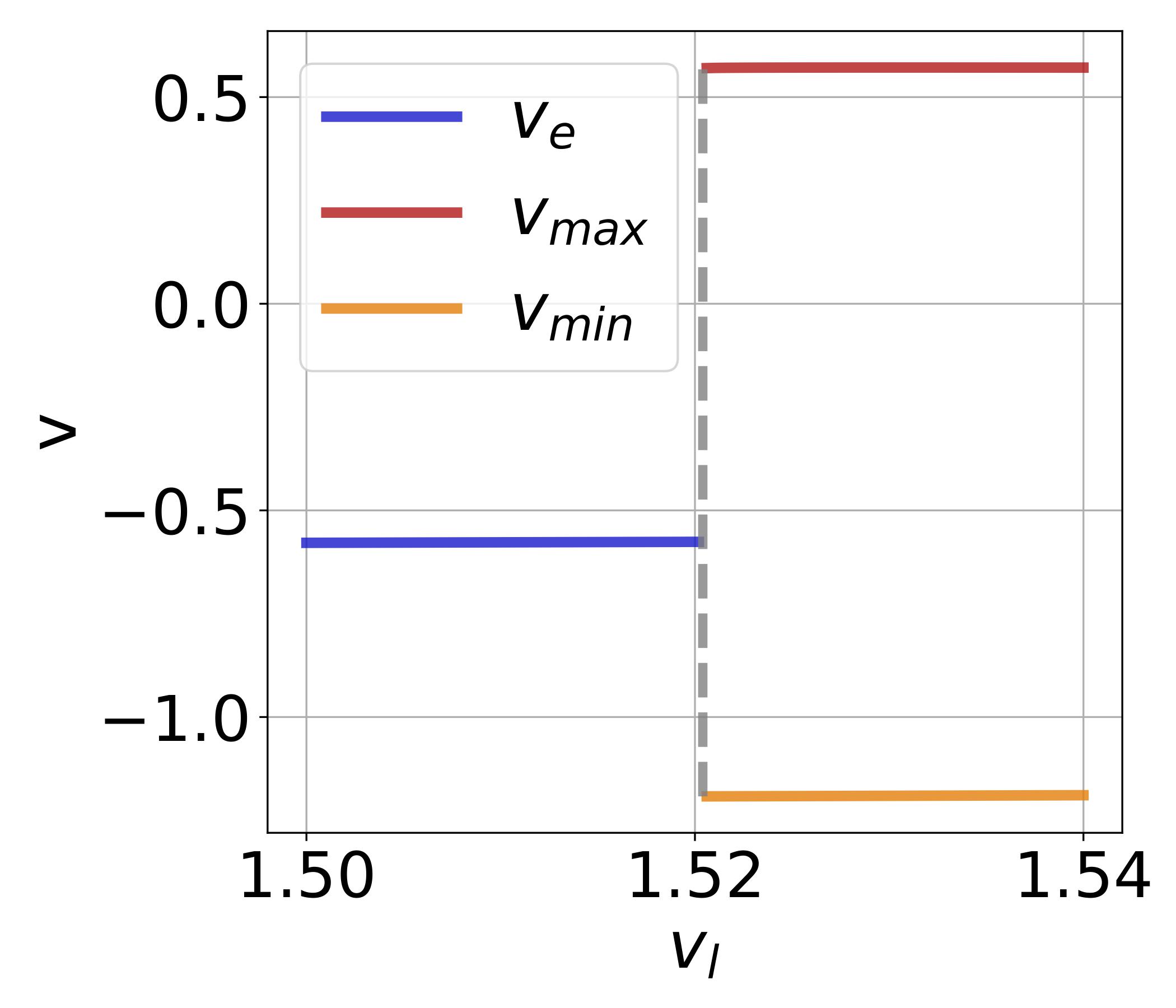}}}
    \subfigure[]{\includegraphics[width=0.21\textwidth]{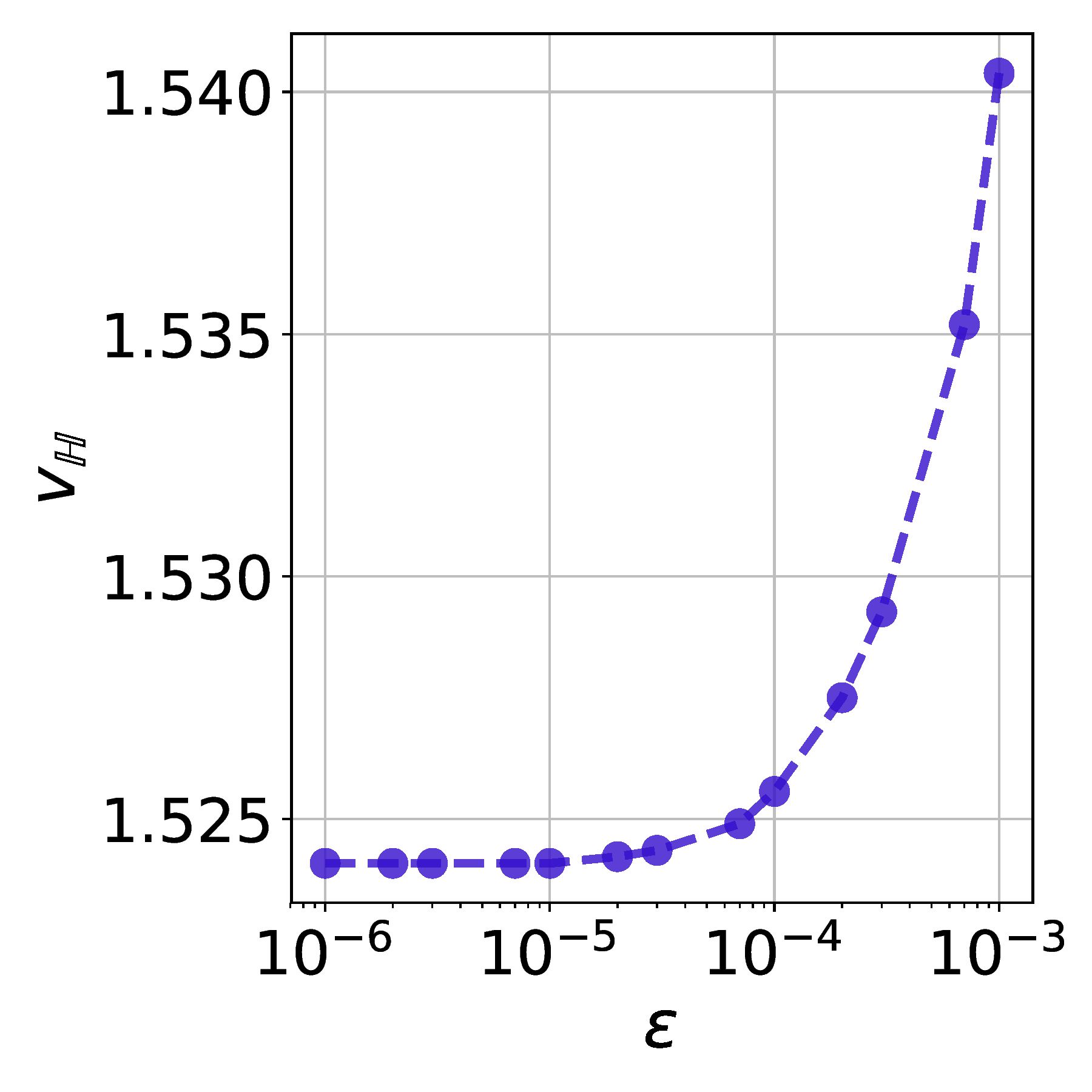}}
    \subfigure[]{\includegraphics[width=0.47\textwidth]{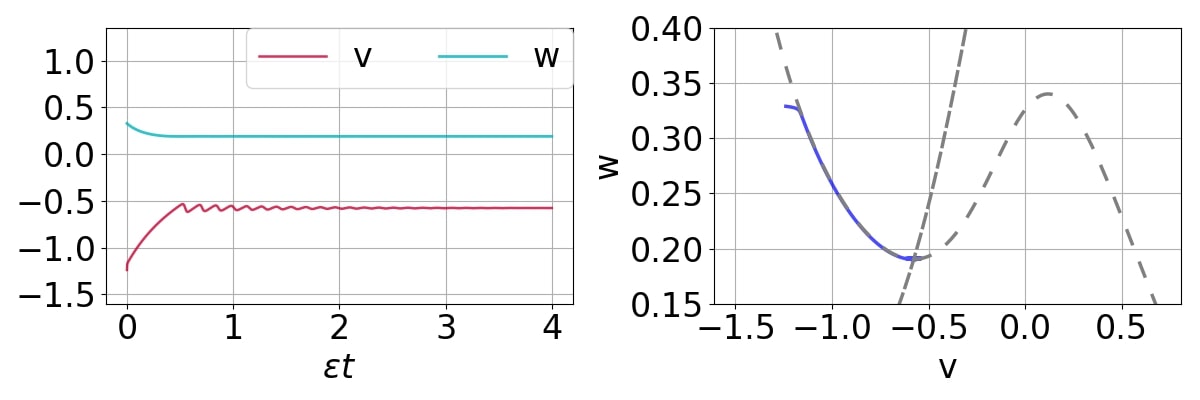}}
    \subfigure[]{\includegraphics[width=0.47\textwidth]{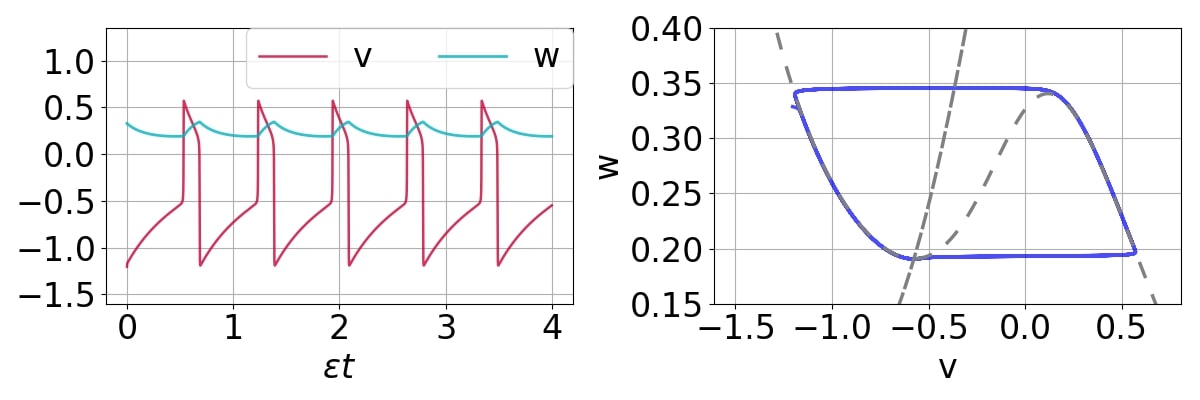}}
    \caption{Panel \textbf{(a)}: Bifurcation diagram of a single isolated ML neuron without autapses against the excitability parameter $v_l$ at a fixed timescale separation parameter $\varepsilon=0.0005$. The Hopf bifurcation value separating the excitable regime (blue line) and oscillatory regime (orange line) is indicated by the dashed-gray vertical line located at $v_l=v_{\mathbb{H}}(\varepsilon=0.0005)=1.52010$. Panel \textbf{(b)}: Variation of the Hopf bifurcation value with the timescale parameter $\varepsilon$. To avoid oscillatory regime due to Hopf bifurcation, we fixed $v_l=1.515$ in all simulations. Panels \textbf{(c)} and \textbf{(d)} show, each, a time series (left) and the associated phase portrait (right) of a trajectory in the excitable regime $v_l=1.515<v_{\mathbb{H}}$ and in the oscillatory regime $v_l=1.525>v_{\mathbb{H}}$, respectively. }
    \label{fig:bifurcation_diagram and variation of hopf}
\end{figure}

In a single isolated ML neuron without autapses, the excitability parameter is $v_l$. Fig.\,\ref{fig:bifurcation_diagram and variation of hopf}\textbf{(a)} shows a bifurcation diagram of a single ML neuron without autapses at a fixed timescale separation parameter $\varepsilon=0.0005$. It is worth noting that the real part of the eigenvalue of the Jacobian matrix associated to a single isolated ML without autapses depends on the timescale separation parameter $\varepsilon$. Hence, the Hopf bifurcation value of the isolated ML neuron without autapses also depends on $\varepsilon$. We observe that by varying the Hopf bifurcation parameter $v_l$ in the interval $[1.50,1.52010)$, the membrane potential $v$ stays at a constant value (i.e., at the unique and stable fixed point $v_e$) represented by the blue horizontal line. At the Hopf bifurcation value $v_l=v_{\mathbb{H}}(\varepsilon=0.0005)=1.52010$, the stable and unique fixed point bifurcates into a stable limit cycle, represented by the dashed-gray vertical line. The top and bottom orange horizontal lines represent the maximum and the minimum values of the limit cycle oscillation, respectively. 

Fig.\,\ref{fig:bifurcation_diagram and variation of hopf}\textbf{(b)} shows the variation of the Hopf bifurcation value $v_{\mathbb{H}}$ with timescale separation parameter $\varepsilon$. We observe that the Hopf bifurcation value is non-linearly proportional to $\varepsilon\ll 1$, and for $10^{-6}\leq \varepsilon \leq 10^{-5}$, the Hopf bifurcation value remains constant at $v_{\mathbb{H}}=1.524$. For this reason, we keep the isolated ML neuron without autapses in the excitable regime by fixing $v_l$ at $v_l=1.515$ throughout this work, so that we are sure that coherent oscillations due to a Hopf bifurcation cannot occur in our simulations.

Figs.\,\ref{fig:bifurcation_diagram and variation of hopf}\textbf{(c)} and \textbf{(d)} show the time series and the corresponding phase portrait of a trajectory when we choose $v_l$ in the excitable regime and in the oscillatory regime, respectively. In the phase portraits, the $S$-shaped curve corresponds to the $v$-nullcline which intersects the $w$-nullcine at a single point, i.e., the unique and stable (unstable in Fig.\,\ref{fig:bifurcation_diagram and variation of hopf}\textbf{(d))} fixed point.

It is worth pointing out that in the deterministic single isolated ML neuron with autapse and the deterministic networks of ML neurons with and without autapses, pairs of new parameters, i.e., $(\kappa^a_e, \tau^a_e)$, $(\kappa^a_c, \tau^a_c)$, $(\kappa_e, \tau_e$), $(\kappa_c, \tau_c)$, $(\kappa^m_e, \tau^m_e)$, $(\kappa^m_c, \tau^m_c)$, enter the bifurcation dynamics. It is therefore very important to also check that, after fixing $v_l=1.515$, these additional set of parameters does not shift the deterministic system into the oscillatory regime via Hopf or saddle-node onto limit cycle bifurcations \cite{scholl2009time,yamakou2019control,yamakou2020optimal}.

We recall that SISR is the occurrence of a limit cycle behavior (coherent oscillations) due to solely the presence of noise and not because of the occurrence of deterministic bifurcations onto limit cycles.
To avoid such deterministic oscillatory regimes, we will always first check that the values of all autaptic, intra-motif, and inter-motif delayed coupling parameters keep the considered system in the excitable regime.

The adiabatic limit $\varepsilon\to0$ in Eq.\,\eqref{eqn:1} reduces Eq.\,\eqref{eqn:dw_single_neuron} to $dw_{p,i}/dt\approx0$. This means that in this limit, the $w_{p,i}$-variables of neurons are frozen, reducing Eq.\,\eqref{eqn:1} into a set of coupled Langevin equations given by
\begin{equation}\label{langevin}
 \frac{dv_{p,i}}{dt} = -\frac{\partial U(v_{p,i},w_{p,i})}{\partial v_{p,i}} + \sigma_{p,i} \frac{dW_{p,i}}{dt},\\   
\end{equation}
where $U(v_{p,i},w_{p,i})$ is the double well potential defined in Eq.\,\eqref{eqn:7}, and in which $w_{p,i}$ is essentially constant.

The conditions (based on large deviation theory \cite{freidlin2001stochastic,freidlin2001stable} and Kramers' law \cite{kramers1940brownian})) necessary for the occurrence of SISR in stochastic slow-fast dynamical systems in the form of Eq.\,\eqref{eqn:1} (the so-called standard form \cite{kuehn2015multiple}) are well established \cite{muratov2005self, deville2007nontrivial,SISRvsCR,yamakou2019control,yamakou2020optimal} and are quite generic for Gaussian noise. 

In Eq.\,\eqref{eqn:6}, we write down these conditions (see, e.g., \cite{yamakou2019control}) for the motif layer $p$ of  Eq.\,\eqref{eqn:1}:
\begin{equation}\label{eqn:6}
\begin{split}
\left\{\begin{array}{lcl}
v_l-v_{\mathbb{H}}<0,\\[1.0mm]
\displaystyle{\lim\limits_{(\sigma_{p,i},\varepsilon)\to(0,0)}\bigg(\frac{\sigma_{p,i}^2}{2}\ln(\varepsilon^{-1})\bigg)\in\Big[\Delta U^l_i(w_e), F_p(\cdot)\Big]},\\[4.0mm]
\displaystyle{\lim\limits_{(\sigma_{p,i},\varepsilon)\to(0,0)}\bigg(\frac{\sigma_{p,i}^2}{2}\ln(\varepsilon^{-1})\bigg)=\mathcal{O}(1)},
\end{array}\right.
\end{split}
\end{equation}
where $F_p$, defined in Eq.\,\eqref{eqn:8}, is a function of the parameters $(\kappa^a_e, \tau^a_e)$, $(\kappa^a_c, \tau^a_c)$, $(\kappa_e, \tau_e)$, $(\kappa_c, \tau_c)$, $(\kappa^m_e, \tau^m_e)$, $(\kappa^m_c, \tau^m_c)$; and $\Delta U^l_i(w_e)$, defined in Eq.\,\eqref{eqn:9}, is the left energy barrier (as opposed to the right energy barrier $U^r_i(w_{p,i})$, both obtained in the adiabatic limit $\varepsilon \to 0$) of a double well potential $U(v_{p,i},w_{p,i})$ (see Eq.\,\eqref{eqn:7}) at the $w_{p,i}$-coordinate of the unique and stable fixed point $(v_e, w_e)$ of Eq.\,\eqref{eqn:1}, in the absence of noise.
\begin{align}\label{eqn:7}
\begin{split}
U(v_{p,i},w_{p,i}) & =-\int\Big[ f(v_{p,i},w_{p,i}) +\kappa^a_{\mathrm{e}}f^a_e(v_{p,i})\\ &+\kappa^a_{c}f^a_{c}(v_{p,i}) + \kappa_{\mathrm{e}}f^G_e(v_{p,i},v_{p,j}) \\ & + \kappa_{c}f^G_{c}(v_{p,i},v_{p,j}) +  \kappa^m_{\mathrm{e}}f^m_e(v_{p,i}, v_{q,i})\\& + \kappa^m_{c}f^m_c(v_{p,i}, v_{q,i}) \Big]dv_{p,i}.
\end{split}
\end{align}
\begin{equation}\label{eqn:8}
\begin{split}
F_p&:=\Big\{(\kappa^a_e, \tau^a_e), (\kappa^a_c, \tau^a_c), (\kappa_e, \tau_e),(\kappa_c, \tau_c),\\&(\kappa^m_e, \tau^m_e), (\kappa^m_c, \tau^m_c) : \Delta U^l_i(w_{p,i}) = \Delta U^r_i(w_{p,i})\Big\},
\end{split}
\end{equation}
where
\begin{equation}\label{eqn:9}
\begin{split}
\left\{\begin{array}{lcl}
\Delta U^l_i(w_{p,i})&:=& U\big(v^{*}_{0}(w_{p,i}),w_{p,i}\big) - U\big(v^{*}_{l}(w_{p,i}),w_{p,i}\big),\\[2.0mm]
\Delta U^r_i(w_{p,i})&:=& U\big(v^{*}_{0}(w_{p,i}),w_{p,i}\big) - U\big(v^{*}_{r}(w_{p,i}),w_{p,i}\big),
\end{array}\right.
\end{split}
\end{equation}
with
\begin{equation}\label{eqn:10}
\begin{split}
v^{*}_{l,0,r}(w_{p,i})&:=\Big\{v_{p,i}: f(v_{p,i},w_{p,i}) +\kappa^a_{\mathrm{e}}f^a_e(v_{p,i})\\ 
& +\kappa^a_{c}f^a_{c}(v_{p,i}) + \kappa_{\mathrm{e}}f^G_e(v_{p,i},v_{p,j})\\
&+\kappa_{c}f^G_{c}(v_{p,i},v_{p,j}) + \kappa^m_{\mathrm{e}}f^m_e(v_{p,i}, v_{q,i})\\
& +\kappa^m_{c}f^m_{c}(v_{p,i}, v_{q,i})=0 \Big\}.
\end{split}
\end{equation}
The sets of solution $v^{*}_{l}(w_{p,i})$, $v^{*}_{0}(w_{p,i})$, and $v^{*}_{r}(w_{p,i})$ in Eq.\,\eqref{eqn:10} are such that $v^{*}_{l}(w_{p,i})<v^{*}_{0}(w_{p,i})<v^{*}_{r}(w_{p,i})$, define the left stable, middle unstable, and right stable branches of the $S$-shaped $v$-nullcline of the ML neuron model, respectively.

The theoretical result given in Eq.\,\eqref{eqn:6} can briefly be interpreted as follow: the first expression (i.e., $v_l\,-\,v_{\mathbb{H}}\,<\,0$) requires the system to be in the excitable regime, i.e., a parameter regime where the zero-noise (deterministic) dynamics does not display a limit cycle nor even its precursor. This condition means that SISR can arise when the parameters are bounded away from bifurcation thresholds (this is in contrast to CR, see, e.g., \cite{deville2005two,yamakou2019control}). The second expression in Eq.\,\eqref{eqn:6} shows that the coherence (regularity) of the spiking created by the noise has a non-trivial dependence on the noise amplitude and the time-scale ratio between fast excitatory variables and slow recovery variables. This expression means that the spiking of the neural system will become more coherent if in the double limit $(\sigma_{p,i},\varepsilon) \to (0,0)$, the quantity $\big[\frac{\sigma_{p,i}^2}{2}\ln(\varepsilon^{-1})\big]$ stays within the interval $\big[\Delta U^l_i(w_e), F_p(\cdot)\big]$. The last expression of Eq.\,\eqref{eqn:6} requires that, in the double limit $(\sigma_{p,i},\varepsilon) \to (0,0)$, the quantity $\big[\frac{\sigma_{p,i}^2}{2}\ln(\varepsilon^{-1})\big]$ be as far away as possible from the boundaries of this interval. This last requirement ensures that the trajectories do not spend too much time in the wells of the double-well potential given in Eq.\,\eqref{eqn:7}, and hence destroy the regularity of the spiking. Therefore, the non-occurrence or strength of  SISR, if it occurs, depends on whether (or to what extend) the chosen values of the system parameters $\{(\kappa^a_e,\tau^a_e), (\kappa^a_c,\tau^a_c),\\ (\kappa_e, \tau_e), (\kappa_c, \tau_c), (\kappa^m_e, \tau^m_e), (\kappa^m_c, \tau^m_c) \}$ satisfies the expressions in Eq.\,\eqref{eqn:6} in the double limit $(\sigma_{p,i},\varepsilon) \to (0,0)$.

Using the theoretical result in Eq.\,\eqref{eqn:6}, we calculate the minimum ($\sigma_{min}$) and maximum ($\sigma_{max}$) noise amplitude between which the degree of SISR is high as follows:
\begin{equation}\label{eqn:13a}
\begin{split}
\left\{\begin{array}{lcl}
\sigma_{min}=\displaystyle{\sqrt{\frac{2\Delta U^l_i(w_e)}{\ln(\varepsilon^{-1})}}},\\[4.0mm]
\sigma_{max}=\displaystyle{\sqrt{\frac{2F(\cdot)}{\ln(\varepsilon^{-1})}}},
\end{array}\right.
\end{split}
\end{equation}
where $\sigma_{min}$ and $\sigma_{max}$ get their dependence on the parameters $(\kappa^a_e, \tau^a_e)$, $(\kappa^a_c, \tau^a_c)$, $(\kappa_e, \tau_e)$, $(\kappa_c, \tau_c)$, $(\kappa^m_e, \tau^m_e)$, and $(\kappa^m_c, \tau^m_c)$ from $U(v_{p,i},w_{p,i})$ and $v^{*}_{l,0,r}(w_{p,i})$. 
Later, we shall return to Eq.\,\eqref{eqn:13a}, when we will use the expressions of $\sigma_{min}$ and $\sigma_{max}$ to provide theoretical explanations to some of our numerical results.

It is also worth noting that the corresponding conditions for the occurrence of SISR in a single isolated neuron with or without autapses and in a single isolated motif layer network without multiplexing can be easily obtained by setting the corresponding autaptic ($\kappa^a_e$ and/or $\kappa^a_c$) and multiplexing ($\kappa^m_e$ and/or $\kappa^m_c$) coupling strengths in Eqs.\eqref{eqn:6}--\eqref{eqn:10} to zero.

To answer the main questions we are interested in (see the introductory section), we fix  $v_l=1.515<v_{\mathbb{H}}$ and choose the relevant coupling strengths and the associated time delays such that Eq.\,\eqref{eqn:1} is in the excitable regime. We also choose a sufficiently small timescale separation parameter, i.e., $\varepsilon=0.0005\ll1$, a weak noise intensity interval, i.e., $0<\sigma_{p,i}\ll1$, and then numerically identify the combined values of $\{(\kappa^a_e, \tau^a_e),
(\kappa^a_c, \tau^a_c),\\(\kappa_e, \tau_e), (\kappa_c, \tau_c), (\kappa^m_e, \tau^m_e), (\kappa^m_c, \tau^m_c)\}$ which satisfy (or at least to some degree) or not the scaling limit conditions in Eq.\,\eqref{eqn:6}. 

\section{Numerical method for integration}\label{Section 4}
In this work, the coefficient of variation ($\mathrm{CV}$) \cite{pikovsky1997coherence, masoliver2017coherence} will be used to measure the degree of coherence of spiking induced via the mechanism of SISR and hence, the extent of satisfaction of Eq.\,\eqref{eqn:6}, when the various synaptic strengths and time delay parameters are varied. $\mathrm{CV}$ is an important statistical measure based on the time intervals between spikes \cite{pikovsky1997coherence, masoliver2017coherence} and which is related to the timing precision of information processing in neural systems \cite{pei1996noise}. When $\mathrm{CV}=0$, the neural system exhibits a deterministic periodic spiking, a value that we cannot reach in our model due to the presence of noise. In the double limit $(\sigma_{p,i},\varepsilon) \to (0,0)$, the coherence of the spiking due to SISR increases as $\mathrm{CV}\rightarrow 0$, i.e., as $\big[\frac{\sigma_{p,i}^2}{2}\ln(\varepsilon^{-1})\big]$ tends to the mid-point of the interval $\big[\Delta U^l_i(w_e), F_p(\cdot)\big]$ as the parameter values change. When $\mathrm{CV}=1$, we only have occasional (rare) spiking, leading to a Poissonian distribution of spiking events which are irregular. When $\mathrm{CV}>1$, we have an occurrence of spikes which is even more irregular than that in a spike train with the Poissonian distribution. In these cases (i.e., when $\mathrm{CV}\geq1$), the quantity $\big[\frac{\sigma_{p,i}^2}{2}\ln(\varepsilon^{-1})\big]$ either lies within the interval $\big[\Delta U^l_i(w_e), F_p(\cdot)\big]$ but very close to its boundaries or outside the interval, especially when $\mathrm{CV}>1$.

We numerically integrated the Eq.\,\eqref{eqn:1} with a step size of $dt=0.008$ for a very long total integration time of $T=3\times10^{5}$. The integration was performed with the second order Runge–Kutta scheme for It\^{o} stochastic differential equations \cite{rossler2009second} using the \textit{itoSRI2} method from the Python package \textit{sdeint}. Moreover, each point on the CV curves was obtained after $6$ realizations of each of these noise intensities.

The $\mathrm{CV}$ of $N$ coupled neurons is defined as~\cite{masoliver2017coherence}:
\begin{align}\label{eqn:12}
\mathrm{CV}=\dfrac{\sqrt{\langle\overline{\mathrm{ISI}^2}\rangle-\langle\overline{\mathrm{ISI}}\rangle^2}}{\langle\overline{\mathrm{ISI}}\rangle},
%\label{eqn:CV_multiple_neurons}
\end{align}
where
\begin{align}\label{eqn:13}
\begin{split}
\left\{\begin{array}{lcl}
\langle\overline{\mathrm{ISI}}\rangle&=&\cfrac{1}{N}\sum\limits_{i=1}^N \langle \mathrm{ISI}_i \rangle,\\[4.0mm]
\langle\overline{\mathrm{ISI}^2}\rangle&=&\cfrac{1}{N}\sum\limits_{i=1}^N \langle \mathrm{ISI}_i^2 \rangle,
\end{array}\right.
\end{split}
\end{align}
in which $\langle \mathrm{ISI}_i \rangle$ and $\langle \mathrm{ISI}_i^2 \rangle$ are the mean and the mean squared (over the total time of simulation $T$) inter-spike intervals of the $i$th neuron, respectively. While $\langle\overline{\mathrm{ISI}}\rangle$ and $\langle\overline{\mathrm{ISI}^2}\rangle$ are the mean (over the total number of neurons $N$) of $\langle \mathrm{ISI}_i \rangle$ and $\langle \mathrm{ISI}_i^2 \rangle$, respectively. The threshold value of the membrane potential variable above which a spike is considered to occur is $v_{\mathrm{th}}=0.0$.

\section{Simulation results and discussion}\label{Section 5}
\subsection{SISR in a single neuron without autapse}
In this subsection, we investigate the degree of SISR in a single isolated ML neuron without autapses in excitable regime (i.e., $v_l=1.515<v_{\mathbb{H}}$) and how it varies with the time scale separation parameter $\varepsilon$ and the noise intensity $\sigma_{1,1}=\sigma$. 
From Eq.\,\eqref{eqn:1} and Eqs.\eqref{eqn:6}--\eqref{eqn:10}, we respectively obtain a single isolated neuron without autapses and the corresponding set of necessary conditions for the occurrence of SISR by setting all coupling strengths to zero, i.e., $\kappa^a_e=\kappa^a_c=\kappa_e=\kappa_c=\kappa^m_e=\kappa^m_c=0$.

\begin{figure}%[ht]
\centering
\subfigure[]{\includegraphics[width=0.49\textwidth]{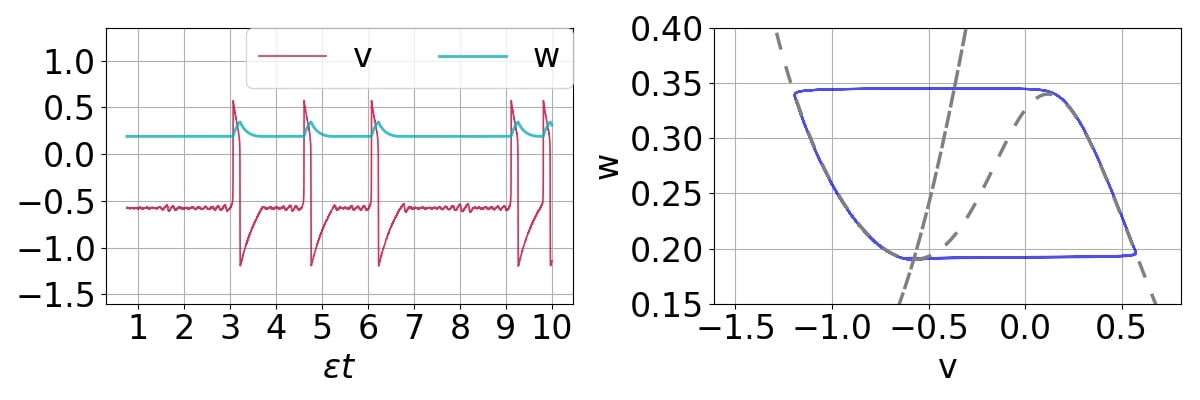}}
\subfigure[]{\includegraphics[width=0.49\textwidth]{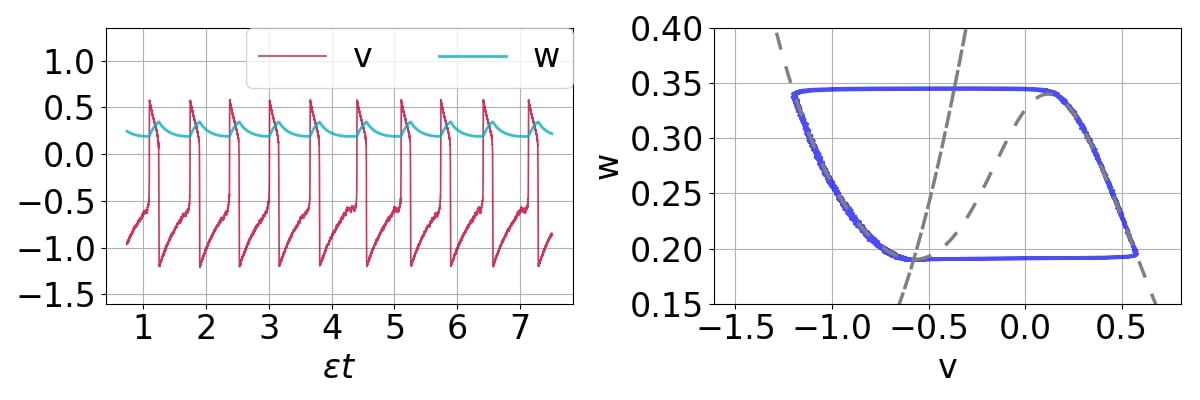}}
\subfigure[]{\includegraphics[width=0.49\textwidth]{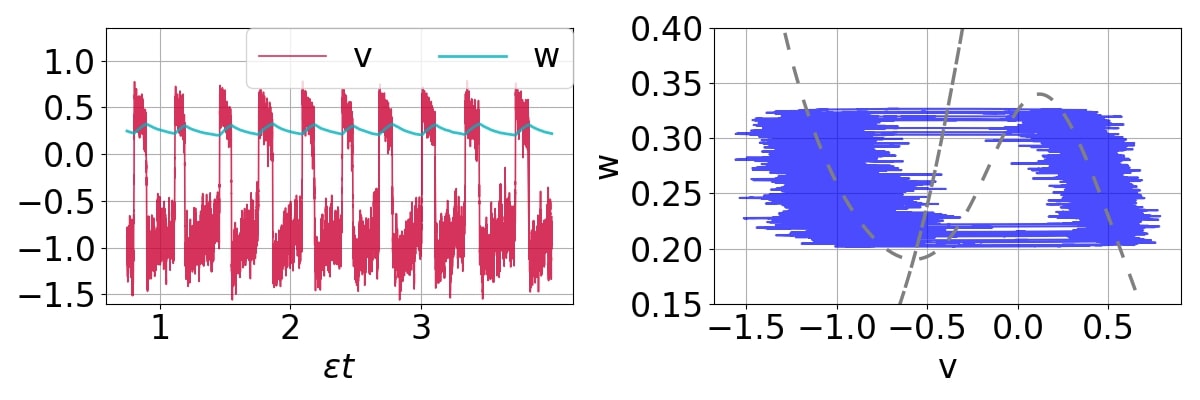}}
\caption{Time series (left) and corresponding phase portraits (right) showing noise-induced coherent oscillations in a single isolated ML neuron in the excitable regime (i.e., $v_l=1.515<v_{\mathbb{H}}$) with $\varepsilon=0.00005$ and for different noise amplitudes:  $\sigma=0.0006$, $0.005$, and $0.12$ in panels \textbf{(a)}, \textbf{(b)}, and \textbf{(c)}, respectively.}
\label{fig:time_series_phase_for_single_ML_SISR1}
\end{figure}

In Fig.\ref{fig:time_series_phase_for_single_ML_SISR1}\textbf{(a)} - \textbf{(c)}, we show sample trajectories in time series and the corresponding phase portrait for increasing noise intensities at $\varepsilon=0.00005$. We observe that as the noise increases, but within the weak limit (i.e., $\sigma\ll1$), the coherence of the spikes is not significantly changed.

\begin{figure}
            \centering
            \includegraphics[width=0.45\textwidth]{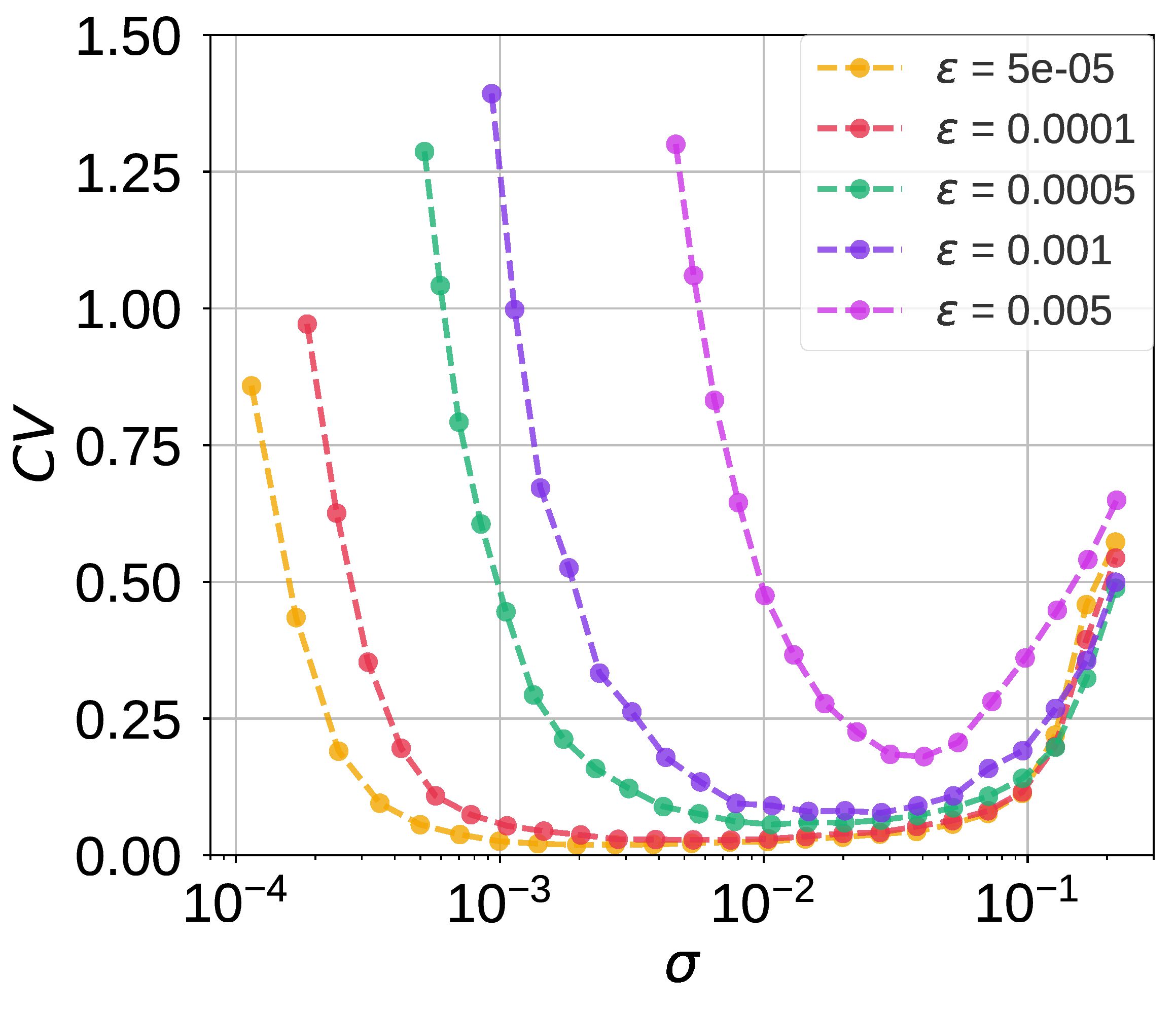}
            \caption{Coefficient of variation $\mathrm{CV}$ against noise amplitude $\sigma$ in a single isolated ML neuron in the excitable regime (i.e., $v_l=1.515<v_{\mathbb{H}}$) for different values of the time scale parameter $\varepsilon$. The intervals of the weak noise intensity in which $\mathrm{CV}<0.2$ shrinks with increasing $\varepsilon$. Furthermore, the minimum $\mathrm{CV}$ values in order of increasing $\varepsilon$ are as follows: $\mathrm{CV}_{\mathrm{min}}= 0.018,~0.028,~0.056,~0.079,~0.18$.}
            \label{fig:CV_ML_epsilons}
\end{figure}

Fig.\ref{fig:CV_ML_epsilons} shows the variation of $\mathrm{CV}$ against the noise intensity $\sigma\ll1$ for several values of the time scale separation parameter $\varepsilon\ll1$. First, we observe that the smaller $\varepsilon$ is, the larger the interval of the noise amplitude $\sigma$ in which the $\mathrm{CV}$ values are the lowest, i.e., typically below 0.2. This is quite remarkable because in these larger intervals of noise where the $\mathrm{CV}$ values are low, one can actually vary the noise intensity without changing the high degree of coherence of the spiking activity due to SISR. On the other hand, the larger $\varepsilon$ is, the higher the minimum value of $\mathrm{CV}$. Thus, as the conditions in Eq.\,\eqref{eqn:6} predict, a high degree of SISR depends on the interplay between the time scale separation parameter and noise intensity in their weak limits $(\sigma,\varepsilon)\to(0,0)$. 

Returning to Eq.\,\eqref{eqn:13a}, we provide a theoretical explanation (based on the expressions of $\sigma_{min}$ and $\sigma_{max}$) to the fact that the left branch of the CV curve in Fig.\,\ref{fig:CV_ML_epsilons} is shifted to the right as $\varepsilon$ increases and while the right branch does not significantly move. Furthermore, we use these theoretical expressions in Eq.\,\eqref{eqn:13a} to accurately calculate the order of magnitude of $\sigma_{min}$ and $\sigma_{max}$ for a single isolated neuron at a given $\varepsilon$. We remind that the explanations and calculations given here for the case of a single isolated neuron also applies to the the rest of the cases investigated in this paper. But as a test-of-principle and for the sake of simplicity, we only show the details for the isolated neuron without autapses.

\begin{figure}
            \centering
            \includegraphics[width=0.45\textwidth]{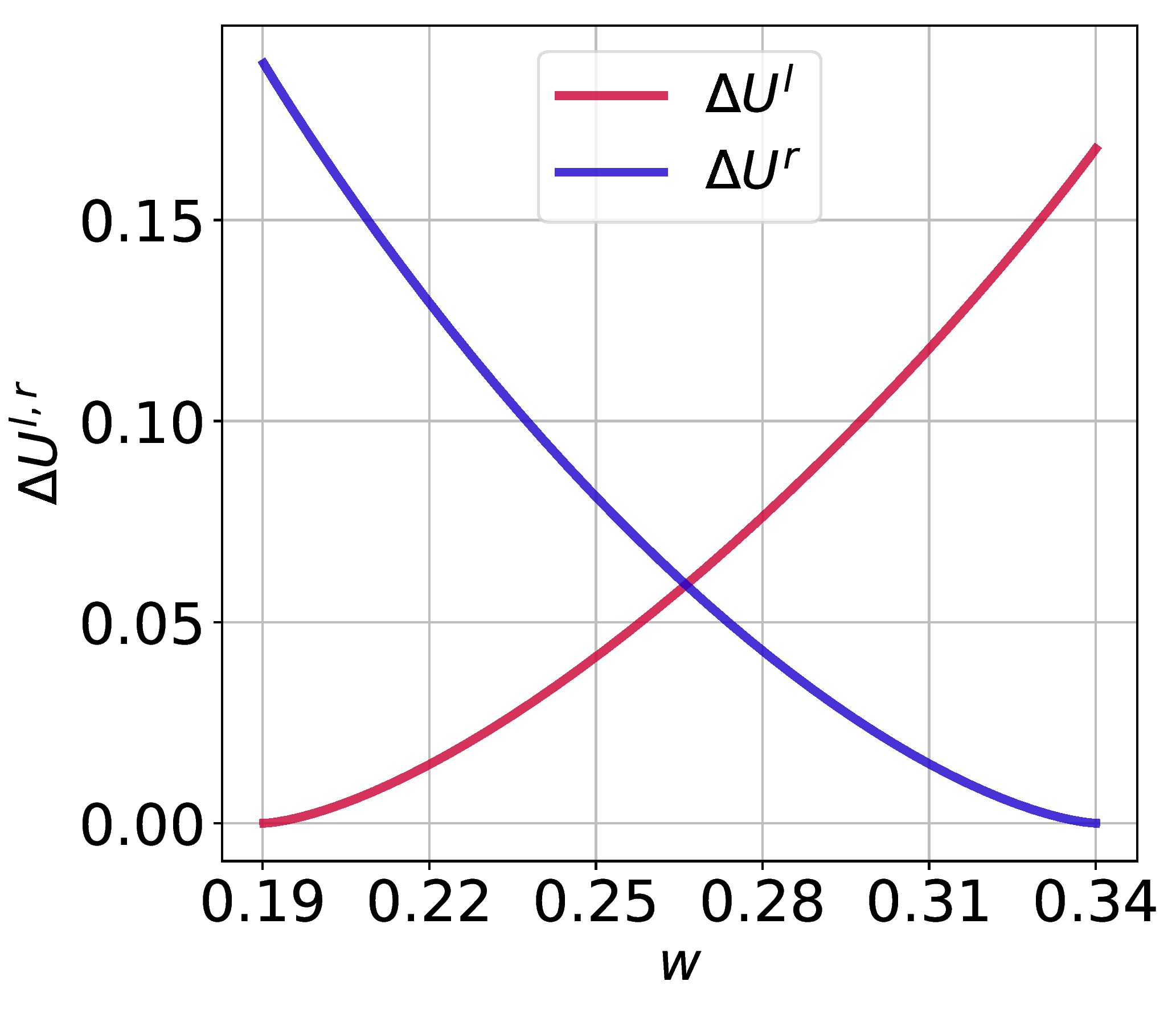}
            \caption{Energy barriers $\Delta U^{l,r}$ of the single isolated ML neuron against the slow variable $w$ in the excitable regime (i.e., $v_l=1.515<v_{\mathbb{H}}$ with $\varepsilon=0.0005$).}
            \label{fig:barrier}
\end{figure}

Why will the CV curves in Fig.\,\ref{fig:CV_ML_epsilons} be shifted to the right as $\varepsilon$ increases? To answer this question, we note that for a fixed set of parameter values, $\Delta U^l(w_e)$ and $F(\cdot)$ in the expressions of $\sigma_{min}$ and $\sigma_{max}$ are also fixed. We further observe that as $\varepsilon$ increases, $\ln(\varepsilon^{-1})$ decreases, and hence $\sigma_{min}$ increases (since $\Delta U^l(w_e)$ is fixed), which, therefore, shifts the left branch of the CV curve to be right as $\varepsilon$ increases. On the other boundary, why will the CV curves in Fig.\,\ref{fig:CV_ML_epsilons} remain almost unchanged as $\varepsilon$ increases? To answer this question, we have to calculate $F(\cdot)$ for the isolated neuron. Fig.\,\ref{fig:barrier} shows the graph of the left $\Delta U^l(w)$ and right $\Delta U^r(w)$ energy barriers given by Eq.\,\eqref{eqn:9}. 
We observe that from the definition of $F(\cdot)$ in Eq.\,\eqref{eqn:8} and Fig.\,\ref{fig:barrier} that $\Delta U^l(w)=\Delta U^r(w)$ at $w=0.2662$ and hence $F(.)=0.059274$ for $\varepsilon=0.0005$. This gives a maximum noise 
$\sigma_{max} = 1.249\times10^{-1}$. With the stable unique fixed point evaluated at $(v_e,w_e)=(-0.5767, 0.19019)$, we calculate left energy barrier at $\Delta U^l(w_e)=1.45\times10^{-6}$ and hence the corresponding minimum noise of 
$\sigma_{min} = 6.0\times10^{-4}$. Comparing $\sigma_{min} = 6.0\times10^{-4}$ and $\sigma_{max} = 1.249\times10^{-1}$ with extreme values
$\sigma$ in Fig.\,\ref{fig:CV_ML_epsilons} when $\varepsilon=0.0005$, we observe that the theoretical results of Eq.\,\eqref{eqn:13a} predict the correct order of magnitude of $\sigma_{min}$ and $\sigma_{max}$. Furthermore, we notice from Fig.\,\ref{fig:barrier} that the value of $F(.)=0.059274$ (i.e., the value of $\Delta U(w)$ when $\Delta U^l(w)=\Delta U^r(w)$) does not change as $w$ varies. This is why the right branches of the CV curves do not change significantly (i.e., they all have the same order of magnitude) as $\varepsilon$ changes.

In the rest of our numerical simulations, we fix the time scale parameter at $\varepsilon=0.0005$. This very small value is chosen for two reasons: (i) The behavior of SISR at a very small value of $\varepsilon$ is qualitatively the same as at relatively larger values (which maybe biologically more relevant), provided that the interplay between the noise amplitude $\sigma$ and $\varepsilon$ is satisfied according to Eq.\,\eqref{eqn:6}. (ii) The phenomenon of SISR is very pronounced at very small values of $\varepsilon$ (if, of course, Eq.\,\eqref{eqn:6} is satisfied), making it easier to understand its behavior.

\subsection{SISR in a single neuron with an electrical autapse}
In this subsection, we investigate the degree of SISR in a single isolated ML neuron with (only) an electrical autapse in the excitable regime (i.e., $v_l=1.515<v_{\mathbb{H}}$) and how it varies with the autaptic coupling strength $\kappa^a_e$, time delay $\tau^a_e$, and the noise intensity $\sigma_{1,1}=\sigma$. In this case, in Eq.\,\eqref{eqn:1} and Eqs.\eqref{eqn:6}--\eqref{eqn:10}, we set $\kappa^a_c=\kappa_e=\kappa_c=\kappa^m_e=\kappa^m_c=0$, except $\kappa^a_e\neq0$.

As we pointed out earlier, time-delayed couplings may invoke a saddle-node onto limit cycles (SNLC) bifurcation, leading to the emergence of self-sustained spiking activity in the autaptic neuron even in the absence of noise \cite{scholl2009time}. SNLC may occur even if the Hopf bifurcation parameter is fixed in the excitable regime (i.e., $v_l=1.515<v_{\mathbb{H}}$) identified in Fig.\,\ref{fig:bifurcation_diagram and variation of hopf}\textbf{(a)}. Thus, it is indispensable to identify and avoid time-delayed coupling values leading to SNLC in the zero-noise dynamics.

\begin{figure}[!ht]
\centering
\includegraphics[width=0.40\textwidth]{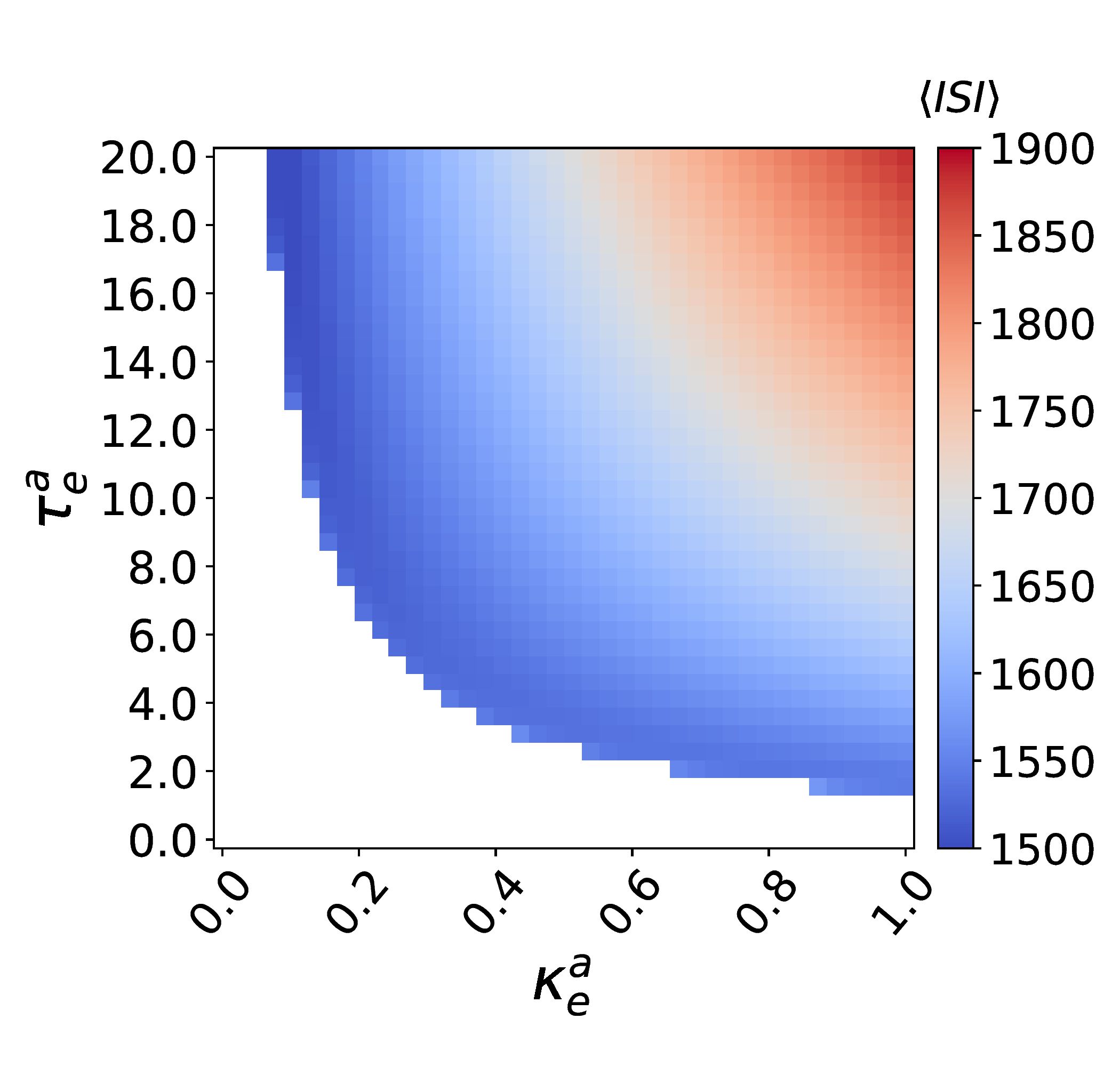}\hspace{1cm}
\caption{The mean inter-spike interval $\langle \mathrm{ISI} \rangle$ is color coded in the $(\kappa^a_e - \tau^a_e)$ plane for the deterministic (i.e., $\sigma_{1,1}=\sigma=0$) ML neuron with an eletrical autapse. The white region represents the excitable regime (i.e., predisposition for SISR) and the colored regions represent the oscillatory regime (i.e., undesired regime) invoked by autaptic time-delayed couplings via SNLC bifurcations. $v_l=1.515<v_{\mathbb{H}}$, $\varepsilon=0.0005$.}
\label{fig:time_series_phase_for_single_ML_SISR}
\end{figure}

Fig.\,\ref{fig:time_series_phase_for_single_ML_SISR} shows a color coded $\mathrm{ISI}$ in a two-parameter ($k^a_e$, $\tau^a_e$) deterministic bifurcation diagram. The white region represents the desired excitable regime (where no spike occurs and thus no $\mathrm{ISI}$), while the colored regions represent the undesired oscillatory regime (with non-zero $\mathrm{ISI}$) induced by SNCL.

\begin{figure}%[!ht]
\centering
\subfigure[$\kappa^a_{\mathrm{e}}=0.05$]{\includegraphics[width=0.265\textwidth]{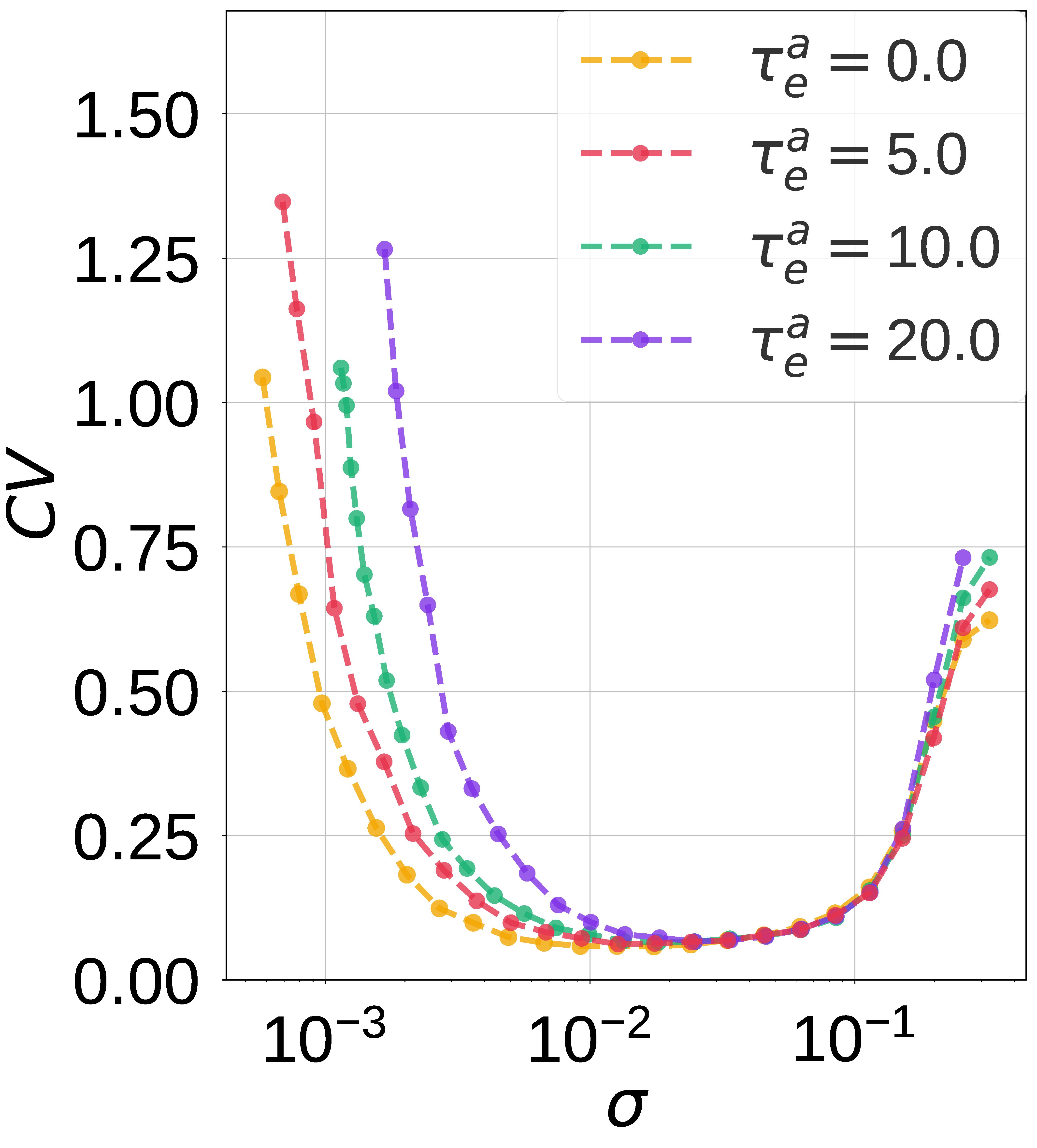}}\hfill
\subfigure[$\kappa^a_{\mathrm{e}}=0.5$]{\includegraphics[width=0.21\textwidth]{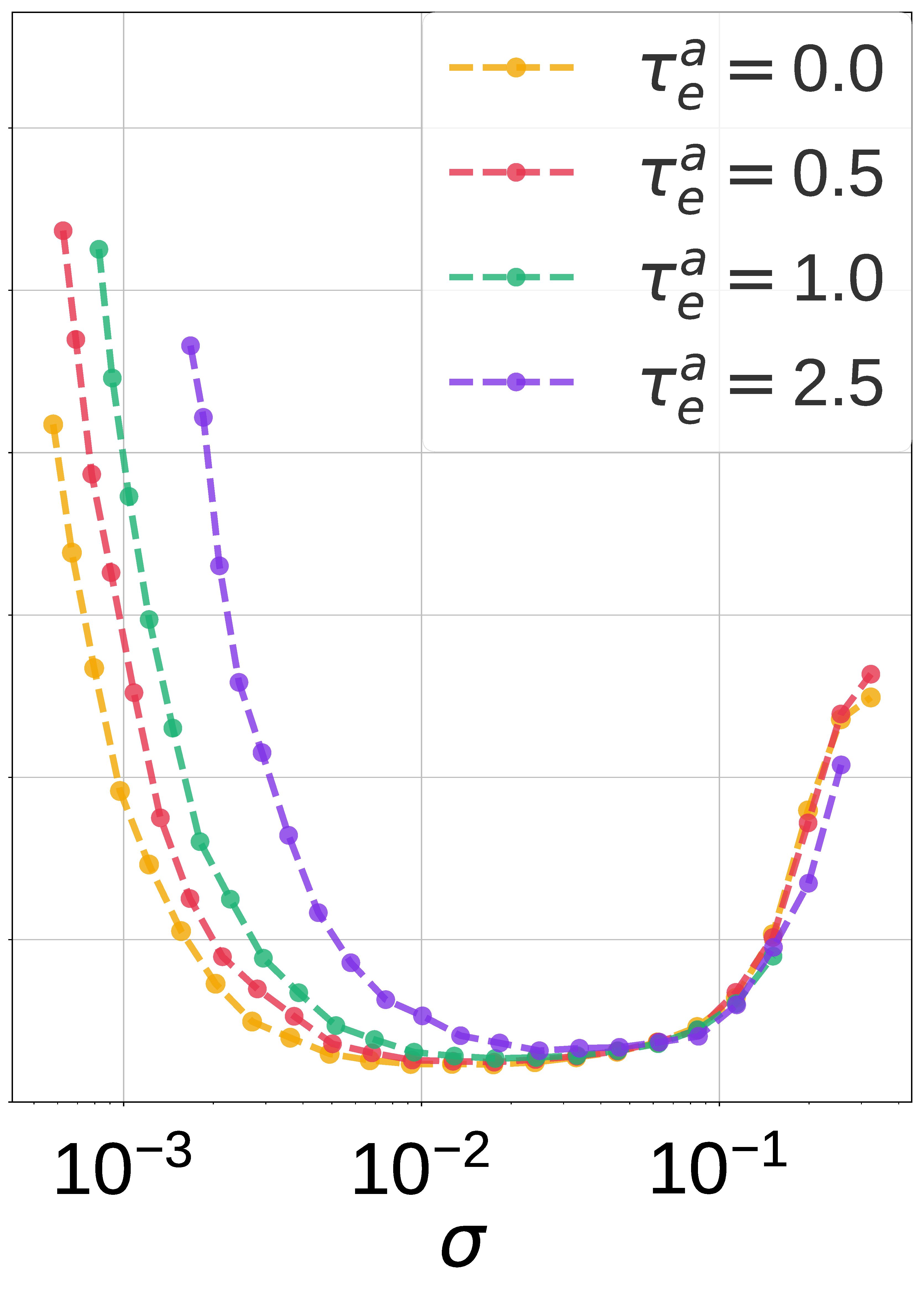}}\hfill
\subfigure[$\tau^a_{\mathrm{e}}=1.0$]{\includegraphics[width=0.265\textwidth]{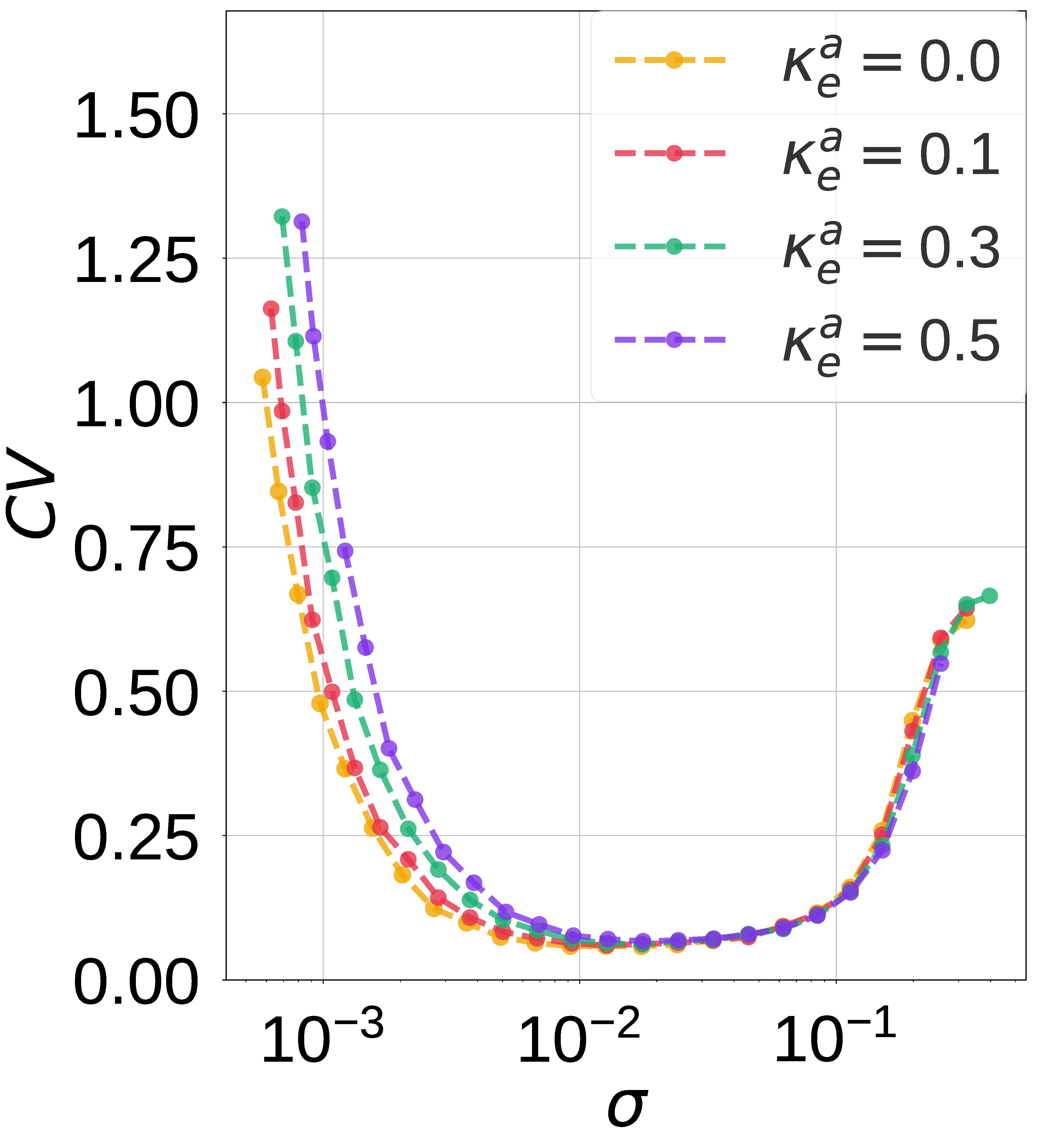}}\hfill
\subfigure[$\tau^a_{\mathrm{e}}=20.0$]{\includegraphics[width=0.21\textwidth]{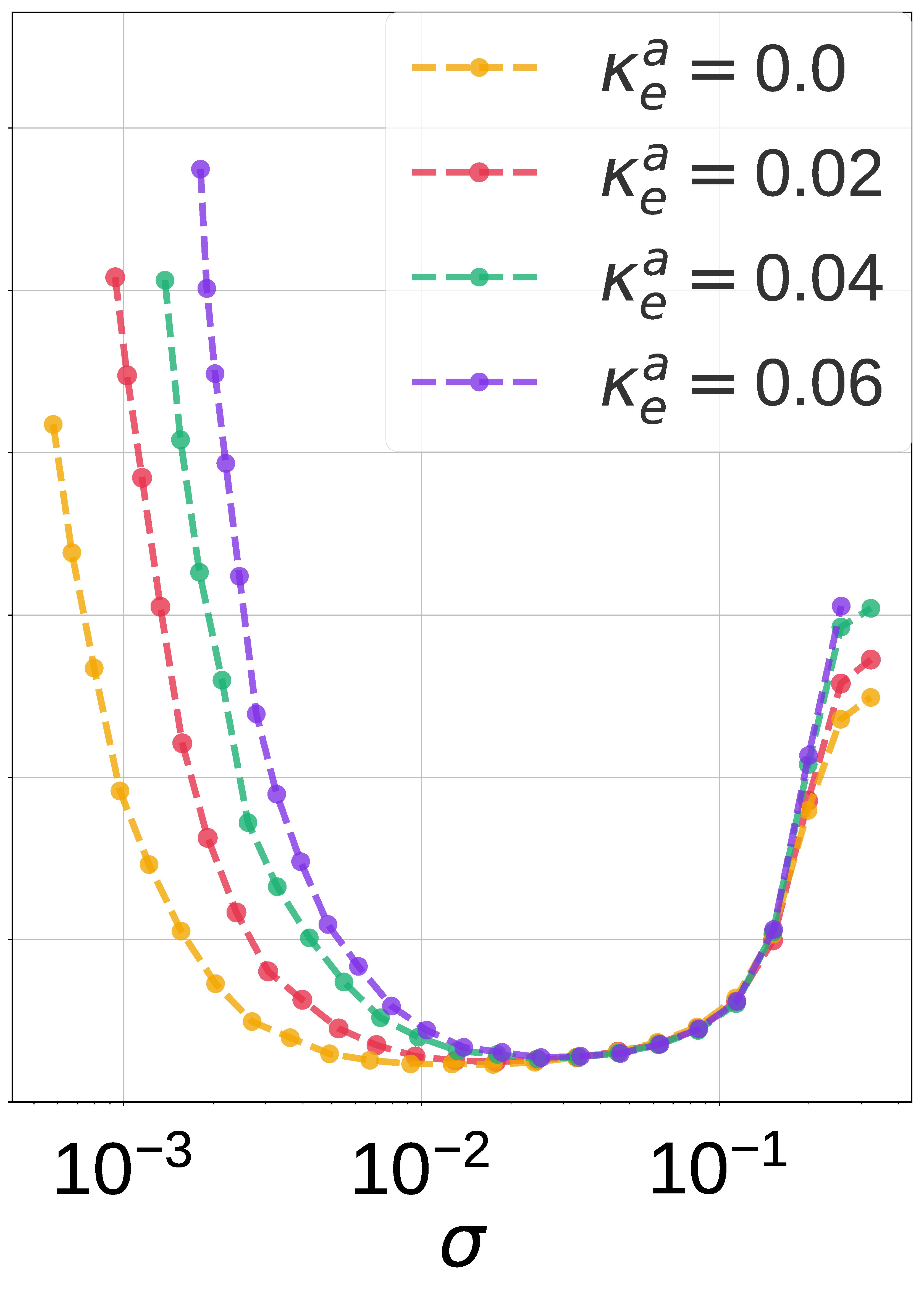}}\hfill
\caption{Coefficient of variation $\mathrm{CV}$ against noise amplitude $\sigma$ for parameter combinations of the electrical autapse $(\kappa^a_{\mathrm{e}},\tau^a_{\mathrm{e}})$ in a single isolated ML neuron. We observe variations in the electrical autaptic parameters do not significantly affect the high degree of SISR, but they can shrink the interval of the noise amplitude in which this degree remains high, by shifting the left branch of the $\mathrm{CV}$ curves to relatively larger noise intensities. $v_l=1.515<v_{\mathbb{H}}$, $\varepsilon=0.0005$.}
\label{fig:CV_el_autapses}
\end{figure}

In Fig.\ref{fig:CV_el_autapses}, we show the variation of  $\mathrm{CV}$ against the noise amplitude $\sigma$ with values of $\kappa^a_e$ and $\tau^a_e$ taken from the excitable regime in Fig.\,\ref{fig:time_series_phase_for_single_ML_SISR}. In Fig.\,\ref{fig:CV_el_autapses}\textbf{(a)}, we choose a weak autaptic coupling $\kappa^a_e=0.05$ and vary the time delay $\tau^a_e\in\{0.0, 5.0, 10.0, 20.0\}$. We observe that in this weak autaptic coupling regime, the time delay has no effect on the high degree of SISR achieved, as all the $\mathrm{CV}$ curves remain at almost the same (low) value. However, as the time delay $\tau^a_e$ becomes longer, the intervals of $\sigma$ in which the $\mathrm{CV}$ curves are the lowest shrink as the left branch of the $\mathrm{CV}$ curves are shifted to the right, i.e., to relatively larger noise intensities.

In Fig.\ref{fig:CV_el_autapses}\textbf{(b)}, where the autaptic strength becomes stronger, i.e., $\kappa^a_e=0.5$, the $\mathrm{CV}$ curves qualitatively behaves as in Fig.\ref{fig:CV_el_autapses}\textbf{(a)}, except that the degree of SISR become very sensitive to small changes in length of time delays: we notice in Fig.\ref{fig:CV_el_autapses}\textbf{(b)}, $\tau^a_e$ varies only between 0.0 and 2.5 in order to have the qualitative behavior in Fig.\ref{fig:CV_el_autapses}\textbf{(a)}.

In Figs.\ref{fig:CV_el_autapses}\textbf{(c)} and \textbf{(d)}, we now fixed $\tau^a_e$ at a short (e.g., $\tau^a_e=1.0$) and long (e.g., $\tau^a_e=20.0$) time delays, respectively, and vary the autaptic coupling strength. We observe that the minimum values of the $\mathrm{CV}$ curves are not significantly changed as the $\kappa^a_e$ changes. Nevertheless, the intervals of $\sigma$ in which the $\mathrm{CV}$ values are the lowest, shrink as the left branches of the $\mathrm{CV}$ curves are again shifted to the right. Moreover, in long time delay regimes such as in Fig.\ref{fig:CV_el_autapses}\textbf{(d)}, the degree of SISR becomes very sensitive to variations of the autaptic coupling strength. Here, $\kappa^a_e$ varies only up to 0.06 in order to have the same qualitative behavior observed in Figs.\ref{fig:CV_el_autapses}\textbf{(c)} where $\kappa^a_e$ varies up to 20.0.

\subsection{SISR in a single neuron with a chemical autapse}
In this subsection, we investigate the degree of SISR in a single isolated ML neuron with (only) an inhibitory chemical autapse and how it varies with the autaptic coupling strength $\kappa^a_{\mathrm{c,inh}}$, time delay $\tau^a_{\mathrm{c,inh}}$, and the noise intensity $\sigma$. 
In this case, in Eq.\,\eqref{eqn:1} and Eqs.\eqref{eqn:6}--\eqref{eqn:10}, we set $\kappa^a_e=\kappa_e=\kappa_c=\kappa^m_e=\kappa^m_c=0$, except $\kappa^a_c\neq0$. It should be noted that we do not consider an excitatory chemical autapse in this case. This is because with this type of autapse, the deterministic ML neuron is always in the oscillatory regime --- the undesired predisposition for SISR. For the whole range of parameter values of the inhibitory chemical autapse used, the isolated ML neuron always remains excitable. 

\begin{figure}%[!ht]
\centering
\subfigure[$\kappa^a_{\mathrm{c,inh}}=0.05$]{\includegraphics[width=0.265\textwidth]{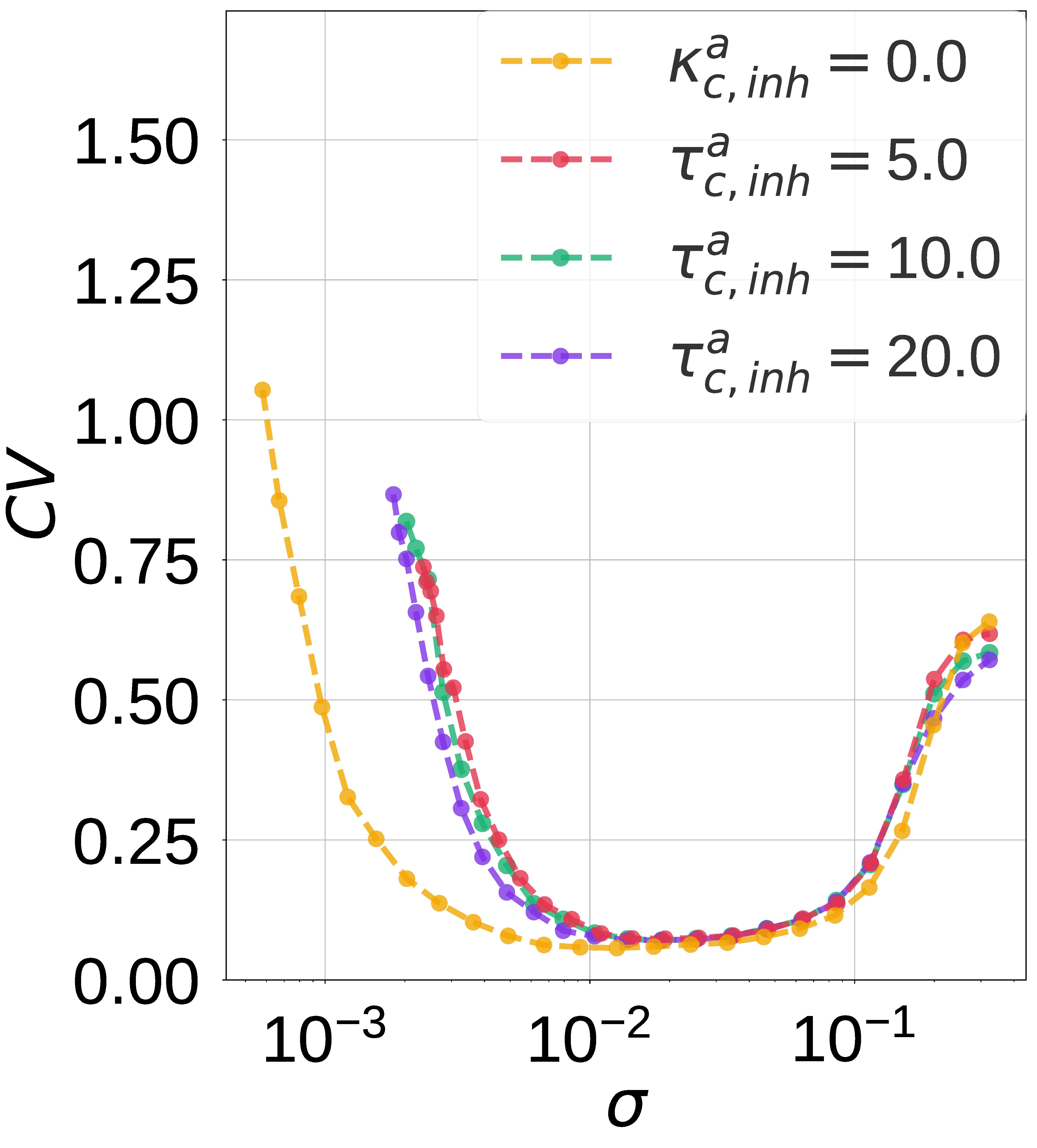}}\hfill
\subfigure[$\kappa^a_{\mathrm{c,inh}}=0.5$]{\includegraphics[width=0.21\textwidth]{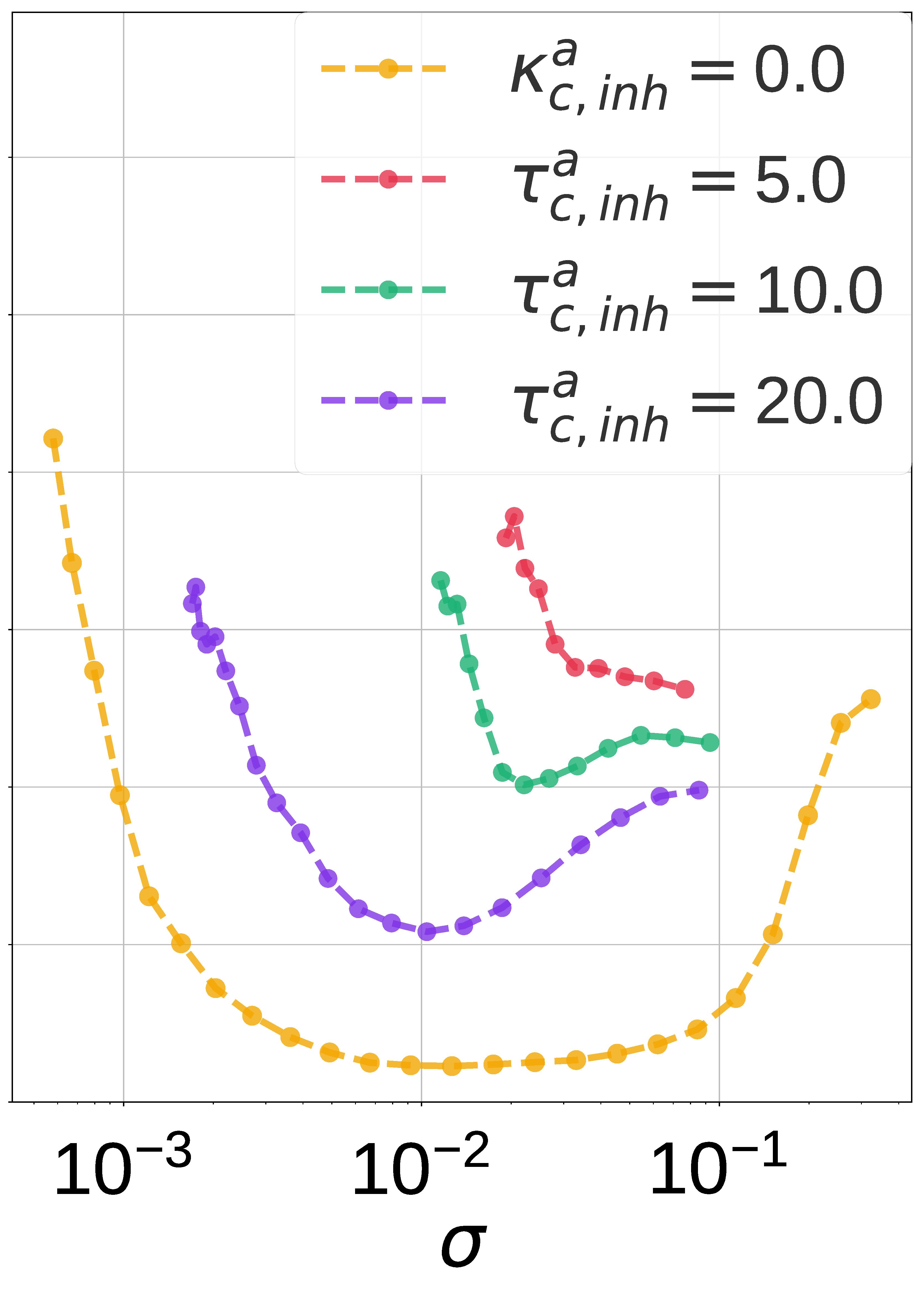}}\hfill
\subfigure[$\tau^a_{\mathrm{c,inh}}=1.0$]{\includegraphics[width=0.265\textwidth]{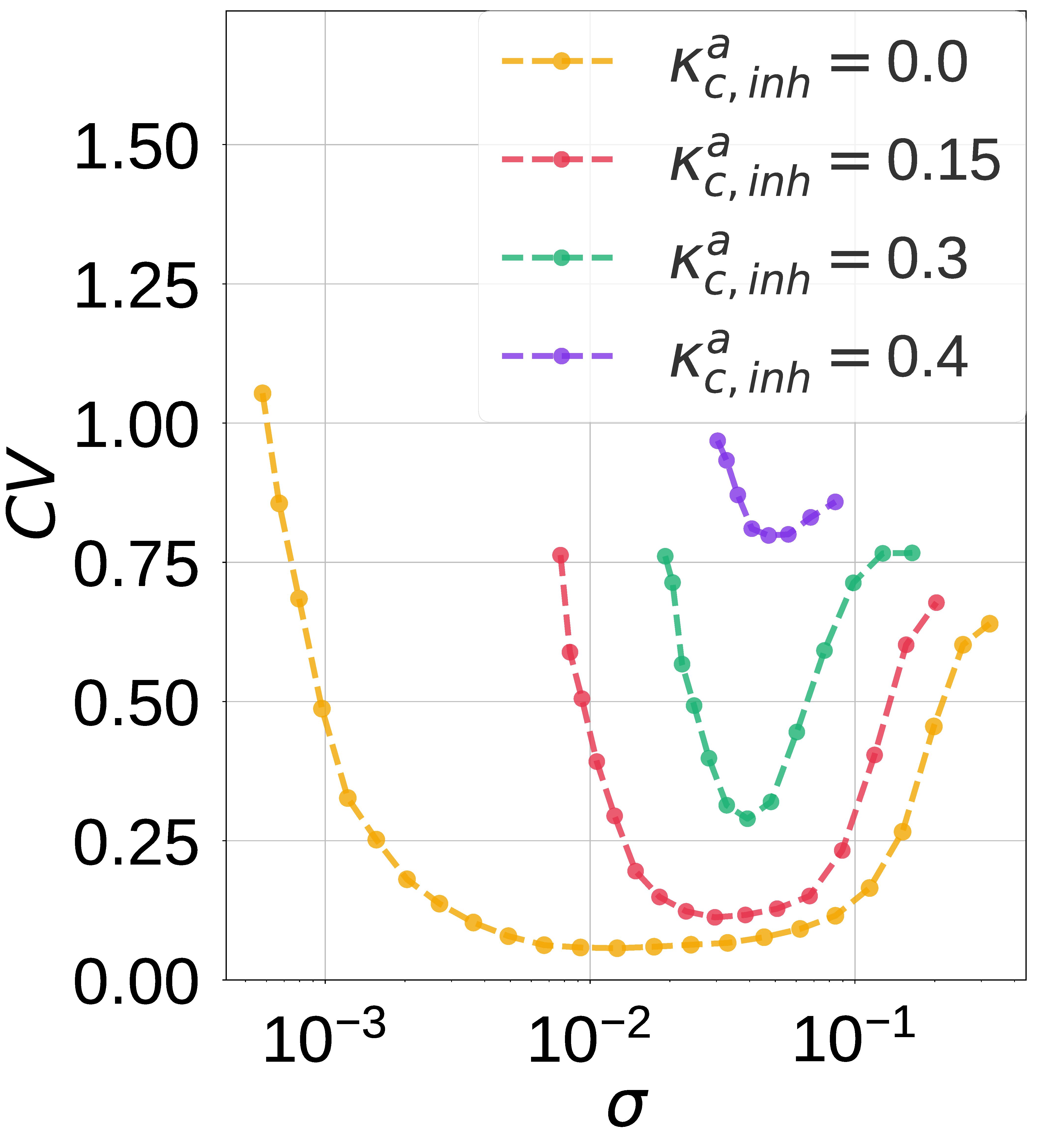}}\hfill
\subfigure[$\tau^a_{\mathrm{c,inh}}=20.0$]{\includegraphics[width=0.21\textwidth]{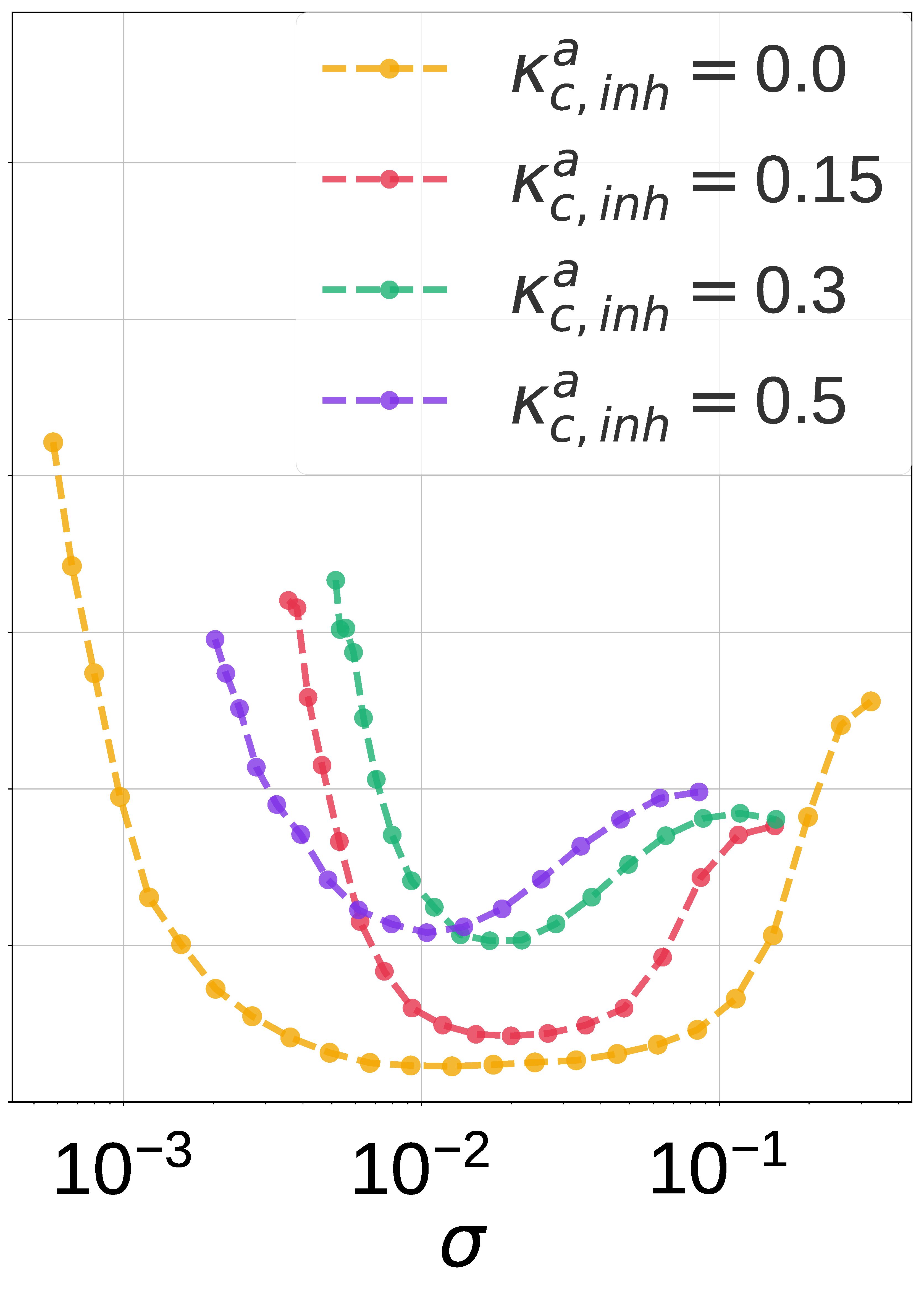}}\hfill
\caption{Coefficient of variation $\mathrm{CV}$ against noise amplitude $\sigma$ for parameter combinations of the inhibitory chemical autapse $(\kappa^a_{\mathrm{c,inh}},\tau^a_{\mathrm{c,inh}})$ in a single isolated ML neuron. We observe that variations in these parameters significantly affect the degree of SISR by shifting the entire $\mathrm{CV}$ curve to higher values and by shrinking, on both ends, the interval of the noise amplitude in which this degree remains relatively high. Longer autaptic time delays enhance SISR, while stronger autaptic couplings destroy SISR. $v_l=1.515<v_{\mathbb{H}}$, $\varepsilon=0.0005$.}
\label{fig:CV_curves_chemical_autapse}
\end{figure}

In Fig.\,\ref{fig:CV_curves_chemical_autapse}, we show the variation of $\mathrm{CV}$ against the noise amplitude $\sigma$ with values of $\kappa^a_{\mathrm{c,inh}}$ and $\tau^a_{\mathrm{c,inh}}$.
In Fig.\,\ref{fig:CV_curves_chemical_autapse}\textbf{(a)}, at a weak autaptic coupling $\kappa^a_{\mathrm{c,inh}}=0.05$, we vary the time delay $\tau^a_{\mathrm{c,inh}}\in\{0.0, 5.0, 10.0, 20.0\}$. In this case, we observe that variations in the time delay have no effect on the high degree of SISR achieved. All the $\mathrm{CV}$ curves remain at almost the same (low) value.

In Fig.\,\ref{fig:CV_curves_chemical_autapse}\textbf{(b)}, where the autaptic strength becomes stronger, i.e., $\kappa^a_{\mathrm{c,inh}}=0.5$, the $\mathrm{CV}$ curves change significantly, both qualitatively and quantitatively. Here, as the time delay increases from a non-zero value, the minimum of the $\mathrm{CV}$ curves get lower, indicating a higher degree of SISR. 

In Fig.\,\ref{fig:CV_curves_chemical_autapse}\textbf{(c)} and \textbf{(d)}, we fix the autaptic time delays at $\tau^a_{\mathrm{c,inh}}=1.0$ and $\tau^a_{\mathrm{c,inh}}=20.0$, respectively, and vary the autaptic strength $\kappa^a_{\mathrm{c,inh}}\in\{0.0, 0.15, 0.3, 0.4, 0.5\}$. We observe that:
$(\mathrm{i})$ SISR is very sensitive to small variations in the autaptic coupling strength $\kappa^a_{\mathrm{c,inh}}$ and $(\mathrm{ii})$  the time delay $\tau^a_{\mathrm{c,inh}}$ and the coupling strength $\kappa^a_{\mathrm{c,inh}}$ have opposite effects on the degree of SISR. While larger values of $\tau^a_{\mathrm{c,inh}}$  increase the degree of SISR, larger values of $\kappa^a_{\mathrm{c,inh}}$ decrease it.

Furthermore, the deterioration of SISR with an inhibitory chemical autapse manifests in two ways:
$(\mathrm{i})$ higher $\mathrm{CV}$ curves and hence lower degree of SISR and $(\mathrm{ii})$ smaller intervals of the noise intensity in which the degree of SISR is relatively high --- notice that the shrinking of the noise interval happens on both the left and the right branch of the $\mathrm{CV}$ curves. Recall from the previous subsection that the deterioration of SISR with an electrical autapse consisted only in a reduction of this interval and also, only from the left branch of the $\mathrm{CV}$ curves.   

\subsection{SISR in a single motif network}
In this subsection, we investigate the degree of SISR in a single isolated motif without autapses and with either only electrical synapses or only chemical synapses between the three neurons and how it varies with the synaptic time-delayed couplings ($\kappa_e$, $\tau_e$) or ($\kappa_c$, $\tau_c$), and the noise intensity $\sigma_{1,1}=\sigma_{1,2}=\sigma_{1,3}=\sigma$. In motifs with electrical synapses, i.e., the topologies in  Fig.\,\ref{fig:network_motifs_topologies}\textbf{(b)}, we set in Eq.\,\eqref{eqn:1} and Eqs.\eqref{eqn:6}--\eqref{eqn:10}, $\kappa^a_e=\kappa^a_c=\kappa_c=\kappa^m_e=\kappa^m_c=0$, except $\kappa_e\neq0$. In the same fashion, for motifs with chemical synapses, i.e., the topologies in  Fig.\,\ref{fig:network_motifs_topologies}\textbf{(c)} and \textbf{(d)}, we set $\kappa^a_e=\kappa^a_c=\kappa_e=\kappa^m_e=\kappa^m_c=0$, except $\kappa_c\neq0$ in Eq.\,\eqref{eqn:1} and Eqs.\eqref{eqn:6}--\eqref{eqn:10}.

\begin{figure}%[!ht]
\centering
\subfigure[]{\includegraphics[width=0.30\textwidth]{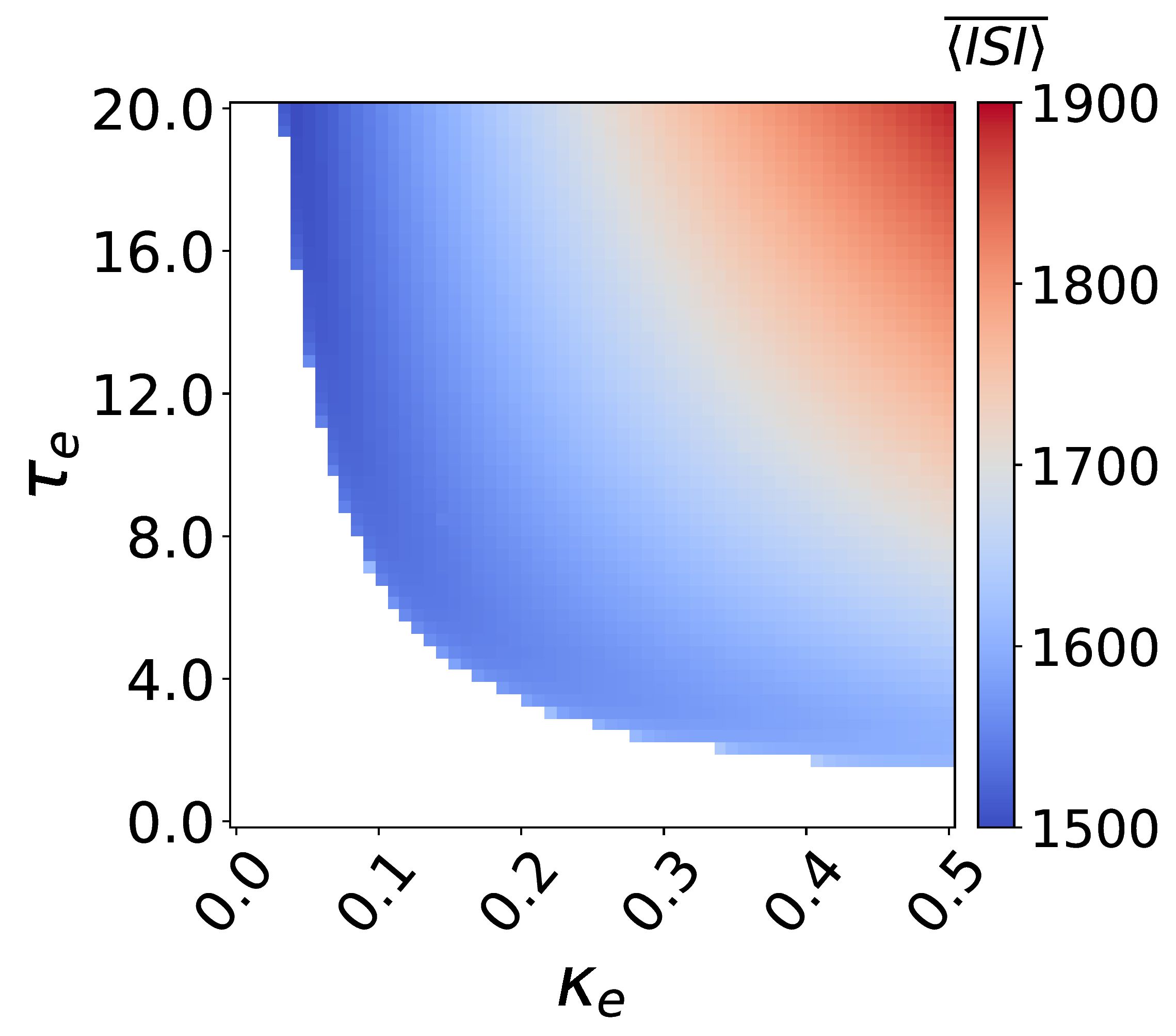}
\raisebox{1.1cm}{\includegraphics[width=0.17\textwidth]{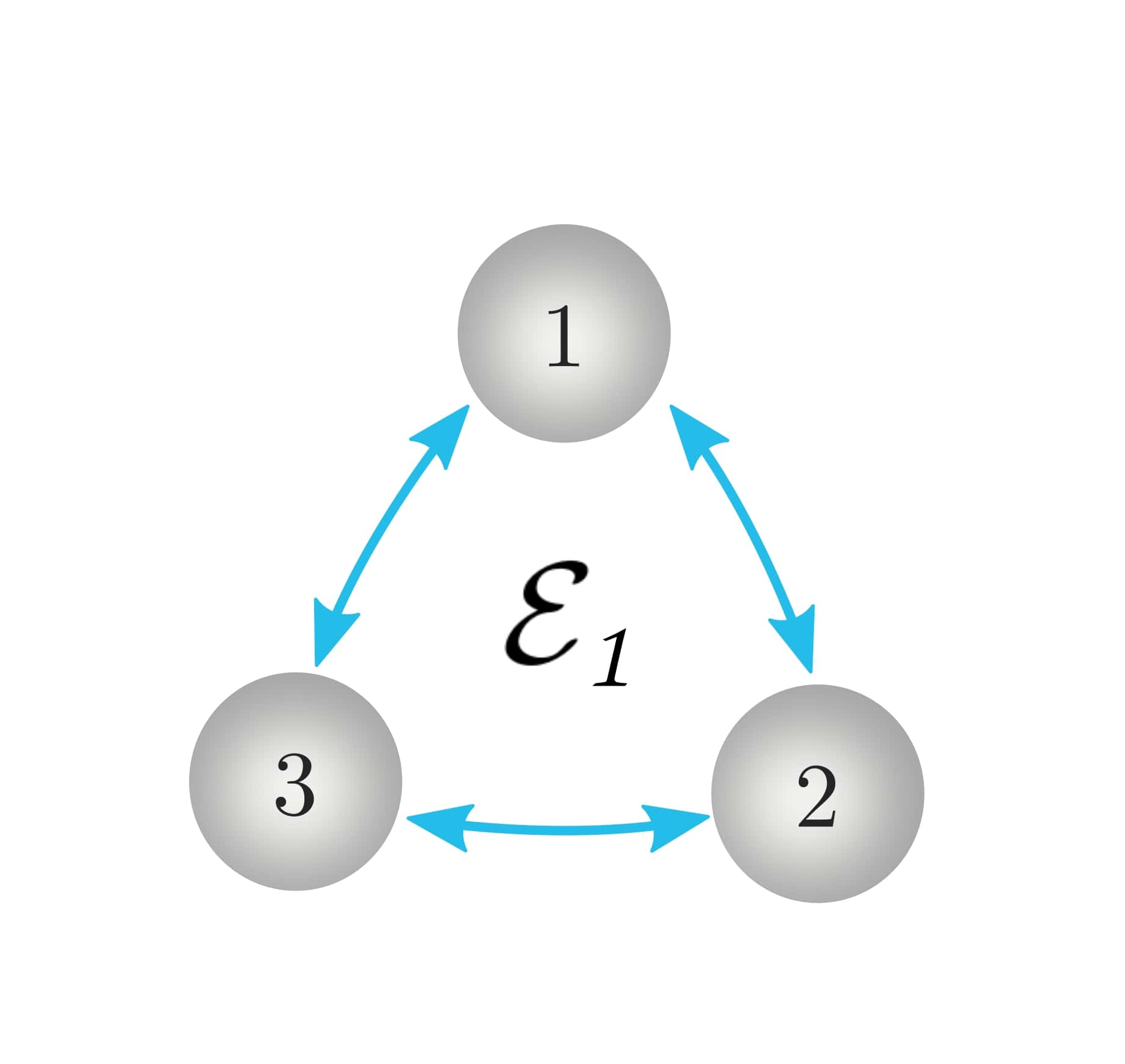}}}\hspace{3em}
\subfigure[]{\includegraphics[width=0.30\textwidth]{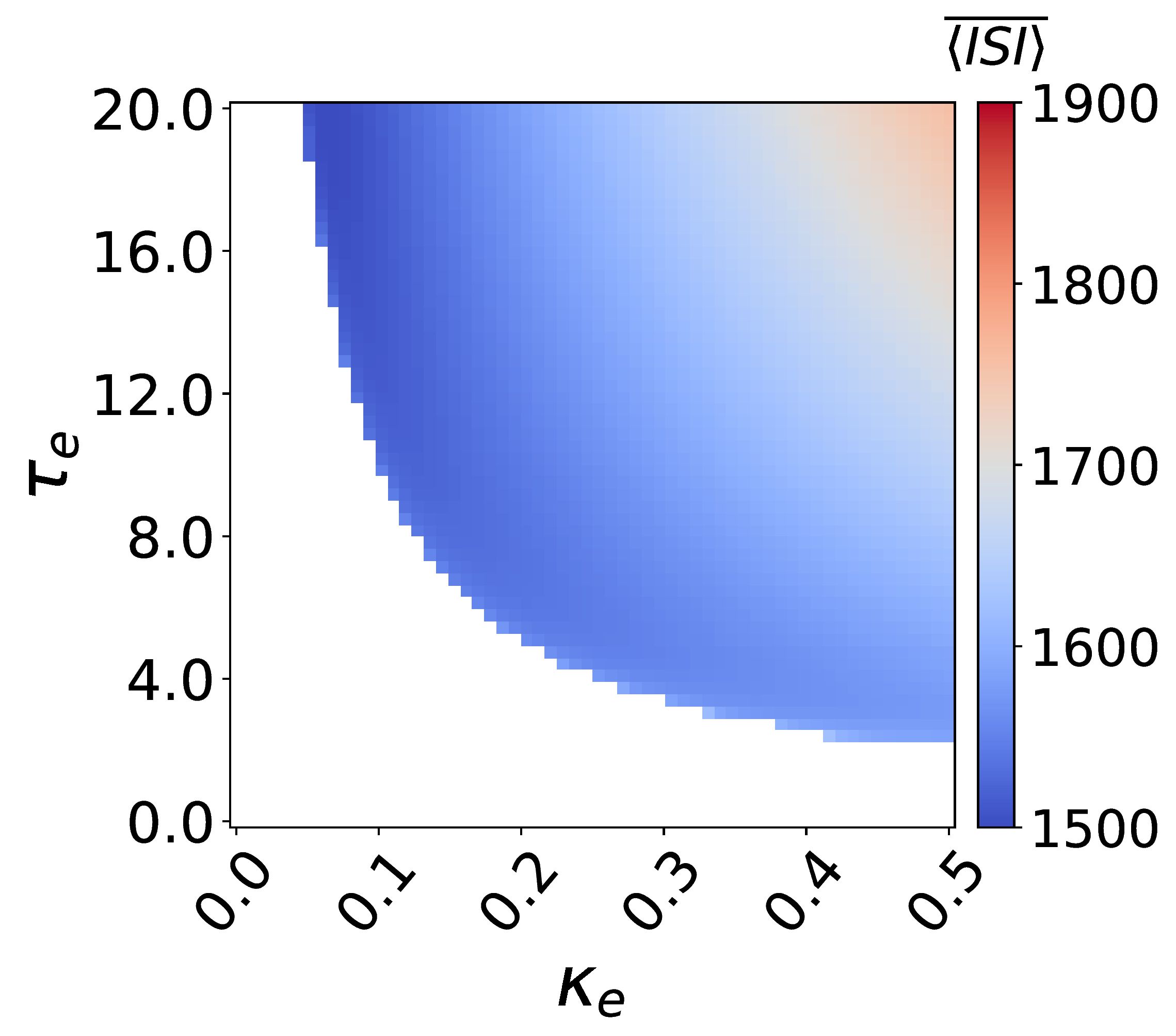}
\raisebox{1.1cm}{\includegraphics[width=0.17\textwidth]{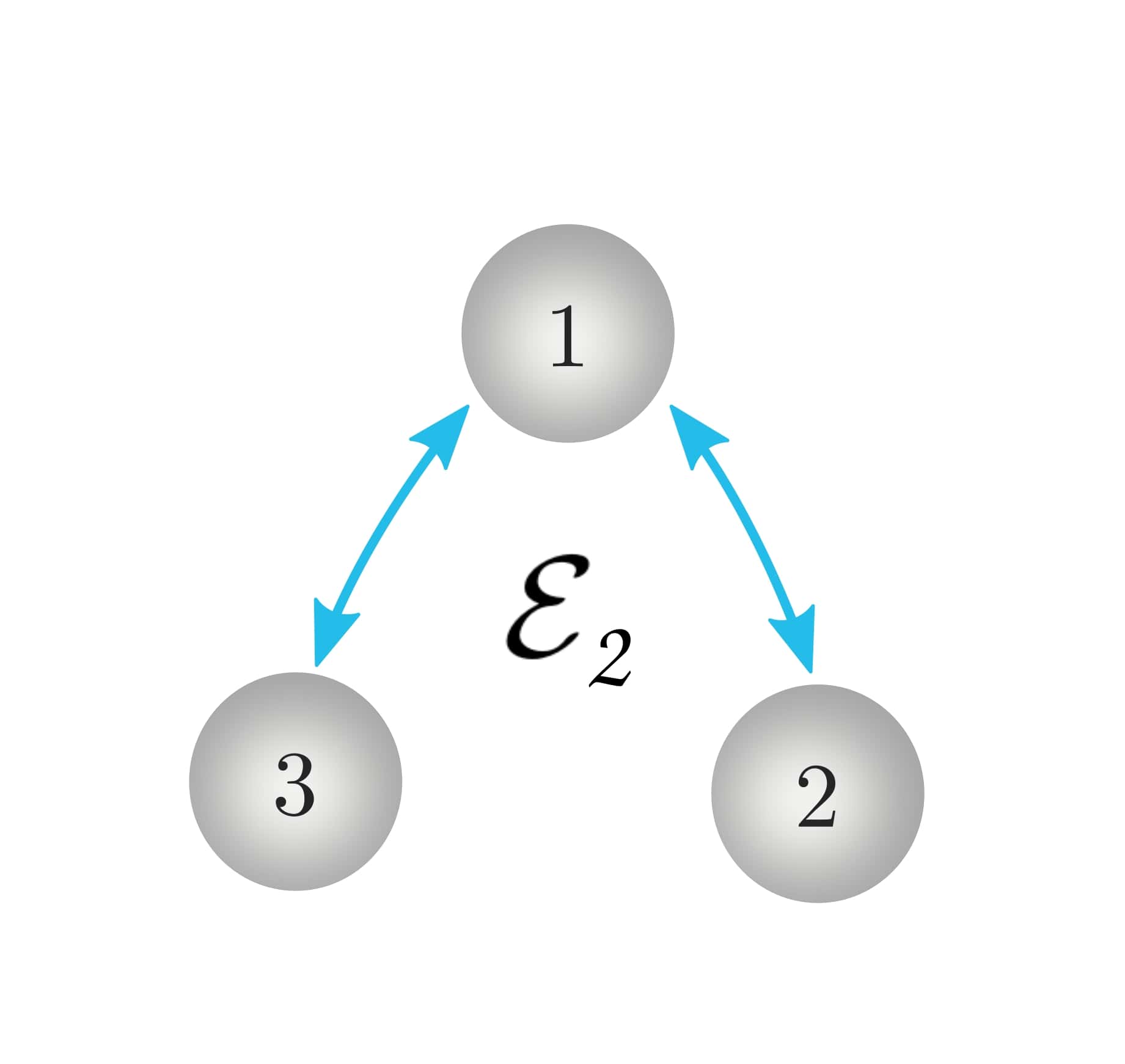}}}
\caption{Excitability maps (left) and corresponding topology (right) of electrically coupled motif layer networks. $\overline{\langle \mathrm{ISI}\rangle}$ is color coded with the white region representing the excitable regime and the colored regions the oscillatory regimes. $v_l=1.515<v_{\mathbb{H}}$, $\varepsilon=0.0005$.}
\label{fig:stability_different_topologies_chem}
\end{figure}

\begin{figure}%[!ht]
\centering
\subfigure[]{\includegraphics[width=0.30\textwidth]{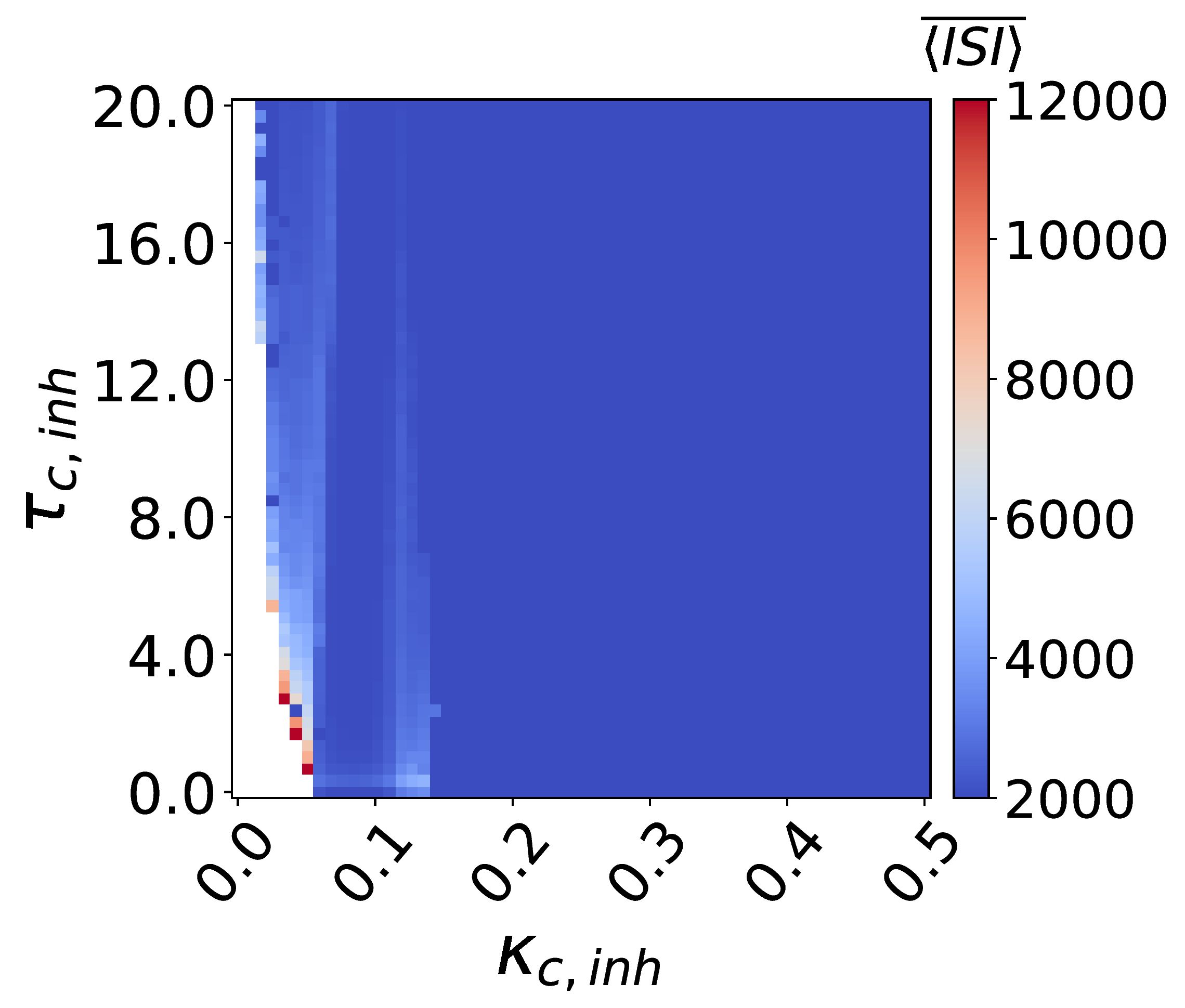}
\raisebox{1.1cm}{\includegraphics[width=0.17\textwidth]{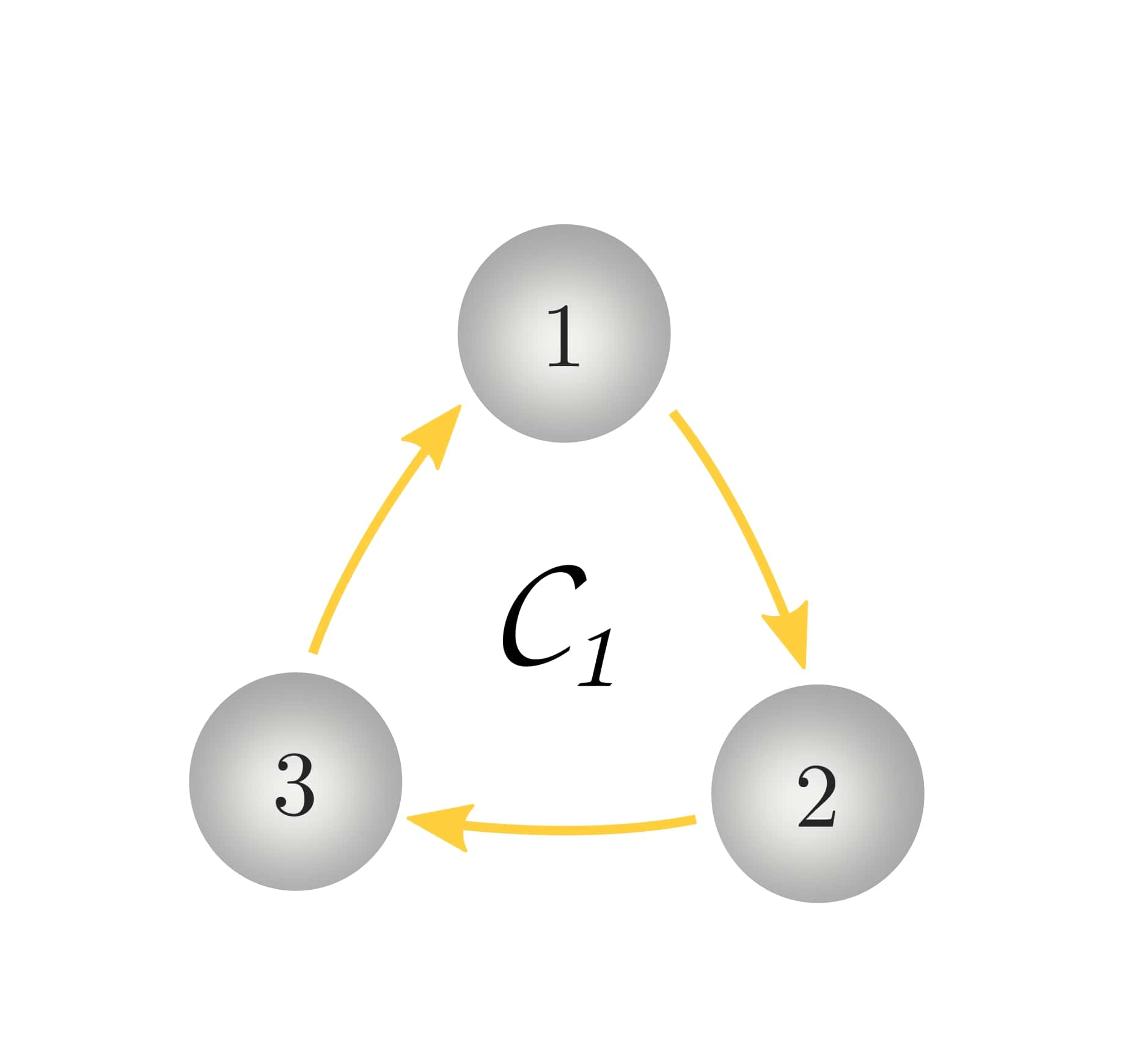}}}\hspace{1.3cm}
\subfigure[]{\includegraphics[width=0.30\textwidth]{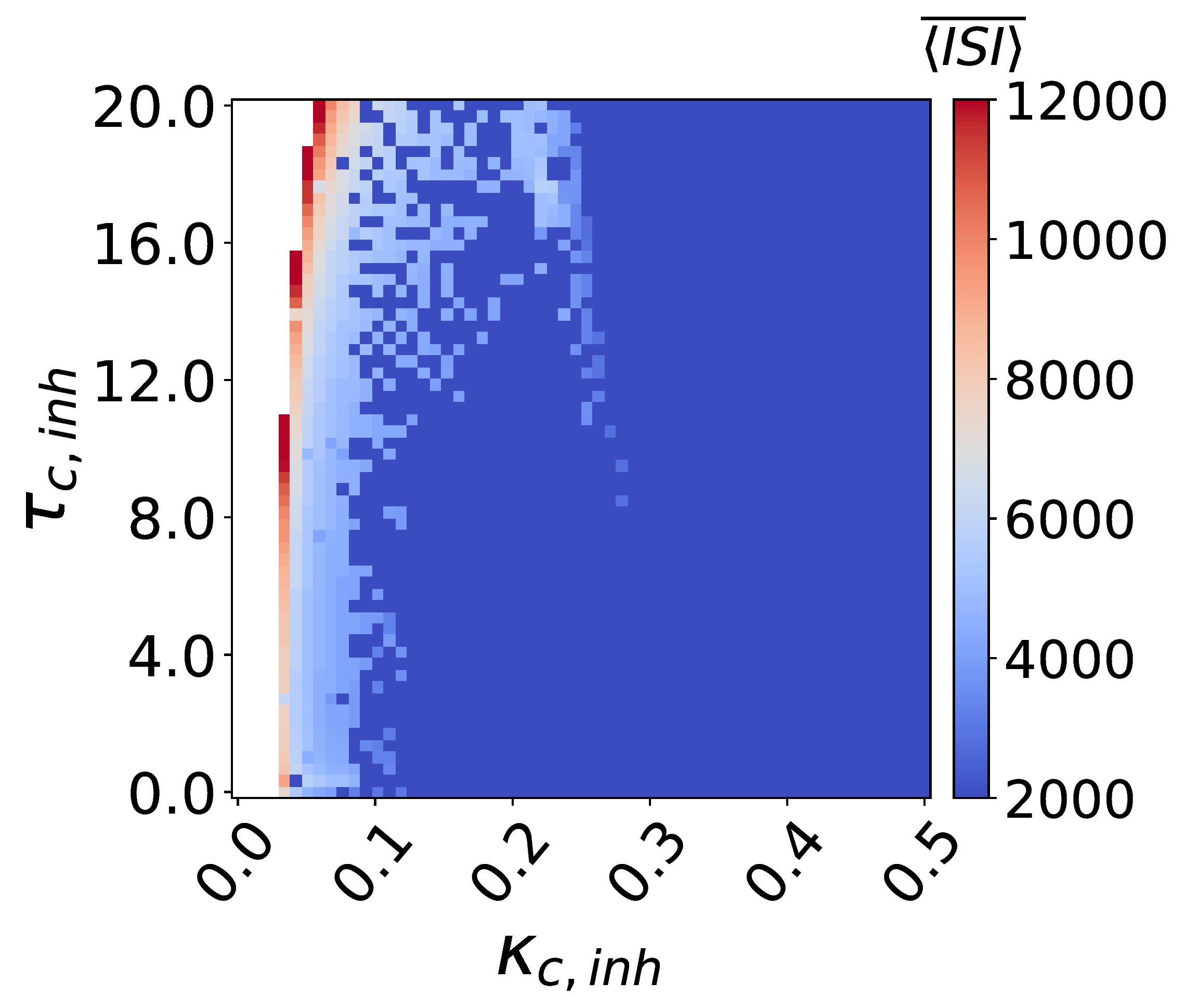}
\raisebox{1.1cm}{\includegraphics[width=0.17\textwidth]{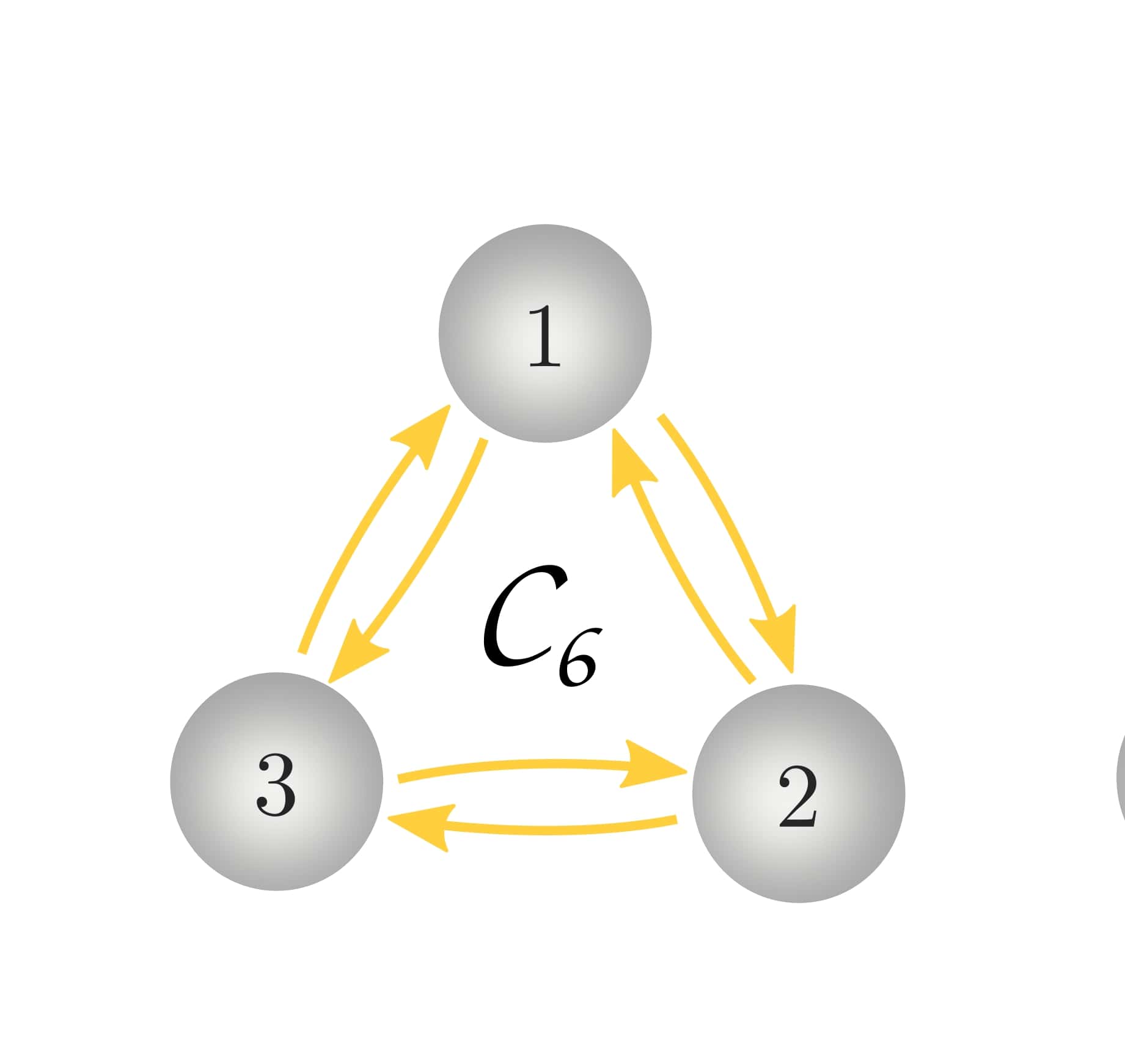}}}\hspace{-0.1cm}
\subfigure[]{\includegraphics[width=0.30\textwidth]{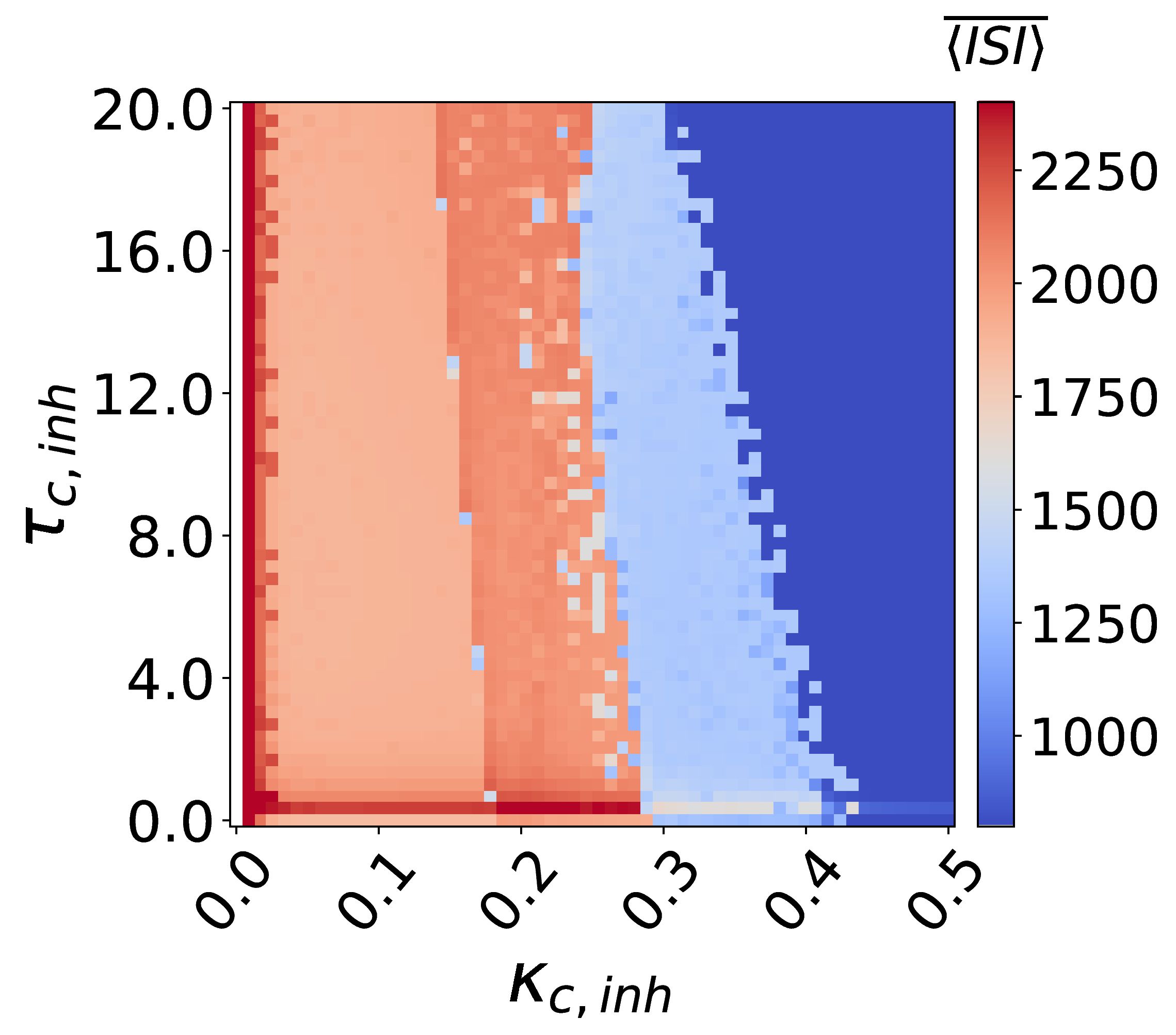}\hspace{0.2cm}
\raisebox{1.1cm}{\includegraphics[width=0.17\textwidth]{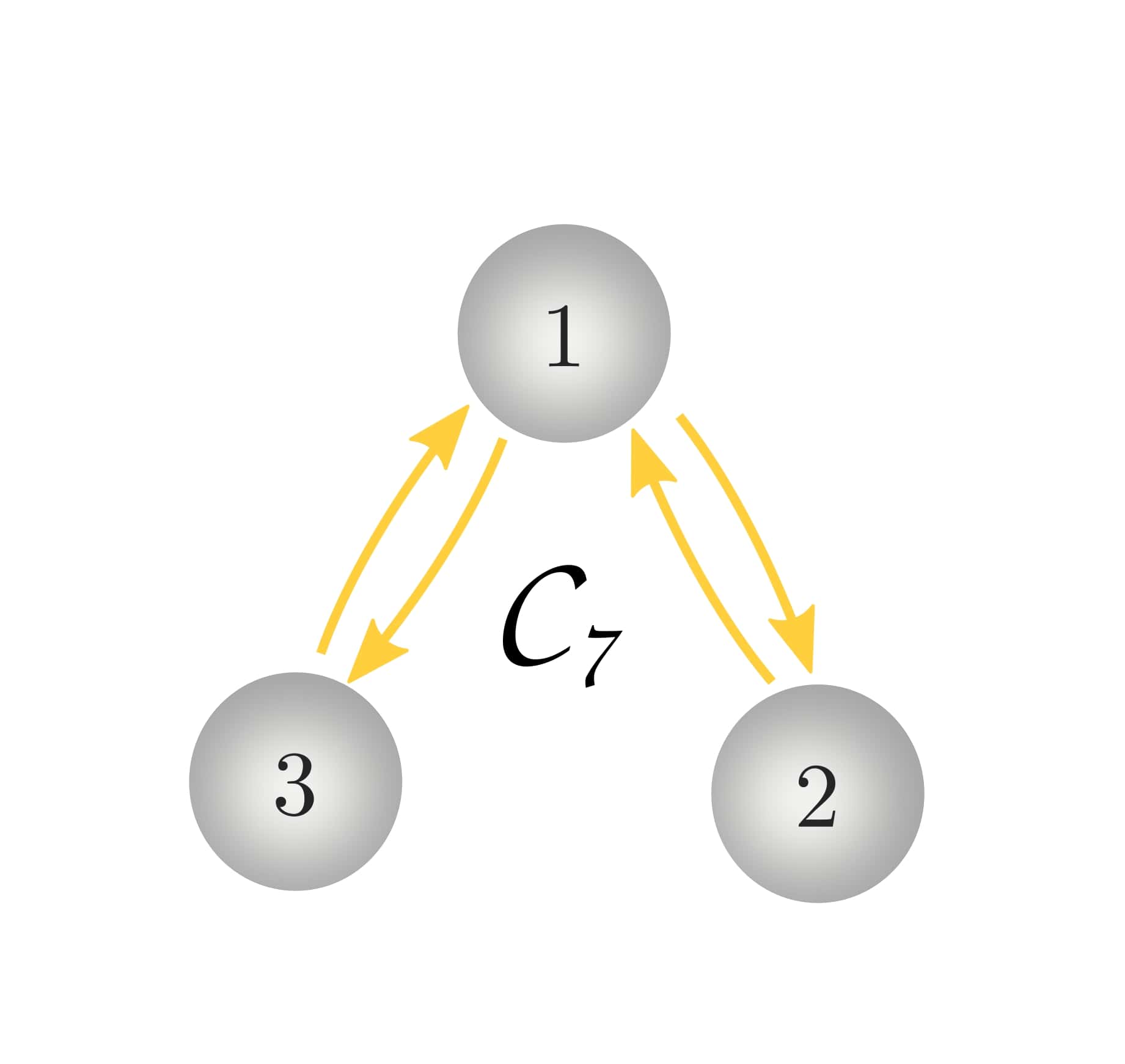}}}
\caption{Excitability maps (left) and corresponding topology (right) of  motif layer networks with inhibitory chemical synapses. $\langle\overline{\mathrm{ISI}}\rangle$ is color coded with the white region representing the excitable regime and the colored regions the oscillatory regimes. Notice that $\mathcal{C}_7$ does not admit excitability. $v_l=1.515<v_{\mathbb{H}}$, $\varepsilon=0.0005$.}
\label{fig:stability_different_topologies_chem2}
\end{figure}

To guarantee the excitability of each of these motifs (where we fix $\varepsilon=0.0005$ and $v_l=1.515$ in each neuron), we compute two-parameter deterministic bifurcation diagrams with respect to the synaptic parameters ($\kappa_e$, $\tau_e$) or ($\kappa_c$, $\tau_c$). Simulations indicate that electrical and inhibitory chemical synapses can set the deterministic motifs into either an excitable or an oscillatory regime, depending on the values of the synaptic coupling strengths ($\kappa_e,\kappa_{\mathrm{c,inh}}\in[0.0,0.5]$) and time delays ($\tau_e, \tau_{\mathrm{c,inh}}\in[0.0,20.0]$), see Fig.\ref{fig:stability_different_topologies_chem} and Fig.\ref{fig:stability_different_topologies_chem2}. 
On the other hand, all the parameter values of the excitatory chemical synapses (figures not shown) set the deterministic motifs into the oscillatory regime. Hence, we do not investigate SISR in motifs layer networks with excitatory chemical synapses. 

Our simulations have also indicated that (figures not shown) the motifs $\mathcal{C}_2$ - $\mathcal{C}_5$, with only chemical inhibitory synapses, are always in the excitable regime for all the synaptic parameter values used. Moreover, Fig.\,\ref{fig:stability_different_topologies_chem2}\textbf{(c)} indicates that the $\mathcal{C}_7$ topology with chemical inhibitory synapses does not admit excitability. Hence, we also exclude the $\mathcal{C}_7$ topology from our investigations of SISR.

\begin{figure}%[H]
\centering
\includegraphics[width=0.270\textwidth]{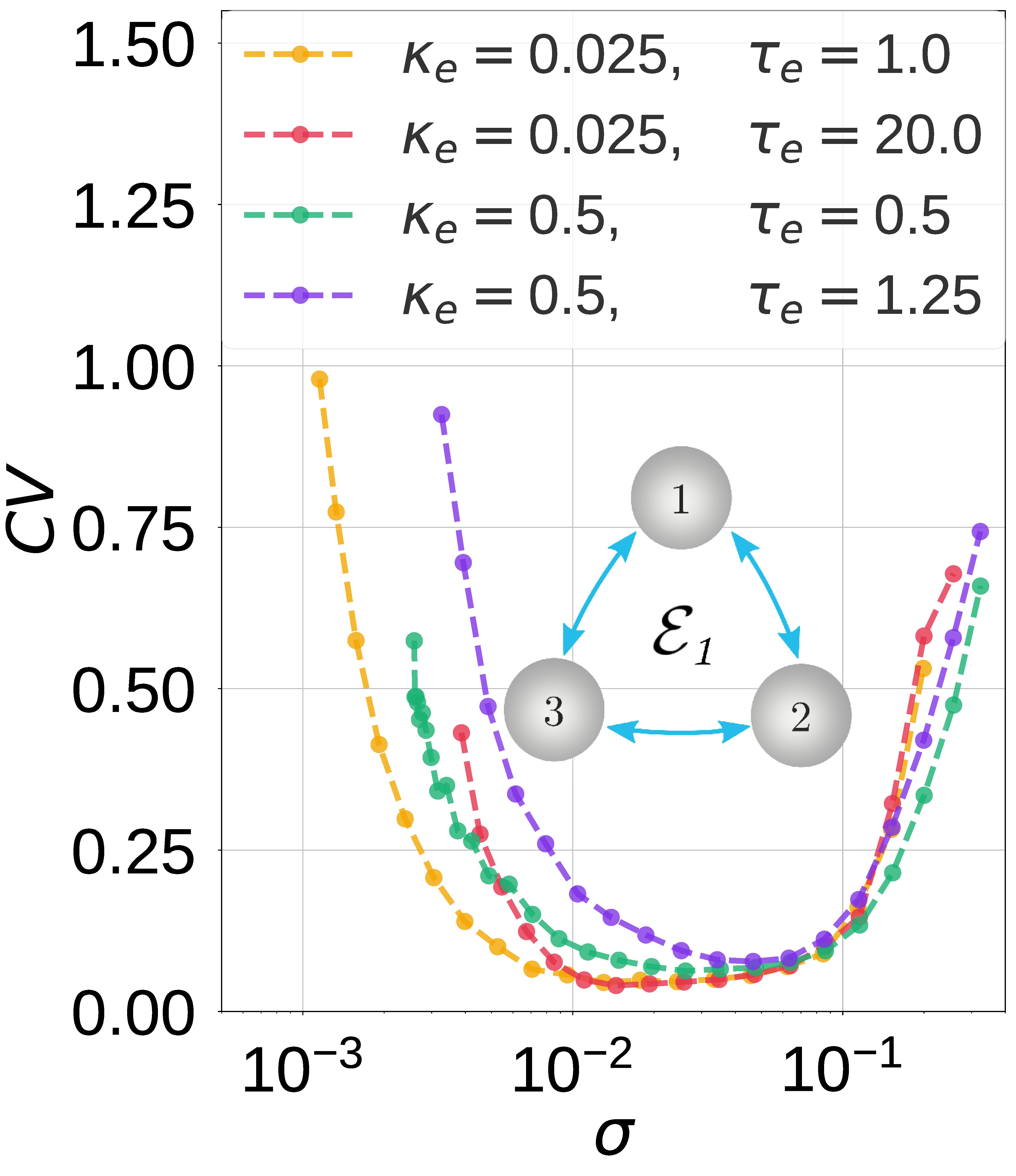}\includegraphics[width=0.215\textwidth]{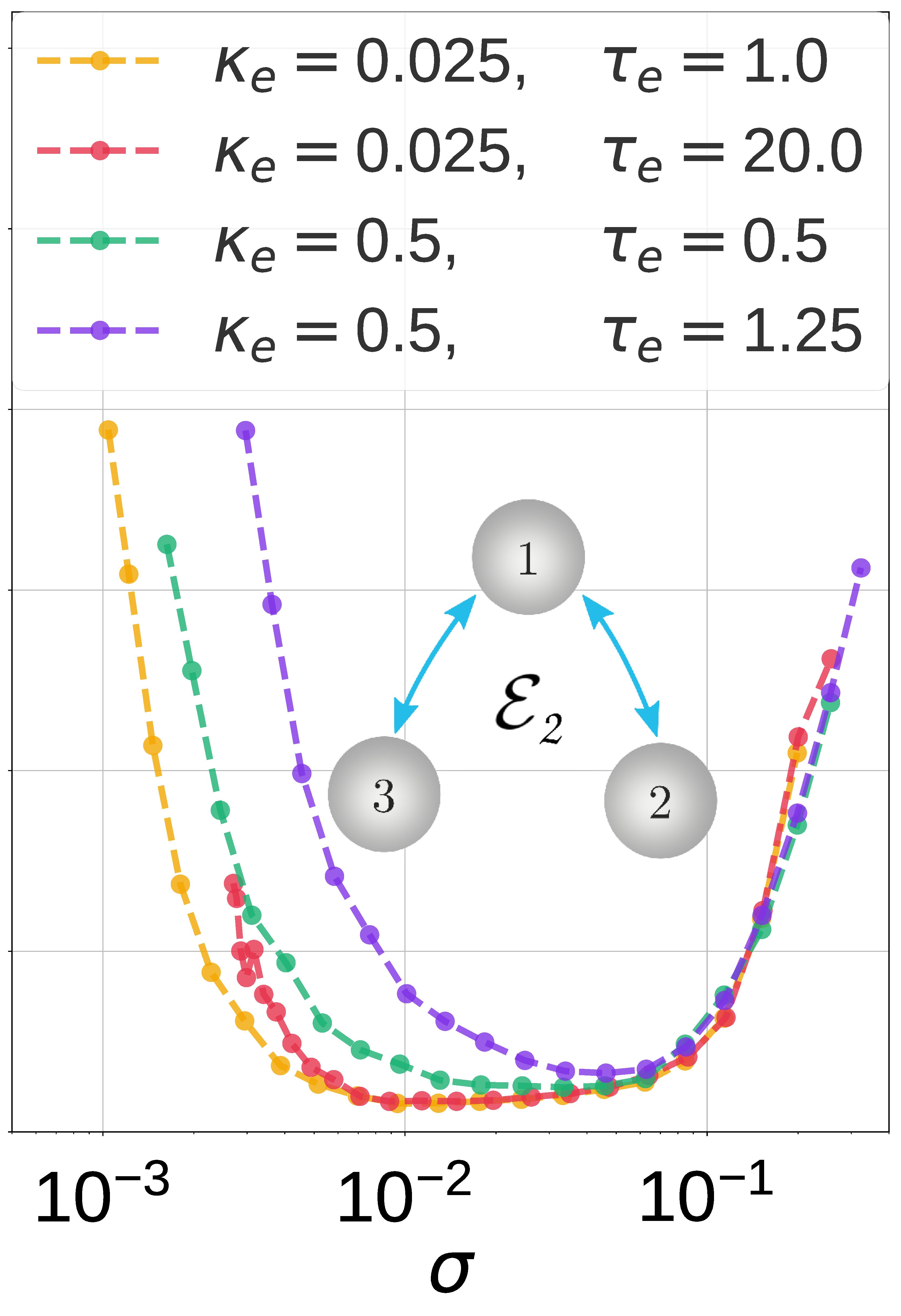}
\caption{Coefficient of variation $\mathrm{CV}$ against noise amplitude $\sigma$ for parameter combinations $(\kappa_{\mathrm{e}},\tau_{\mathrm{e}})$ in the electrical motifs topology indicated. $v_l=1.515<v_{\mathbb{H}}$, $\varepsilon=0.0005$.}
\label{fig:CV_curves_3NMs}
\end{figure}

\begin{figure}%[!t]
\centering
\includegraphics[width=0.270\textwidth]{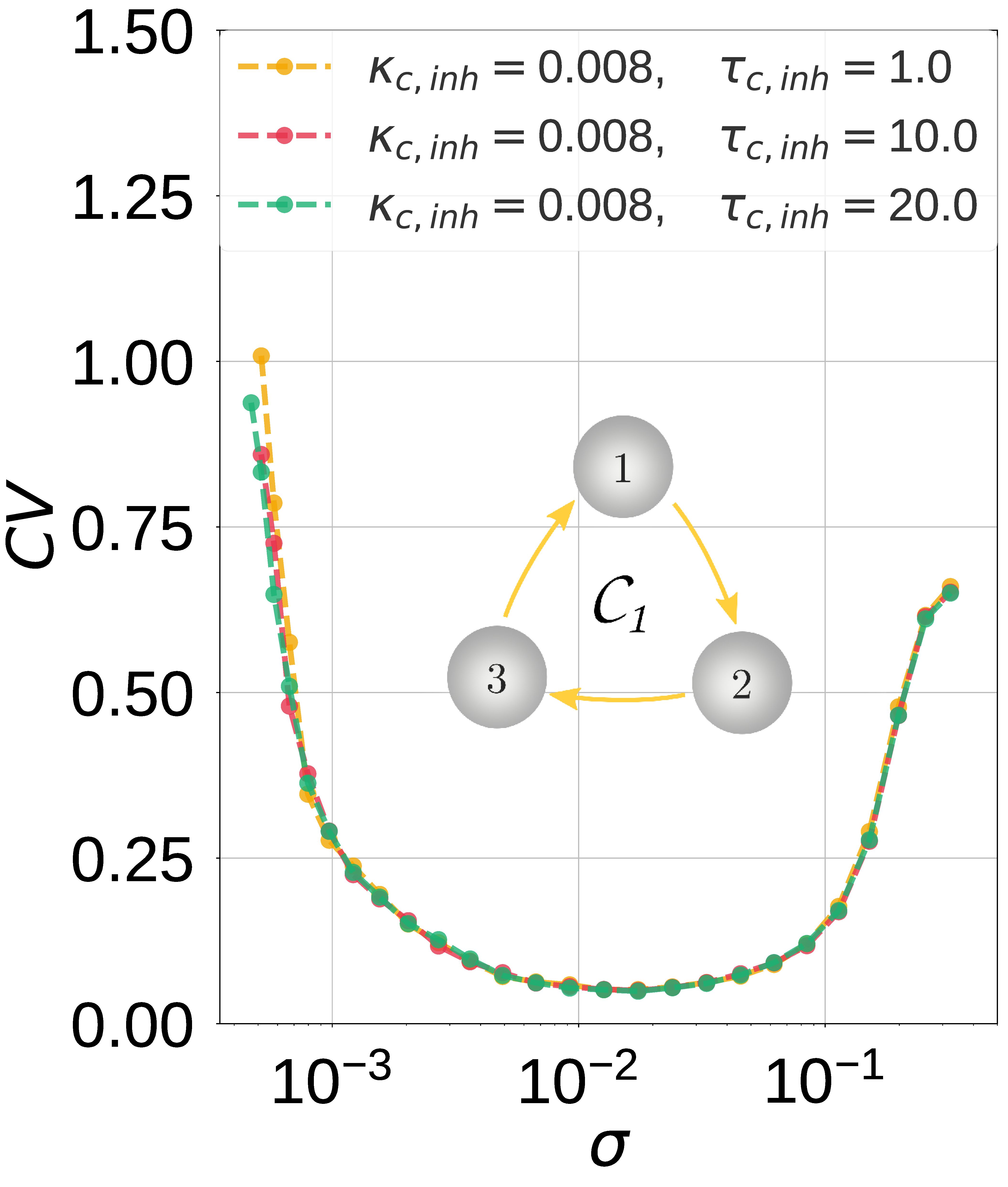}\includegraphics[width=0.215\textwidth]{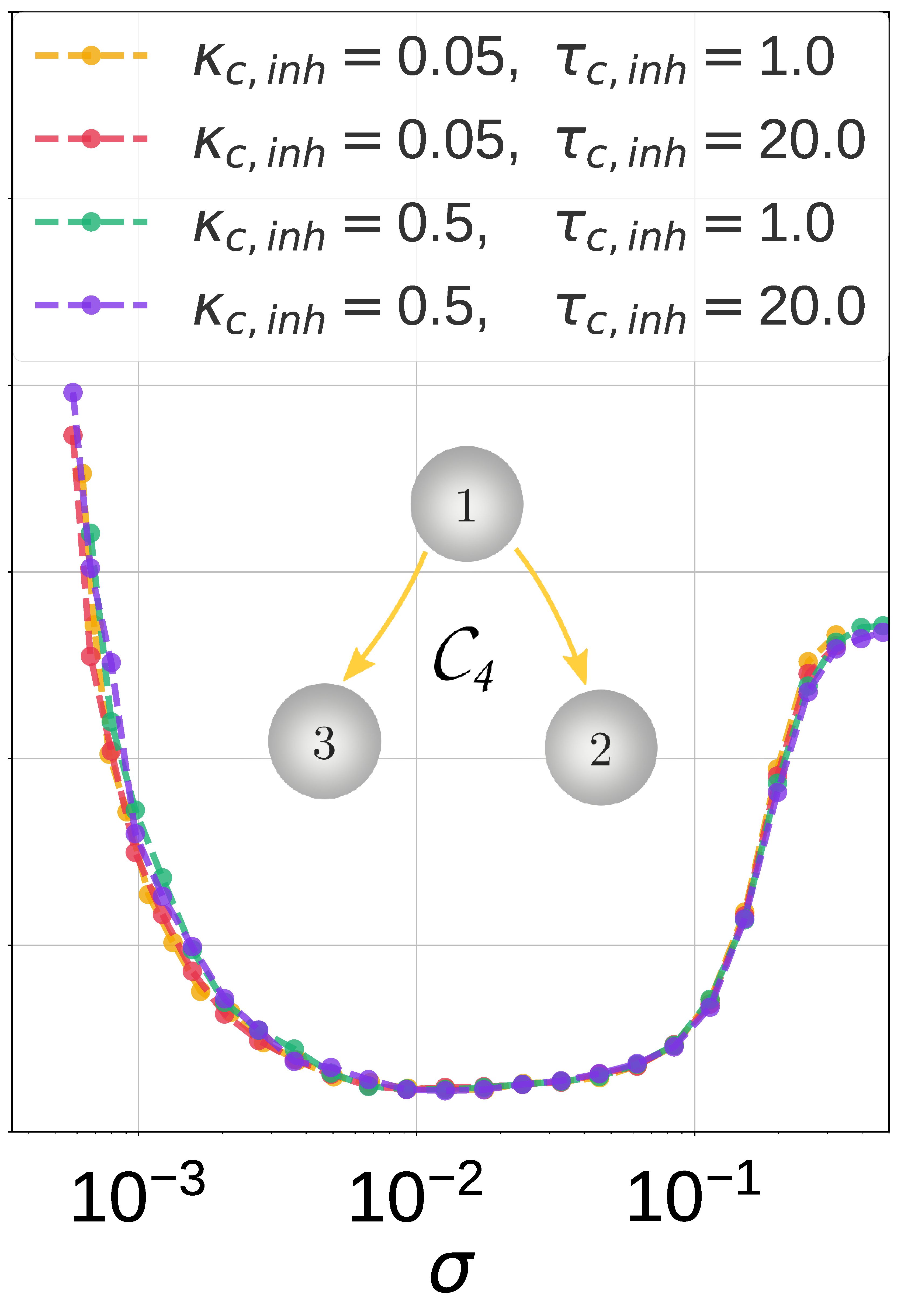}\\
\includegraphics[width=0.270\textwidth]{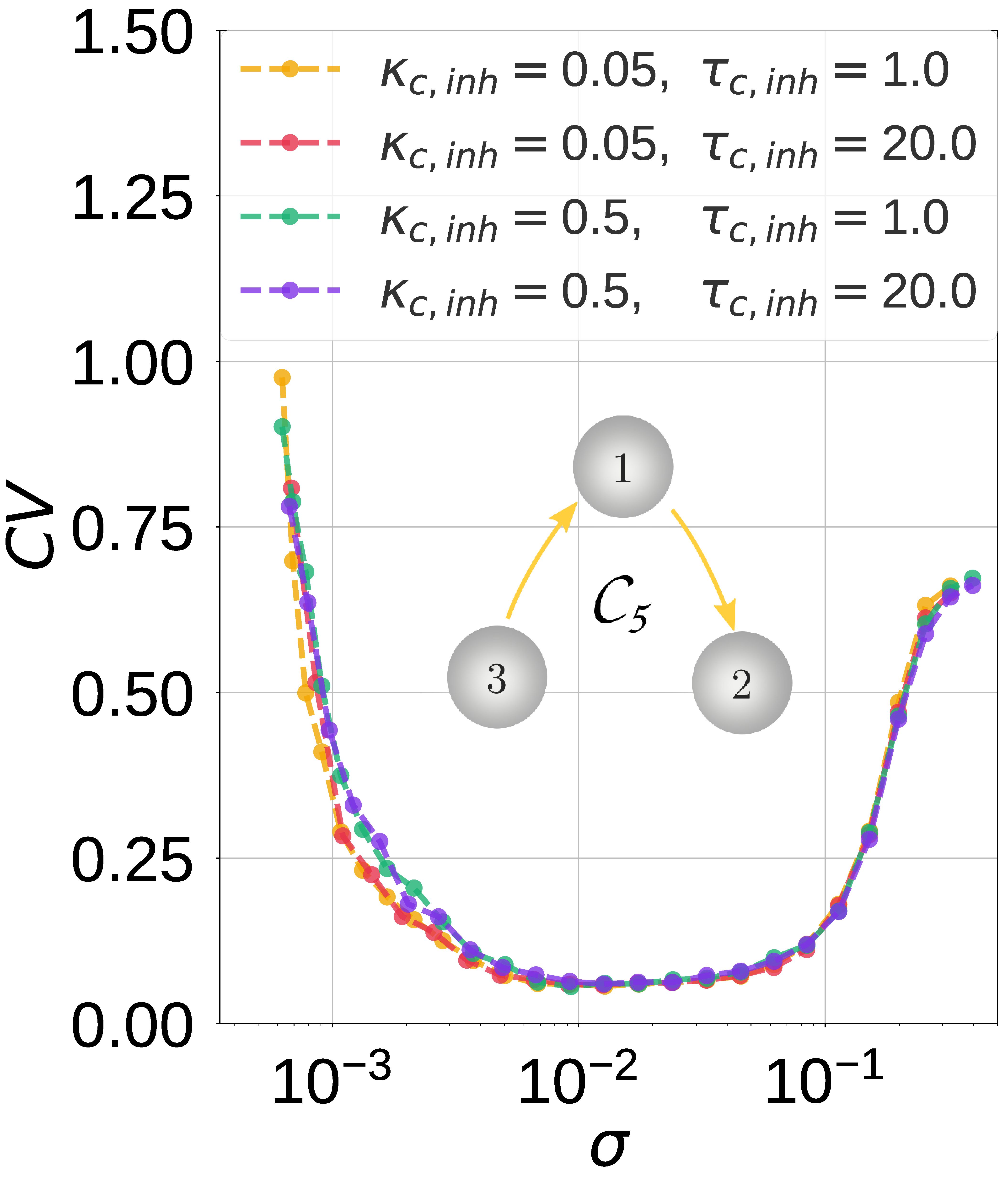}\includegraphics[width=0.215\textwidth]{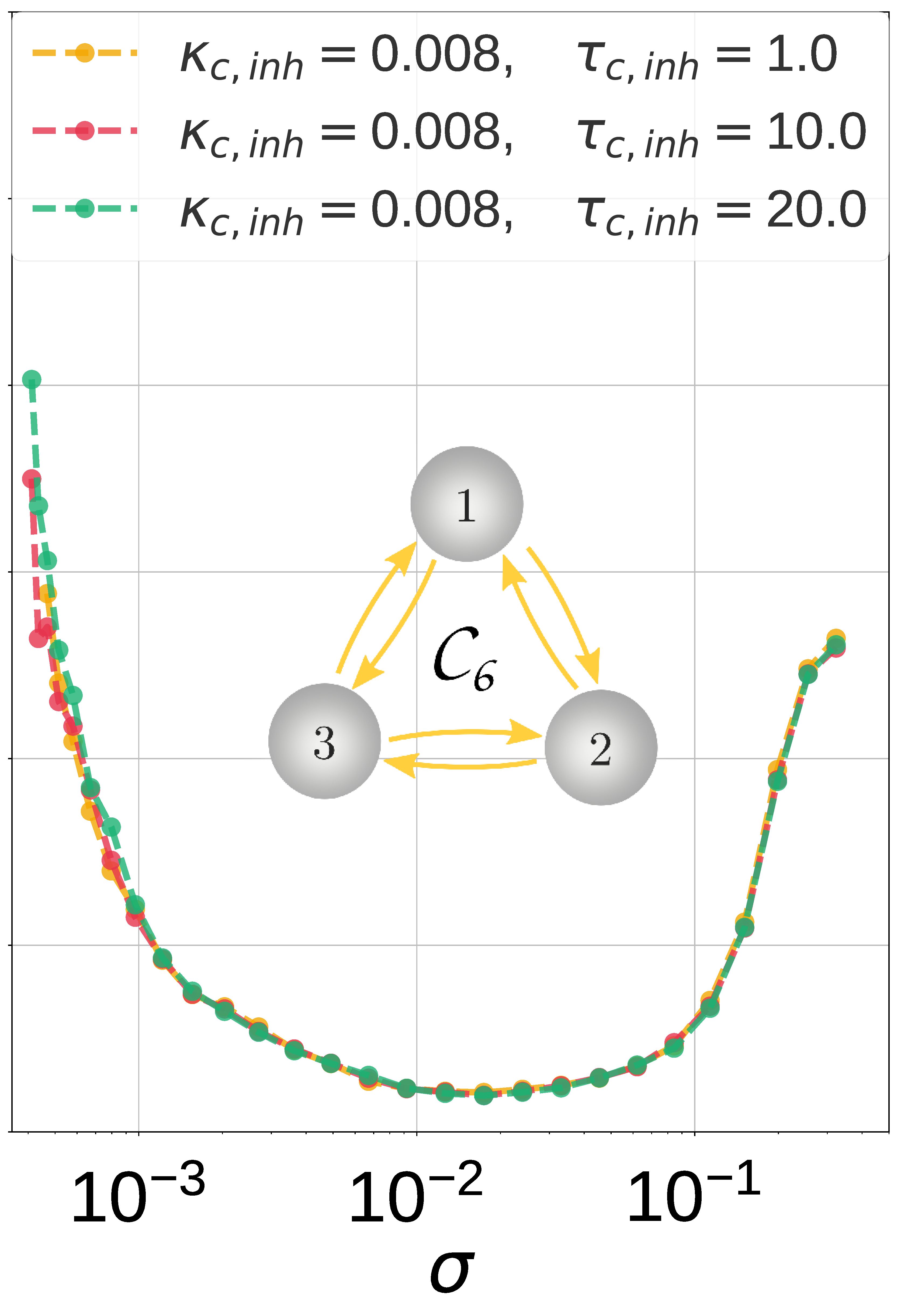}
\caption{Coefficient of variation $\mathrm{CV}$ against noise amplitude $\sigma$ for parameter combinations $(\kappa_{\mathrm{c,inh}},\tau_{\mathrm{c,inh}})$ in the inhibitory chemical motifs topology indicated. In these motifs, SISR is more robust to parametric changes than in the motifs considered in Fig.\ref{fig:CV_curves_3NMs3}.  $v_l=1.515<v_{\mathbb{H}}$, $\varepsilon=0.0005$.}
\label{fig:CV_curves_3NMs2}
\end{figure}

\begin{figure}%[!ht]
\centering
\includegraphics[width=0.270\textwidth]{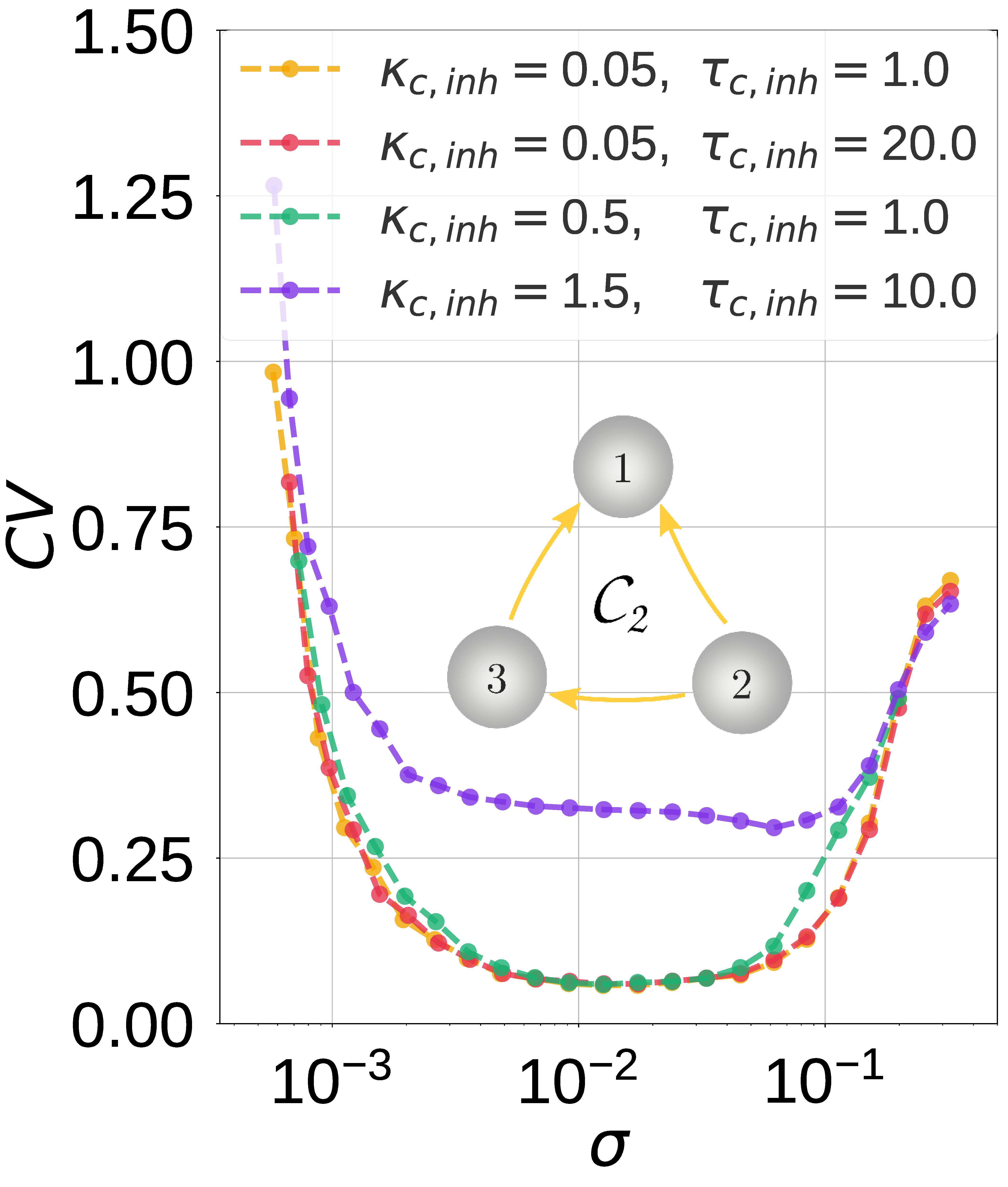}\includegraphics[width=0.215\textwidth]{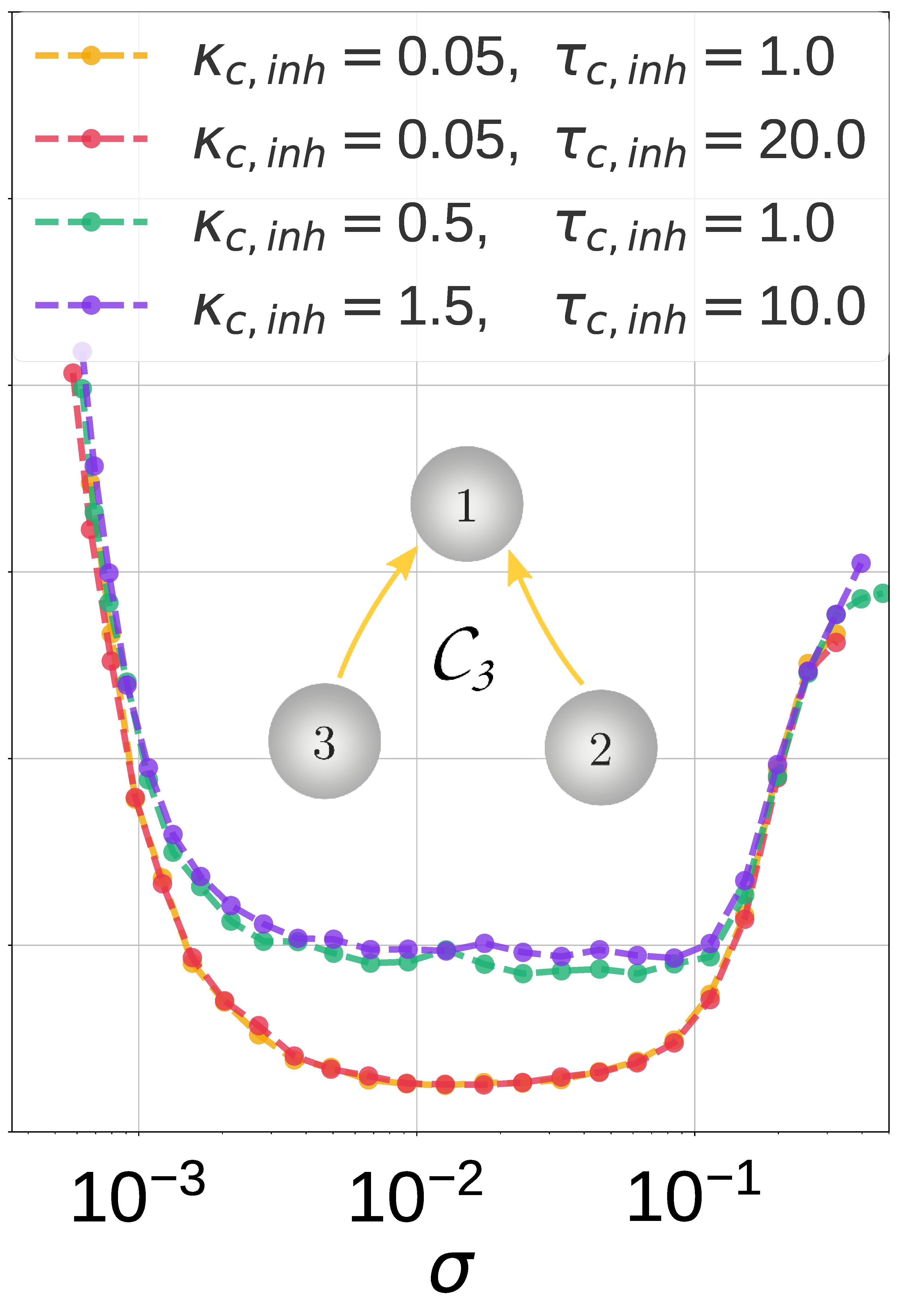}
\caption{Coefficient of variation $\mathrm{CV}$ against noise amplitude $\sigma$ for parameter combinations $(\kappa_{\mathrm{c,inh}},\tau_{\mathrm{c,inh}})$ in the inhibitory chemical motifs topology indicated. In these motifs the degree is SISR can be significantly deteriorated (see the purple curves in each panel) as opposed to the degree in the motifs in  $v_l=1.515<v_{\mathbb{H}}$, $\varepsilon=0.0005$.}
\label{fig:CV_curves_3NMs3}
\end{figure}

In Figs.\ref{fig:CV_curves_3NMs} - \ref{fig:CV_curves_3NMs3}, we present the variations of the $\mathrm{CV}$ curves against noise intensity for the indicated motif topology and different intra-motif time-delayed coupling parameter values. 

In Fig.\,\ref{fig:CV_curves_3NMs}, with the electrical motifs $\mathcal{E}_1$ and $\mathcal{E}_2$, respectively, we observe that weaker coupling strengths $\kappa_{\mathrm{e}}$ and shorter time delays $\tau_e$ lead to a higher degree of SISR, especially at weaker noise intensities. There are no significant differences in the degree of SISR in $\mathcal{E}_1$ and $\mathcal{E}_2$.

In Fig.\ref{fig:CV_curves_3NMs2}, with the inhibitory chemical motifs $\mathcal{C}_1$, $\mathcal{C}_4$, $\mathcal{C}_5$, and $\mathcal{C}_6$, we mainly observe that the high degree of SISR achieved in the corresponding motifs are quite robust to parametric changes in the synapses --- the $\mathrm{CV}$ curves remain very low and the intervals of the noise intensity in which these high degrees are achieved remain unchanged as the parameters change.

In  Fig.\,\ref{fig:CV_curves_3NMs3}, the $\mathrm{CV}$ curves of motifs $\mathcal{C}_2$ and $\mathcal{C}_3$ show significant fluctuations in the degree of SISR as the parameters change. In these two cases, stronger coupling strengths and longer time delays lead to a lower degree of SISR. See, e.g., the purple $\mathrm{CV}$ curves in panels of Fig.\,\ref{fig:CV_curves_3NMs3}, where $\kappa_{\mathrm{c,inh}}=1.5$ and  $\tau_{\mathrm{c,inh}}=10.0$. We observe that these curves are shifted upwards to higher $\mathrm{CV}$ values compared to other curves, thus reducing the degree of SISR. The minimum $\mathrm{CV}$ values of these purple curves are: $\mathrm{CV}_{min}=0.293$ in the left panel and $\mathrm{CV}_{min}=0.232$ in the right panel of Fig.\,\ref{fig:CV_curves_3NMs3}. 

Comparing the degree of SISR in isolated neurons and motif networks, we can conclude that the degree of SISR in motifs can be as good as in the isolated neuron, but not better. As we can see from  Figs.\,\ref{fig:CV_curves_3NMs} - \ref{fig:CV_curves_3NMs3}, as the strength of synaptic couplings increases, the interval of the noise amplitude within which the degree of SISR is high decreases and the minimum of the CV curves rises to a small but significant value. But as the strength of these synaptic couplings decreases, this interval of noise increases and the minimum of the CV descends to lower values. Eventually, when the synaptic couplings decrease to vanishingly small values (e.g., $\kappa_{c,inh}=0.008$) or even to zero (in which case we have isolated neurons) the degree of SISR in the motifs becomes very close to the degree of SISR in the isolated neuron and identical when the synaptic coupling becomes zero. Therefore, we conclude that the motif can at most do as well as an isolated neuron, but not better.

A natural and interesting question to investigate is whether this significantly poor degree of SISR can be enhanced in the motifs $\mathcal{C}_2$ and $\mathcal{C}_3$. In the sequel, we present two enhancement strategies for SISR in these motifs.

\subsection{Autaptic enhancement of SISR in single motifs}
In this subsection, we present one enhancement strategy of SISR in single motifs that is based on the use of autapses. To illustrate the efficacy of this strategy, we use the motifs $\mathcal{C}_2$ and $\mathcal{C}_3$ from  Fig.\,\ref{fig:CV_curves_3NMs3}, with time-delayed coupling fixed at $\kappa_{\mathrm{c,inh}}=1.5$ and $\tau_{\mathrm{c,inh}}=10.0$. With this setting and in the absence of autapses, the degree of SISR is relatively poor, i.e.,  $\mathrm{CV}_{min}=0.293$ in the motif $\mathcal{C}_2$ and $\mathrm{CV}_{min}=0.232$ in the motif $\mathcal{C}_3$ as indicated by the purple curves in the panels of Fig.\,\ref{fig:CV_curves_3NMs3}.

\begin{figure}%[!ht]
\centering
{\hspace{0.80cm}{\includegraphics[width=0.175\textwidth]{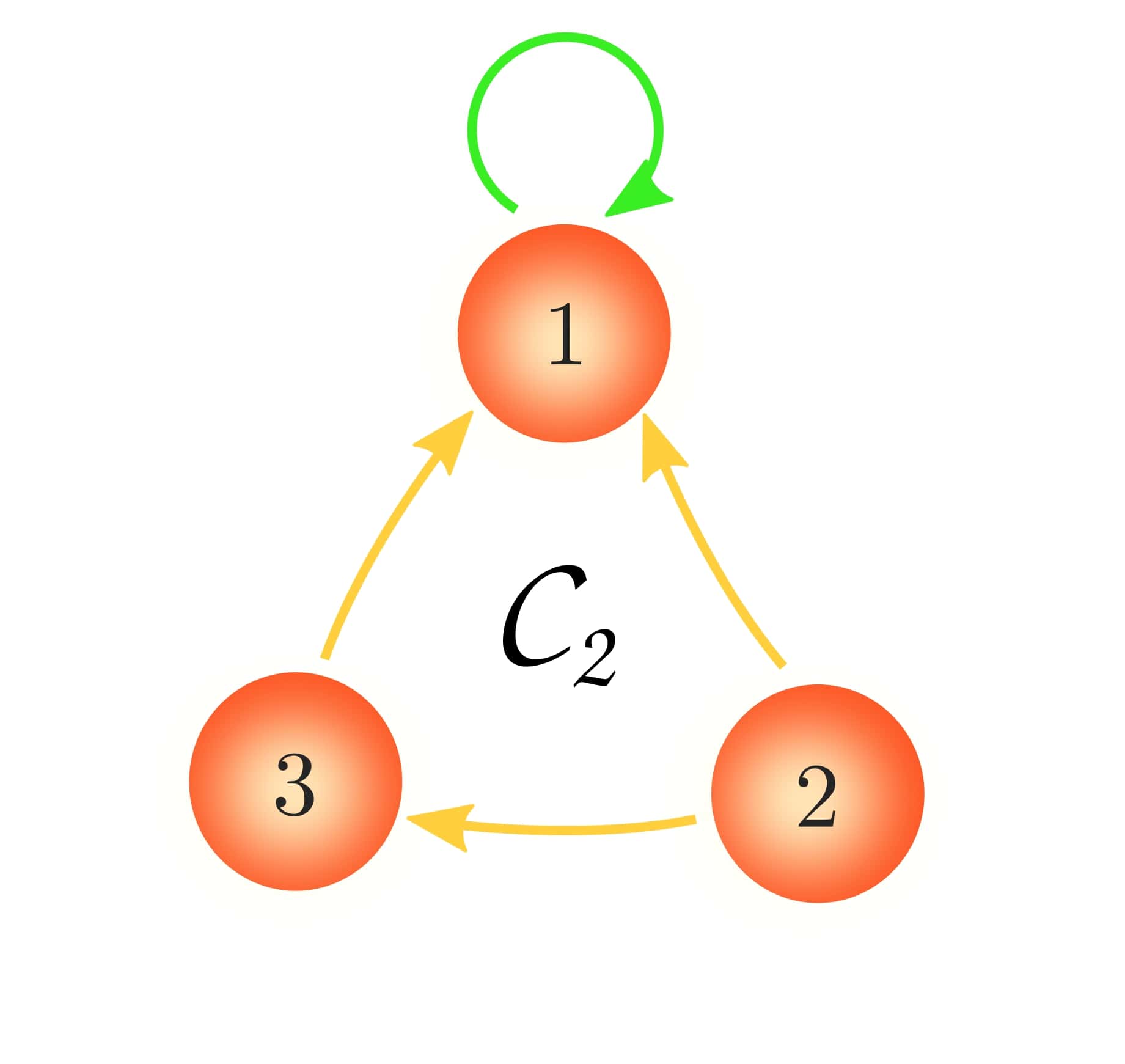}}}\vspace{-0.6cm}
\subfigure[]{\includegraphics[width=0.22\textwidth]{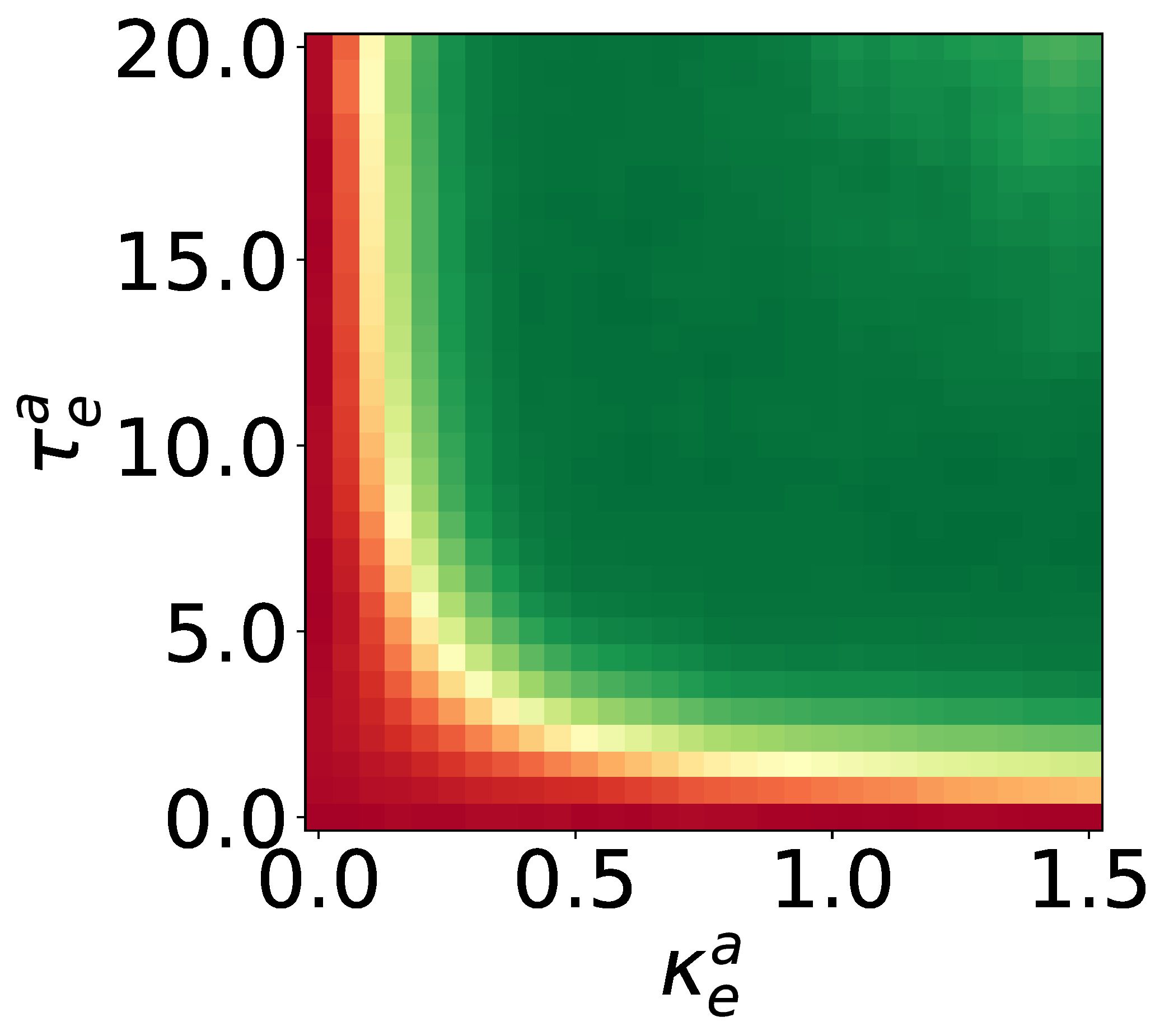}}\subfigure[]{\includegraphics[width=0.25\textwidth]{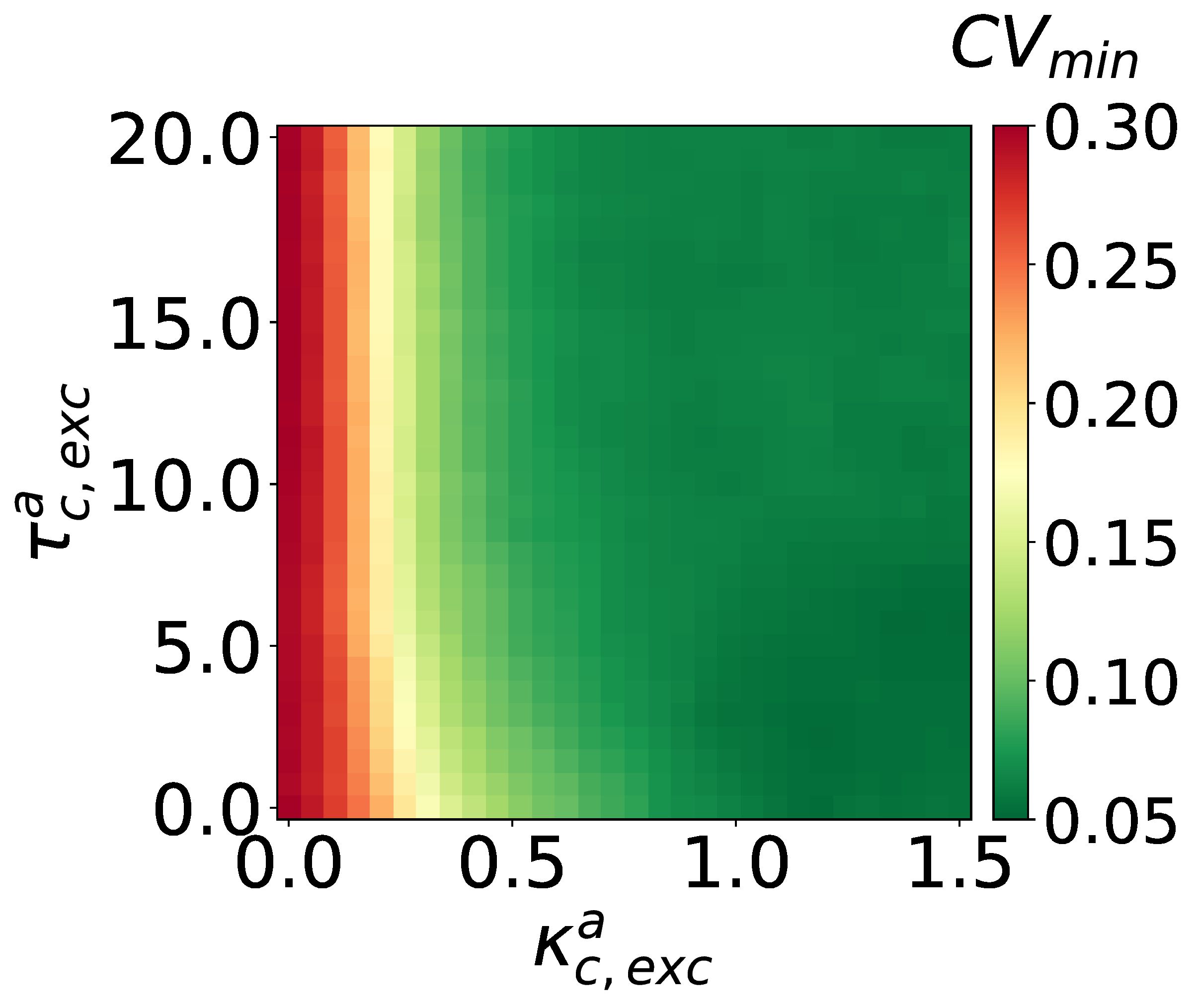}}
\caption{Minimum coefficient of variation $\mathrm{CV}_{\mathrm{min}}$ for the motif $\mathcal{C}_2$ against: \textbf{(a)} electrical autaptic parameters $(\kappa^a_e,\tau^a_e)$ and \textbf{(b)} excitatory chemical autaptic parameters $(\kappa^a_{\mathrm{c,exc}},\tau^a_{\mathrm{c,exc}})$.  There is an enhancement in the degree of SISR from $\mathrm{CV}_{\mathrm{min}}=0.293$ in the absence of autapses to $\mathrm{CV}_{\mathrm{min}}\approx0.06$ in their presence. $\kappa_{\mathrm{c,inh}}=1.5$, $\tau_{\mathrm{c,inh}}=10.0$, $v_l=1.515<v_{\mathbb{H}}$, $\varepsilon=0.0005$.} 
\label{fig:CV_min_autaptic_effect1}
\end{figure}

\begin{figure}%[!ht]
\centering
{\hspace{0.80cm}{\includegraphics[width=0.175\textwidth]{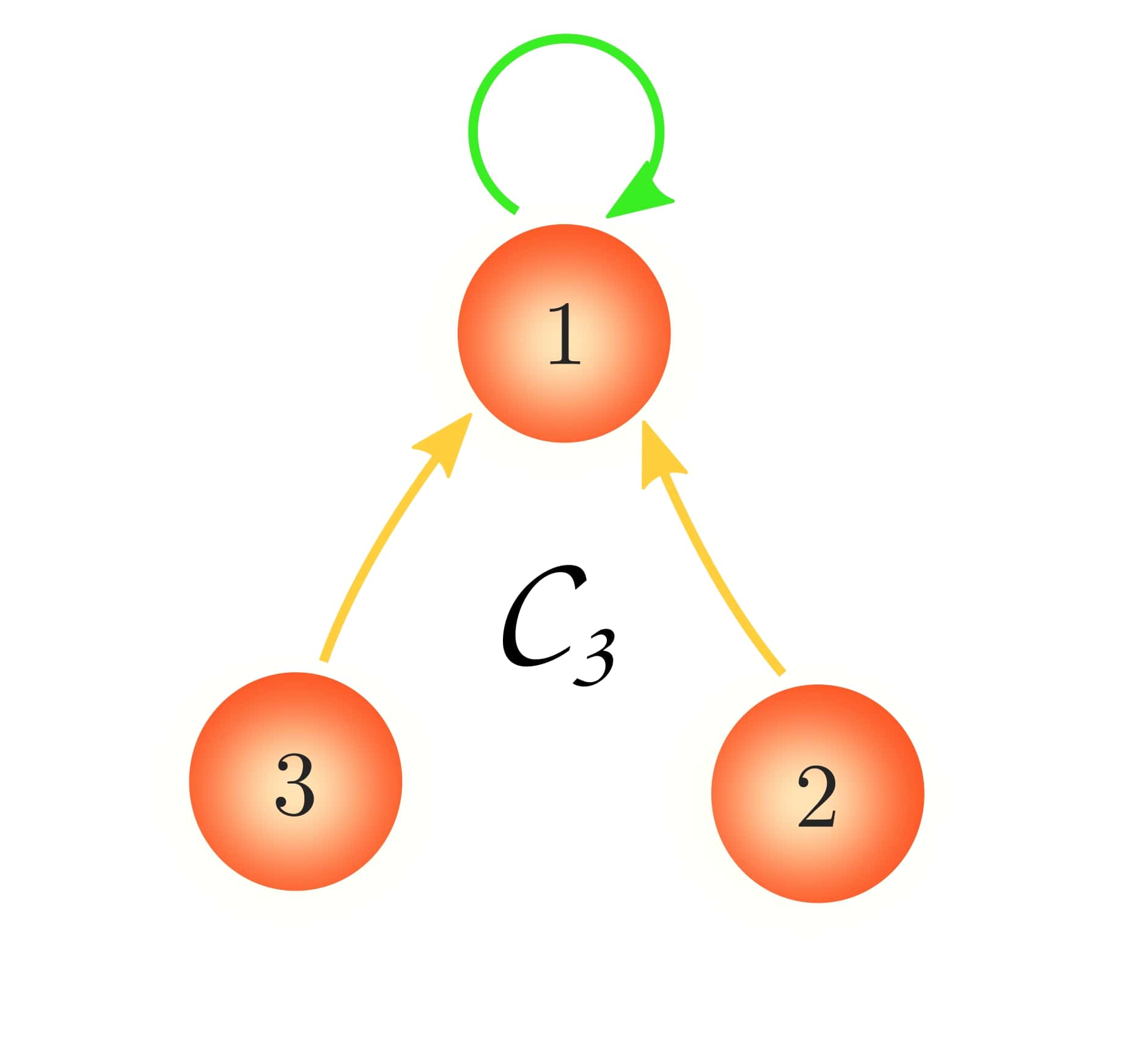}}}\vspace{-0.6cm}
\raisebox{0.00cm}{\subfigure[]{\includegraphics[width=0.22\textwidth]{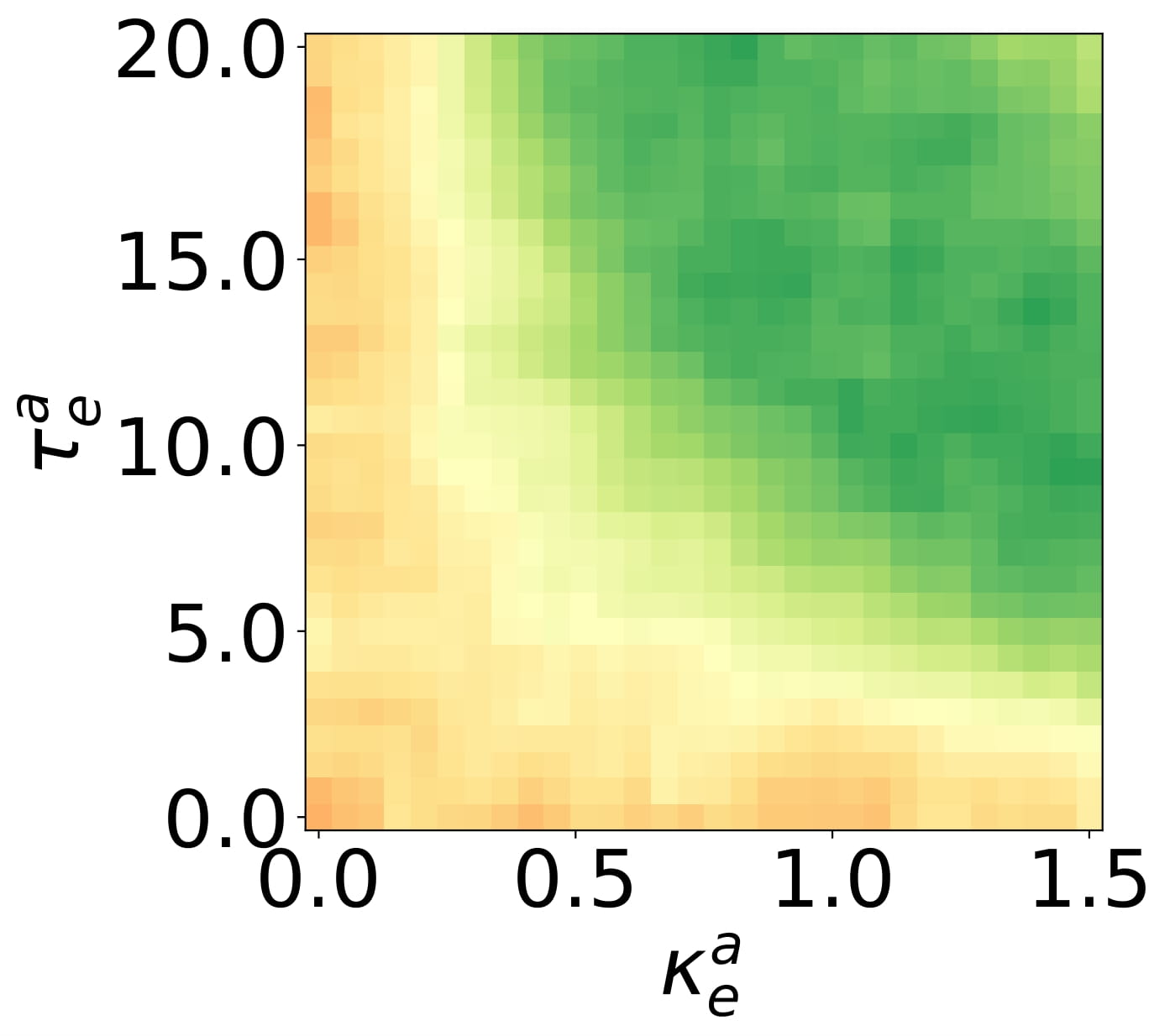}}}\subfigure[]{\includegraphics[width=0.25\textwidth]{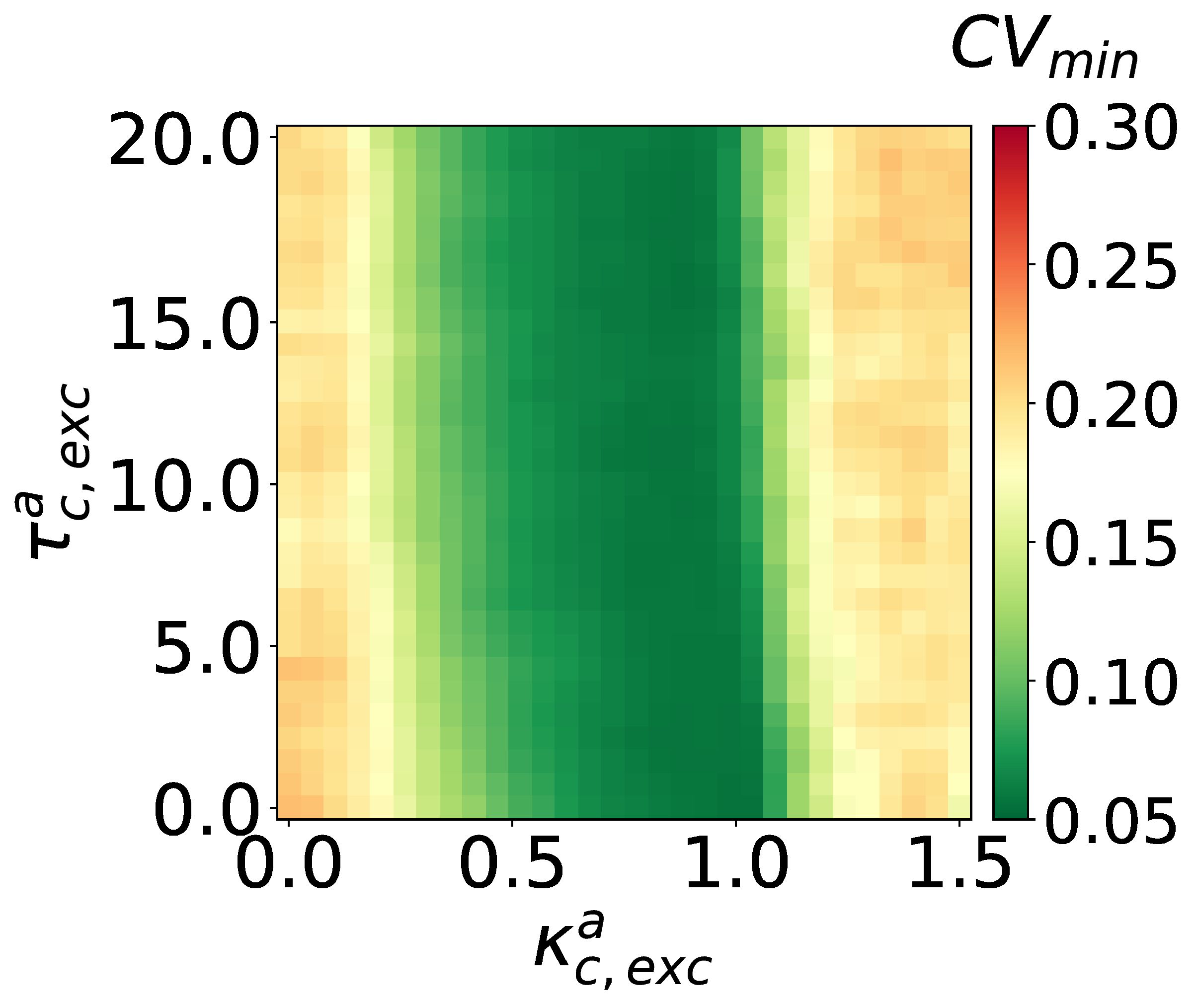}}
\caption{Minimum coefficient of variation $\mathrm{CV}_{\mathrm{min}}$ for the motif $\mathcal{C}_3$ against: \textbf{(a)} electrical autaptic parameters $(\kappa^a_e,\tau^a_e)$ and \textbf{(b)} excitatory chemical autaptic parameters $(\kappa^a_{\mathrm{c,exc}},\tau^a_{\mathrm{c,exc}})$.  There is an enhancement in the degree of SISR from $\mathrm{CV}_{\mathrm{min}}=0.232$ in the absence of autapses to $\mathrm{CV}_{\mathrm{min}}\approx 0.09$ and $\mathrm{CV}_{\mathrm{min}}\approx 0.06$ in \textbf{(a)} and \textbf{(b)}, respectively, in the presence of autapses. $\kappa_{\mathrm{c,inh}}=1.5$, $\tau_{\mathrm{c,inh}}=10.0$, $v_l=1.515<v_{\mathbb{H}}$, $\varepsilon=0.0005$.} 
\label{fig:CV_min_autaptic_effect2}
\end{figure}

The goal of this strategy is to lower the value of these  $\mathrm{CV}_{\mathrm{min}}$ using an autapse with appropriate parameter values. Our simulations have indicated (not shown) that an inhibitory chemical autapse is not effective with this strategy. The reason is that this type of autapse puts the motifs into the (undesired) oscillatory regime in the absence of noise. Inserting an electrical autapse (with $\kappa^a_{\mathrm{e}}\in [0.0,1.5]$, $\tau^a_{\mathrm{e}}\in[0.0,20.0]$) or an excitatory chemical autapse (with $\kappa^a_{\mathrm{c,exc}}\in [0.0,1.5]$, $\tau^a_{\mathrm{c,exc}}\in [0.0,20.0]$) on neuron number one of these motifs, kept the motifs in the (desired) excitable regime. Furthermore, our extensive numerical investigations have indicated that this enhancement strategy is most efficient when the electrical autapse or the excitatory chemical autapse is attached only to the neuron with the highest in-degree, i.e., the neuron number one of the motifs. Thus, we only show the results of this case in Fig.\,\ref{fig:CV_min_autaptic_effect1} for motif $\mathcal{C}_2$ and in Fig.\,\ref{fig:CV_min_autaptic_effect2} for motif $\mathcal{C}_3$. It is worth noting that this result on the enhancement of SISR by an autapse attached to the neuron with the highest in-degree may not be generally robust and may not apply to other types of motif or larger neural networks not considered in this study.

Fig.\,\ref{fig:CV_min_autaptic_effect1}\textbf{(a)} and \textbf{(b)}, respectively, show the color coded variation of the $\mathrm{CV}_{\mathrm{min}}$ of the motif $\mathcal{C}_2$ with respect to the time-delayed electrical or excitatory chemical autaptic couplings (represented by the green self-feedback loop) attached to neuron number one of the motif. We observe that these autapses can significantly improve the degree of SISR in the motif $\mathcal{C}_2$. But, in the presence autapses, $\mathcal{C}_2$ with $\kappa_{\mathrm{c,inh}}=1.5$ and $\tau_{\mathrm{c,inh}}=10.0$ has a very low $\mathrm{CV}_{\mathrm{min}}\approx0.06$, i.e., a much higher degree of SISR. We further notice that the excitatory chemical autapse outperforms the electrical autapse in this enhancement strategy. That is, the former autapse provides a larger range of parameters values in which a high degree of SISR is achieved than the latter, especially at stronger couplings and shorter time delays.

In Fig.\,\ref{fig:CV_min_autaptic_effect2}\textbf{(a)} and \textbf{(b)}, respectively, we show the color coded variation of the $\mathrm{CV}_{\mathrm{min}}$ of the motif $\mathcal{C}_3$ against the time-delayed electrical and excitatory chemical autaptic couplings. In this case, we observe that the degree the of SISR is considerably enhanced, with a drop from $\mathrm{CV}_{\mathrm{min}}=0.232 $ to $\mathrm{CV}_{\mathrm{min}}\approx0.09$ for an electrical autapse and $\mathrm{CV}_{\mathrm{min}}\approx0.06$ for an  excitatory chemical autpase. However, in terms of parameter ranges, the electrical autapse is more efficient than the excitatory chemical autapse for stronger couplings $\kappa^a_e>1.0$ and longer delays $\tau^a_e>5.0$.  The excitatory chemical autapse is better at enhancing SISR for intermediate  $0.25<\kappa^a_{c,exc}<1.0$ coupling strengths and all time delays. Moreover, we can also see that with this autaptic enhancement strategy, the degree of SISR is best enhanced in motif topology $\mathcal{C}_2$ than $\mathcal{C}_3$. 

\subsection{Enhancement of SISR based on  multiplexing}
\begin{figure}%[ht!]
\centering
\hspace{0.9cm}\subfigure[$\mathcal{C}_2$~~$\mathrm{controls}$~~$\mathcal{C}_2$]{\raisebox{0.155cm}{\raisebox{0.2cm}{\includegraphics[width=0.15\textwidth]{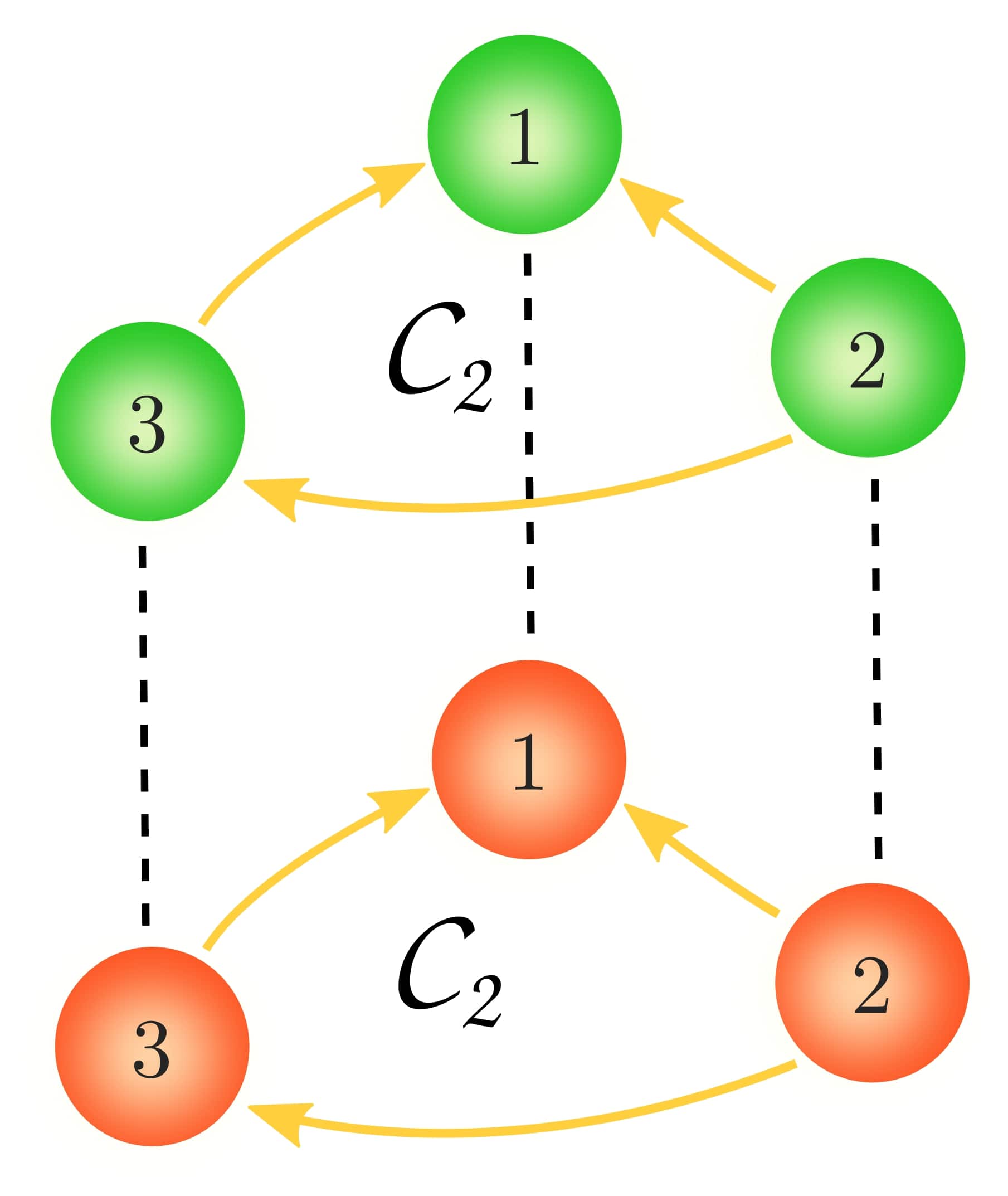}}}}\hspace{0.25cm}
\subfigure[excitability map of electrical multiplexing]{\includegraphics[width=0.25\textwidth]{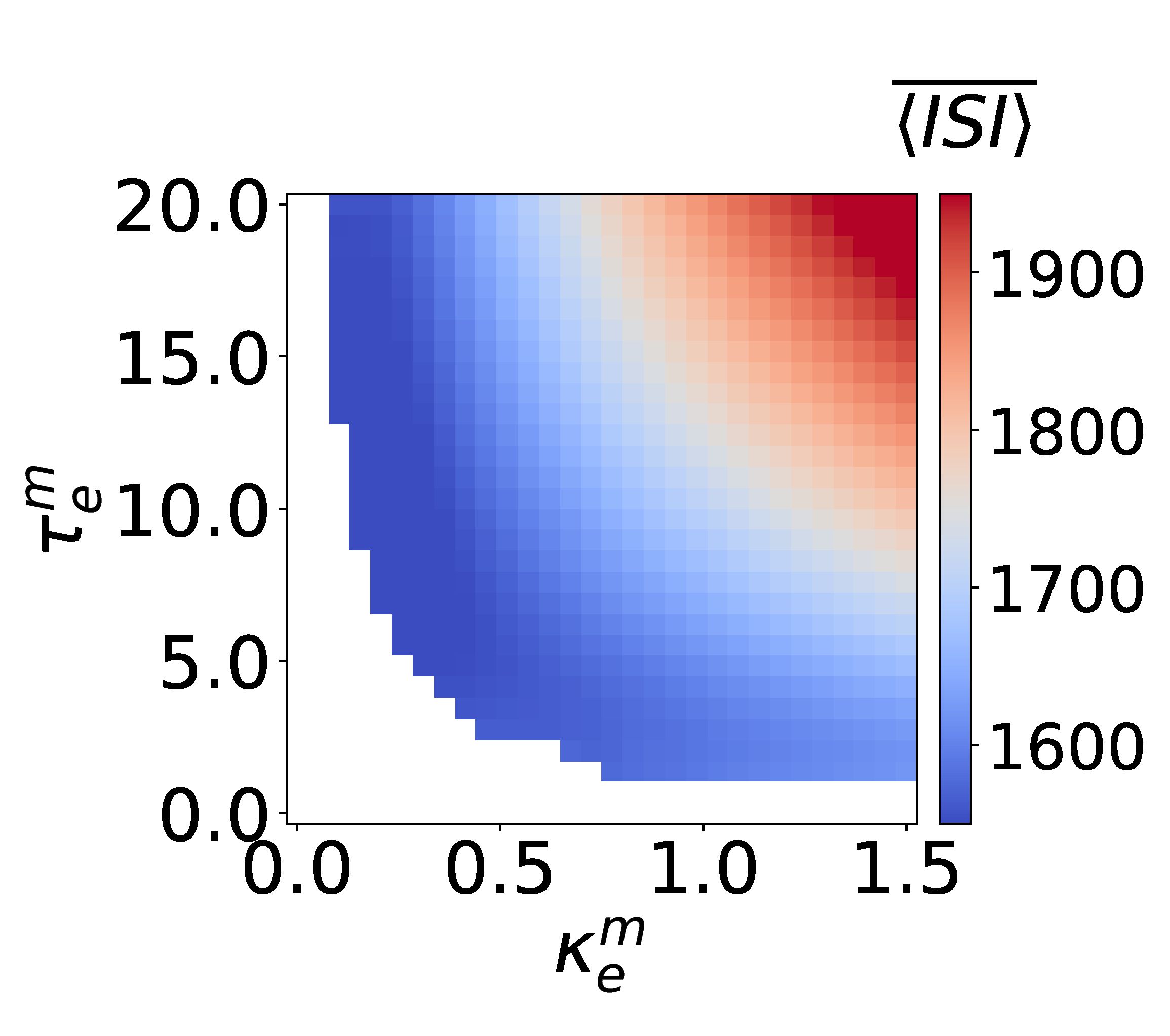}}
\subfigure[$\mathcal{C}_2$~$\xrightarrow{\mathrm{electrical}}$~$\mathcal{C}_2$]{\includegraphics[width=0.21\textwidth]{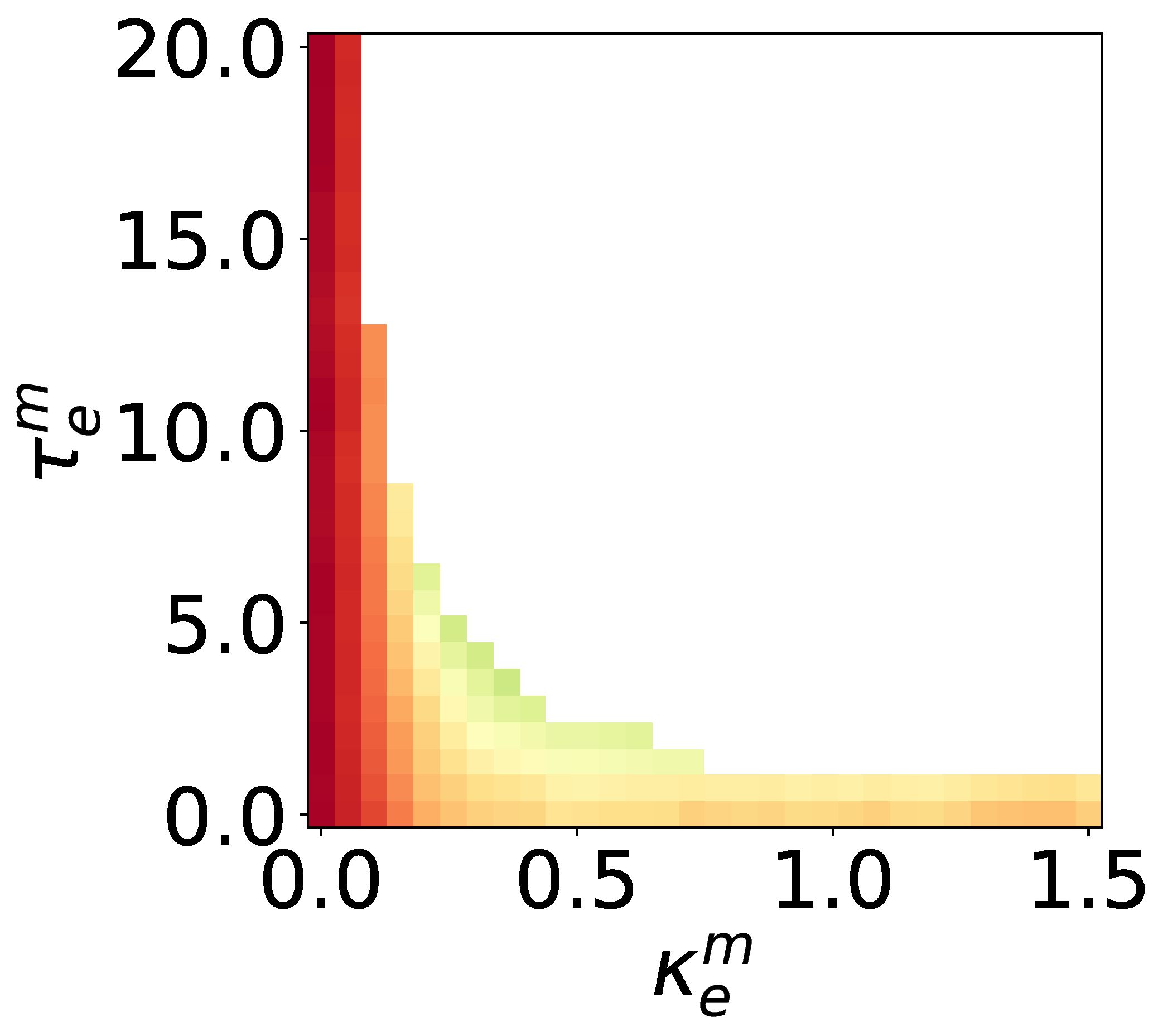}}
\subfigure[$\mathcal{C}_2$~$\xrightarrow{\mathrm{inhibitory~ chemical}}$~$\mathcal{C}_2$]
{\includegraphics[width=0.25\textwidth]{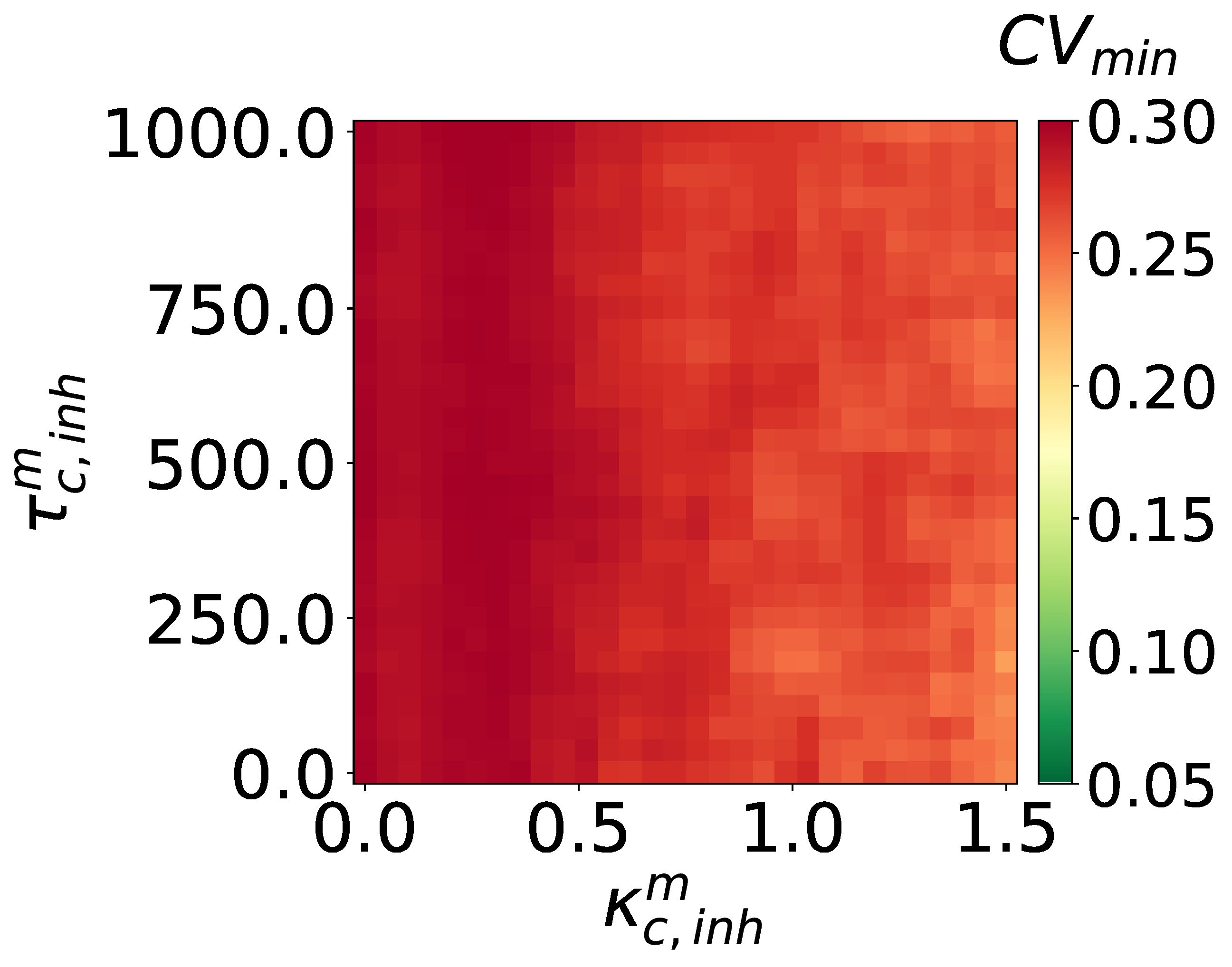}}
\caption{Excitability map in \textbf{(b)} showing $\langle\overline{\mathrm{ISI}}\rangle$ of the multiplexing scheme in \textbf{(a)} with electrical synapses in the absence of noise. $\mathrm{CV}_{\mathrm{min}}$ for the lower (in red) motif layer $\mathcal{C}_2$ against: electrical $(\kappa^m_e,\tau^m_e)$ and inhibitory chemical $(\kappa^m_{\mathrm{c,inh}},\tau^m_{\mathrm{c,inh}})$ multiplexing parameters in \textbf{(c)} and \textbf{(d)} respectively. 
There is an enhancement in the degree of SISR from $\mathrm{CV}_{\mathrm{min}}=0.293$ in the absence of multiplexing to $\mathrm{CV}_{\mathrm{min}}\approx 0.12$ and $\mathrm{CV}_{\mathrm{min}}\approx 0.25$ in \textbf{(c)} and \textbf{(d)} in the presence of multiplexing, respectively. $(\kappa_{\mathrm{c,inh}},\tau_{\mathrm{c,inh}})=(1.5,10.0)$ in the lower motif, $(\kappa_{\mathrm{c,inh}},\tau_{\mathrm{c,inh}})=(0.05,20.0)$ in the upper motif. $v_l=1.515<v_{\mathbb{H}}$, $\varepsilon=0.0005$.}
\label{fig:CV_min_SISR_improvement1}
\end{figure}

\begin{figure}%[ht!]
\centering
\hspace{0.9cm}\subfigure[$\mathcal{C}_2$~~$\mathrm{controls}$~~$\mathcal{C}_3$]{\raisebox{0.155cm}{\raisebox{0.2cm}{\includegraphics[width=0.15\textwidth]{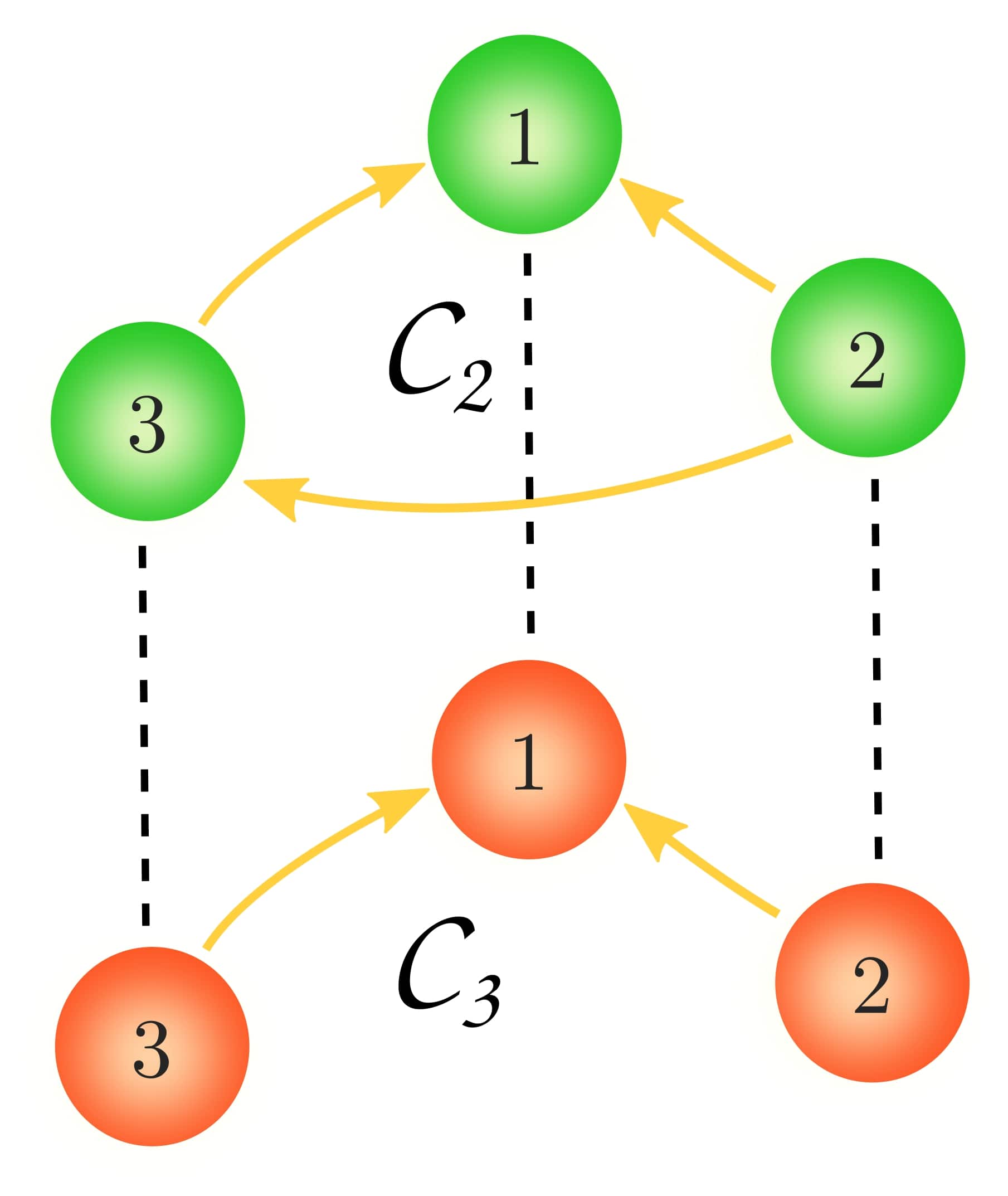}}}}\hspace{0.25cm}
\subfigure[excitability map of electrical multiplexing]{\includegraphics[width=0.25\textwidth]{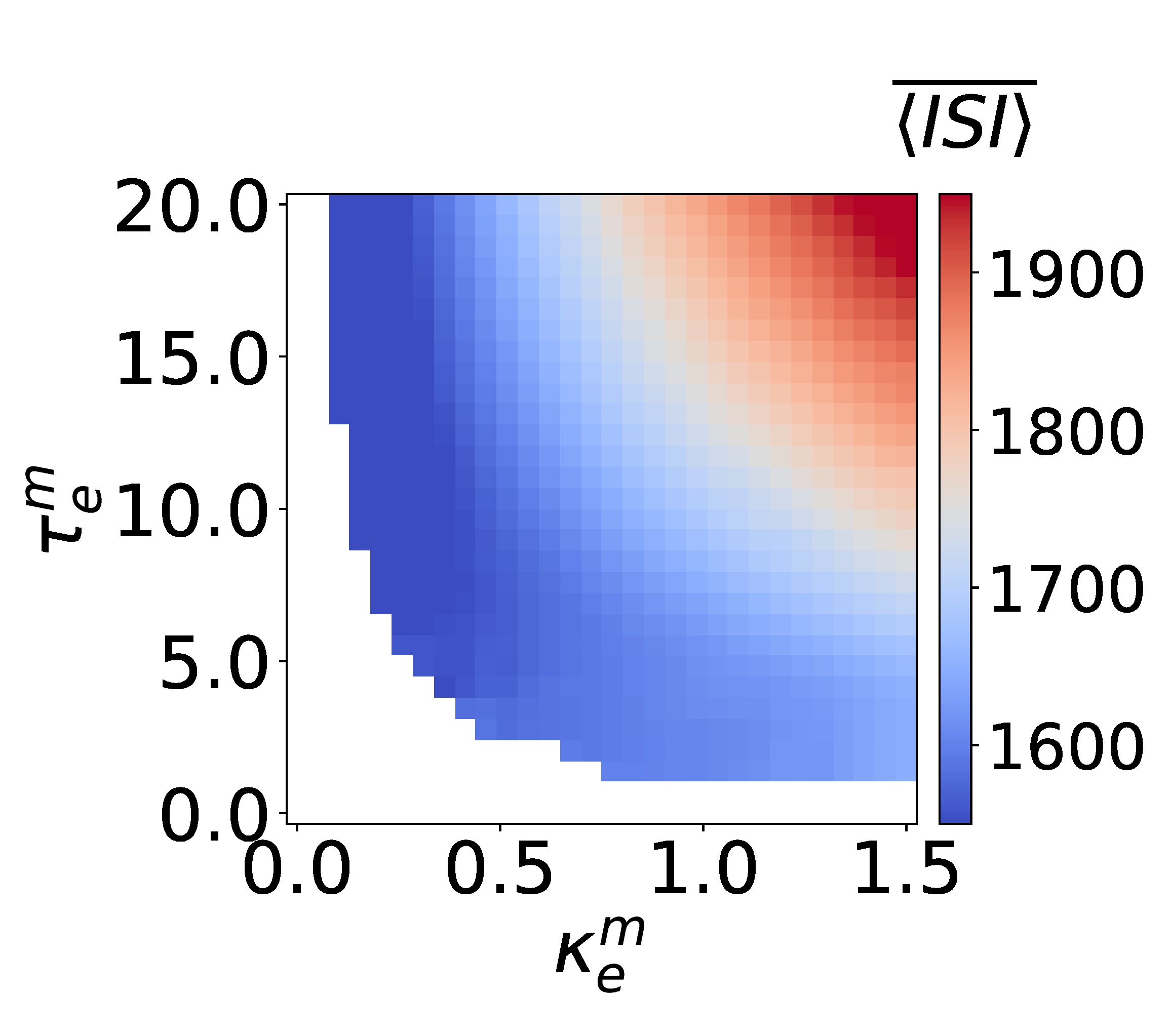}}
\subfigure[$\mathcal{C}_2$~$\xrightarrow{\mathrm{electrical}}$~$\mathcal{C}_3$]{
\includegraphics[width=0.21\textwidth]{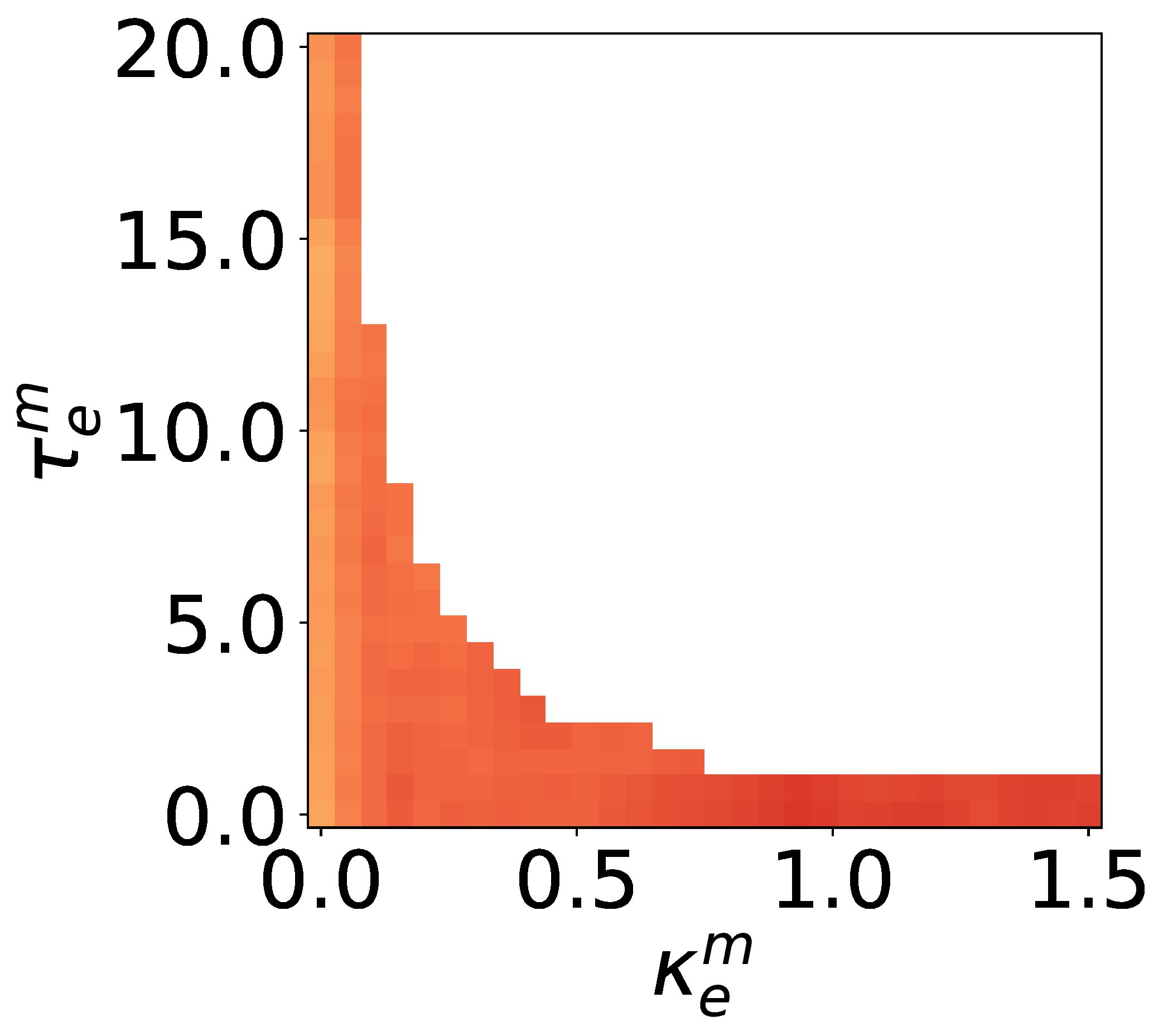}}
\subfigure[$\mathcal{C}_2$~$\xrightarrow{\mathrm{inhibitory~ chemical}}$~$\mathcal{C}_3$]{\includegraphics[width=0.25\textwidth]{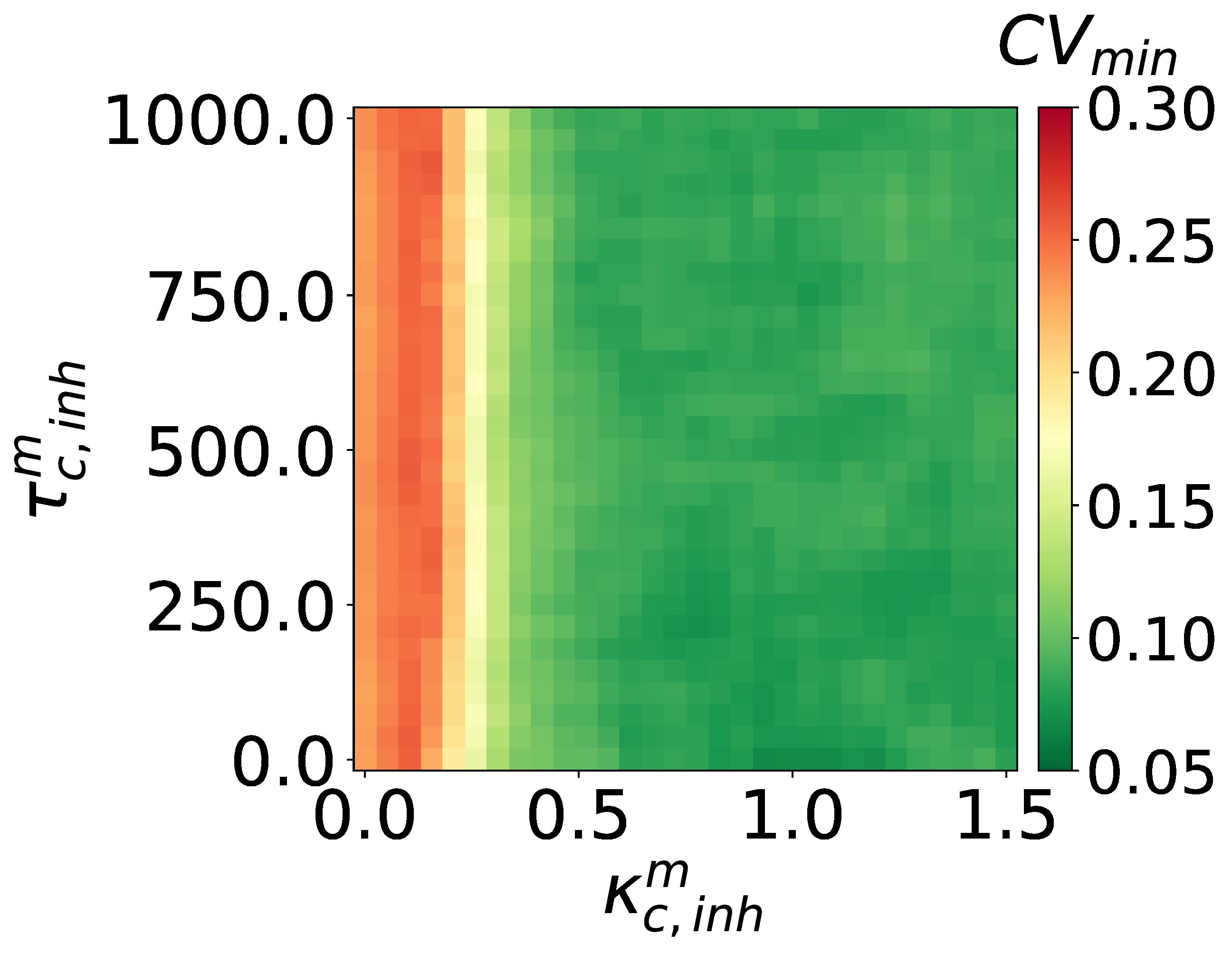}}\hfill
\caption{Excitability map \textbf{(b)} showing $\langle\overline{\mathrm{ISI}}\rangle$ of the multiplexing scheme in \textbf{(a)} with electrical synapses in the absence of noise. $\mathrm{CV}_{\mathrm{min}}$ for the motif layer $\mathcal{C}_3$ against:  electrical $(\kappa^m_e,\tau^m_e)$ and inhibitory chemical $(\kappa^m_{\mathrm{c,inh}},\tau^m_{\mathrm{c,inh}})$ multiplexing parameters in \textbf{(c)} and \textbf{(d)}, respectively. There is no enhancement in the degree of SISR in \textbf{(c)} and an enhancement from $\mathrm{CV}_{\mathrm{min}}=0.232$ in the absence of multiplexing to $\mathrm{CV}_{\mathrm{min}}\approx 0.058$ in \textbf{(d)} in the presence of multiplexing. $(\kappa_{\mathrm{c,inh}},\tau_{\mathrm{c,inh}})=(1.5,10.0)$ in the lower motif, $(\kappa_{\mathrm{c,inh}},\tau_{\mathrm{c,inh}})=(0.05,20.0)$ in the upper motif. $v_l=1.515<v_{\mathbb{H}}$, $\varepsilon=0.0005$.}
\label{fig:CV_min_SISR_improvement2}
\end{figure}

In this subsection, we present another enhancement strategy of  SISR in a motif, based on multiplexing. That is, we connect two motif layers into a multiplex network, where each neuron in one layer is only connected to the replica neuron in the other layer;
see Fig.\ref{fig:CV_min_SISR_improvement1}\textbf{(a)} and Fig.\ref{fig:CV_min_SISR_improvement2}\textbf{(a)}. In each of these figures, the upper motif layer in green is such that, in isolation, the degree of SISR is very high, i.e., $\mathrm{CV}_{\mathrm{min}}=0.058$ in Fig.\ref{fig:CV_min_SISR_improvement1}\textbf{(a)} and Fig.\ref{fig:CV_min_SISR_improvement2}\textbf{(a)}. The lower motif layer in red is the one with a poor degree of SISR when it is in isolation (i.e., $\mathrm{CV}_{\mathrm{min}}=0.293$ in Fig.\ref{fig:CV_min_SISR_improvement1}\textbf{(a)} and $\mathrm{CV}_{\mathrm{min}}=0.232$ in Fig.\ref{fig:CV_min_SISR_improvement2}\textbf{(a)}), and in which we want to enhance the degree of SISR by connecting it to the upper motif in a multiplexing manner. 

In Fig.\,\ref{fig:CV_min_SISR_improvement1}, we represent the efficiency of this enhancement strategy of SISR in the $\mathcal{C}_2$ motif layer in red when it is multiplexed to another $\mathcal{C}_2$ motif layer in green, with electrical inter-motif connections or single unidirectional inhibitory chemical inter-motif connections. Excitatory chemical connections were not used in this strategy because they will induce self-sustained deterministic oscillations. Furthermore, not all time-delayed coupling values of electrical inter-motif connections would set the system into the desired excitable regime. We can see in Fig.\,\ref{fig:CV_min_SISR_improvement1}\textbf{(b)} the range of values of the inter-motif connection parameters --- $\kappa^m_e$ and $\tau^m_e$ --- (the white region) in which the entire multiplex network remains in the excitable regime. Only the excitable values of $\kappa^m_e$ and $\tau^m_e$ are used in the enhancement strategy.
When we use the $\mathcal{C}_2$-$\mathcal{C}_2$ network with electrical multiplexing, we can identify multiplexing parameters from Fig.\,\ref{fig:CV_min_SISR_improvement1}\textbf{(c)} that improve the degree of SISR in the red motif from $\mathrm{CV}_{\mathrm{min}}=0.293$ when it is in isolation to  $\mathrm{CV}_{\mathrm{min}}=0.12$ when it is multiplexed with itself. However, this enhancement strategy is not as efficient when multiplexing connections are mediated by inhibitory chemical couplings, see Fig.\,\ref{fig:CV_min_SISR_improvement1}\textbf{(d)}, where the degree of SISR is improved by just a little, i.e., $\mathrm{CV}_{\mathrm{min}}=0.25$.

In Fig.\,\ref{fig:CV_min_SISR_improvement2}\textbf{(c)} with the $\mathcal{C}_2$-$\mathcal{C}_3$ configuration, we now try to enhance SISR in the $\mathcal{C}_3$ motif. The multiplexing enhancement strategy does not work if the multiplexing is mediated by electrical connections. But in Fig.\,\ref{fig:CV_min_SISR_improvement2}\textbf{(d)}, when inter-motif couplings are mediated by inhibitory chemical connections, the enhancement strategy becomes very efficient. In this case we have $\mathrm{CV}_{\mathrm{min}}=0.058$ in $\mathcal{C}_3$.

We note that in this multiplexing enhancement strategy of SISR, the motif in which we aim at enhancing SISR (i.e., the red lower motifs in Fig.\,\ref{fig:CV_min_SISR_improvement1}\textbf{(a)} and Fig.\,\ref{fig:CV_min_SISR_improvement2}\textbf{(a)} which have a relatively poor degree of SISR in isolation) is connected in a multiplexed fashion to another motif in which the degree of SISR is very high in isolation. It is worth pointing out here that during the enhancement of SISR in the red lower motifs in Fig.\ref{fig:CV_min_SISR_improvement1}\textbf{(d)} and Fig.\ref{fig:CV_min_SISR_improvement2}\textbf{(d)} (for the time-delayed couplings values indicated), the high degree of SISR in the green upper motifs is not affected, except in the case where we have electrical multiplexing between the motifs. In this case (see, e.g., Fig.\,\ref{fig:CV_min_SISR_improvement1}\textbf{(c)}), while the degree of SISR is enhanced in the red lower motif (i.e., $\mathrm{CV}_{\mathrm{min}}$ goes from $0.293$ when it is in isolation to $0.12$ when it is multiplexed to the green upper motif), the high degree of SISR in the green upper motif is significantly deteriorated (i.e., $\mathrm{CV}_{\mathrm{min}}$ goes from $0.058$ in isolation to $0.19$ when it is multiplexed). Hence, in a $\mathcal{C}_2$-$\mathcal{C}_2$ multiplex network configuration, inhibitory chemical couplings between the motifs is the way to go if we want to enhance SISR in one motif without significantly deteriorating SISR in the other motif. 

\section{Summary and conclusions}\label{Section 6}
In this work, the phenomenon of self-induced stochastic resonance (SISR) in the Morris-Lecar (ML) neuron model was systematically investigated. First, we established the analytical conditions necessary for the occurrence of SISR in a motif layer network of ML neurons. Then, from our extensive numerical simulations, we found that:
\begin{itemize}
\item In a single isolated ML neuron without autapses, decreasing the time scale separation parameter $\varepsilon$ between the membrane potential $v$ and the ionic current $w$ variable leads to $(\mathrm{i})$ a significant increase in the degree of SISR and $(\mathrm{ii})$ an increase in the interval of the noise intensity in which the degree of SISR remains high.
\item In a single isolated ML neuron with (only) an electrical autapse, the degree of SISR is not significantly sensitive to variations of the autaptic parameters ($\kappa^a_e$, $\tau^a_e$). However, at stronger autaptic coupling strengths, the degree of SISR is relatively more sensitive to small changes in time delays than at weaker coupling strengths. This behavior is also observed with longer and shorter time delays as the coupling strength changes.
\item In a single isolated ML neuron with (only) an inhibitory chemical autapse, the degree of SISR and the interval of the noise intensity in which a relatively high degree of SISR can be achieved is found to be very sensitive to changes in the autaptic parameters  ($\kappa^a_{\mathrm{c,inh}}$, $\tau^a_{inh}$). While stronger autaptic coupling strengths deteriorate SISR, longer autaptic time delays enhances SISR.
\item In single-motif layer networks with electrical connections $\mathcal{E}_1$ and $\mathcal{E}_2$, we observed that the weaker coupling strengths $\kappa_{\mathrm{e}}$ and shorter time delays $\tau_e$ are, the higher the degree of SISR, especially at weaker noise intensities. On the other hand, single-motif layer networks with inhibitory chemical connections show different behaviors. In the inhibitory chemical motifs $\mathcal{C}_1$, $\mathcal{C}_4$, $\mathcal{C}_5$, and $\mathcal{C}_6$, the high degrees of SISR are robust to  changes in the synaptic parameters as the $\mathrm{CV}$ values remain very low and the intervals of the noise intensity in which these high degrees are achieved remain unchanged. However, SISR in the inhibitory chemical motifs $\mathcal{C}_2$ and $\mathcal{C}_3$ show sensitivity to changes in the coupling strength $\kappa_{\mathrm{c,inh}}$ and time delay $\tau_{\mathrm{c,inh}}$, i.e., the larger $\kappa_{\mathrm{c,inh}}$ and $\tau_{\mathrm{c,inh}}$ are, the more deteriorated is SISR.    
\item It is shown that the poor degree of SISR in the single-motif layer networks $\mathcal{C}_2$ and $\mathcal{C}_3$ can be enhanced using two different strategies. In the enhancement strategy based on autapses, it was shown that electrical autapses with stronger  couplings ($\kappa^a_{\mathrm{e}}>0.25$) and longer time delays ($\tau^a_{\mathrm{e}}>5.0$) or excitatory chemical autapses with stronger ($\kappa^a_{\mathrm{c,exc}}>0.25$) and longer time delays ($\tau^a_{\mathrm{c,exc}}\ge 0.0$) enhances the degree of SISR in $\mathcal{C}_2$ from $\mathrm{CV}=0.293$ to $\mathrm{CV}=0.06$. 
In the $\mathcal{C}_3$ motif network, a similar behavior was observed, but the autapse-based enhancement strategy of SISR is found to be relatively better in the motif $\mathcal{C}_2$ than $\mathcal{C}_3$ in the sense that $\mathcal{C}_2$ can achieve a lower $\mathrm{CV}$ and larger range of autaptic parameter values in which these low $\mathrm{CV}$ values can be attained.
\item For the enhancement strategy of SISR in a motif based on the multiplexing of this motif with another motif, it was found that only electrical multiplexing of two $\mathcal{C}_2$ motifs can enhance the degree of SISR in one of the motifs. In the $\mathcal{C}_2-\mathcal{C}_3$ multiplex configuration, only inhibitory chemical connections between the motifs can enhance the degree of SISR in $\mathcal{C}_3$.
\end{itemize}
Although inhibitory connections between groups of neurons are generally thought to support competitive learning~\cite{rumelhart1985feature,grossberg1987competitive,rabinovich2001dynamical,savin2010independent,krotov2019unsupervised}, our finding that inhibitory chemical point-to-point connections between different motifs can enhance the degree of SISR in one of the motifs additionally suggests a putative mechanism for dynamically adjusting neural dynamics to maintain optimal information processing.

In this paper, we treated the input noise process as solely Gaussian. Looking forward, we must be cognizant that Gaussian white noise is only one possible type of a noise which can induce resonance. Stochastic processes with a non-Gaussian distribution are well-known to more accurately model the dynamics of real biological neurons~\cite{wu2017levy}. 
In~\cite{segev2002long}, a plot of interspike interval and interevent interval distributions indicates that neurons and neural network activities are characterized by a non-Gaussian heavy-tail interval distribution, thereby providing a solid reason as to why it makes sense to consider non-Gaussian noise such as L\'{e}vy noise in the study of neural systems. Therefore, the mechanism via which noise with a non-Gaussian distribution and a temporal correlation (i.e., colored noise) can induce SISR is worth investigating in future research. In particular, the additional timescale brought into the system by the temporal correlation may come along with new interesting dynamics.
 
\section*{Data availability statement}
The simulation data that support the findings of this study are available within the article.
%%%%%%%%%%%%%%%%%%%%%%%%%%%%%%%%%%%%%%%%%%%%%%%%%%%%%%%%%%%%%%%%%%%%%%%%%%%%%%%%%%%%%%%%%%%%%%%%%%%%%%%%%%%%%%%%%
\section*{Acknowledgment}
This work was funded by the Deutsche Forschungsgemeinschaft (DFG, German Research Foundation) via the grants KR 5148/2-1 to PK -- Project No. 436456810 and YA 764/1-1 to MEY -- Project No. 456989199. FB thanks the Chair for Dynamics, Control and Numerics (Alexander von Humboldt Professorship), Department of Data Science, Friedrich-Alexander-Universit\"at, Erlangen-Nürnberg, Germany, for the warm hospitality. 

%\section*{References}
%\bibliographystyle{elsarticle-num-names}
%\medskip
%\bibliographystyle{unsrt}
%\bibliography{refs_final}

\begin{thebibliography}{114}
\providecommand{\natexlab}[1]{#1}
\providecommand{\url}[1]{\texttt{#1}}
\providecommand{\urlprefix}{URL }
\expandafter\ifx\csname urlstyle\endcsname\relax
  \providecommand{\doi}[1]{doi:\discretionary{}{}{}#1}\else
  \providecommand{\doi}[1]{doi:\discretionary{}{}{}\begingroup
  \urlstyle{rm}\url{#1}\endgroup}\fi
\providecommand{\bibinfo}[2]{#2}

\bibitem[{McDonnell and Ward(2011)}]{mcdonnell2011benefits}
\bibinfo{author}{M.~D. McDonnell}, \bibinfo{author}{L.~M. Ward},
  \bibinfo{title}{The benefits of noise in neural systems: bridging theory and
  experiment}, \bibinfo{journal}{Nature Reviews Neuroscience}
  \bibinfo{volume}{12}~(\bibinfo{number}{7}) (\bibinfo{year}{2011})
  \bibinfo{pages}{415--425}.

\bibitem[{Longtin(1993)}]{longtin1993stochastic}
\bibinfo{author}{A.~Longtin}, \bibinfo{title}{Stochastic resonance in neuron
  models}, \bibinfo{journal}{Journal of statistical physics}
  \bibinfo{volume}{70}~(\bibinfo{number}{1-2}) (\bibinfo{year}{1993})
  \bibinfo{pages}{309--327}.

\bibitem[{Patel and Kosko(2008)}]{patel2008stochastic}
\bibinfo{author}{A.~Patel}, \bibinfo{author}{B.~Kosko},
  \bibinfo{title}{Stochastic resonance in continuous and spiking neuron models
  with Levy noise}, \bibinfo{journal}{IEEE Transactions on Neural Networks}
  \bibinfo{volume}{19}~(\bibinfo{number}{12}) (\bibinfo{year}{2008})
  \bibinfo{pages}{1993--2008}.

\bibitem[{Gang et~al.(1993)Gang, Ditzinger, Ning, and
  Haken}]{gang1993stochastic}
\bibinfo{author}{H.~Gang}, \bibinfo{author}{T.~Ditzinger},
  \bibinfo{author}{C.-Z. Ning}, \bibinfo{author}{H.~Haken},
  \bibinfo{title}{Stochastic resonance without external periodic force},
  \bibinfo{journal}{Physical Review Letters}
  \bibinfo{volume}{71}~(\bibinfo{number}{6}) (\bibinfo{year}{1993})
  \bibinfo{pages}{807}.

\bibitem[{Gutkin et~al.(2007)Gutkin, Jost, and Tuckwell}]{gutkin2007transient}
\bibinfo{author}{B.~Gutkin}, \bibinfo{author}{J.~Jost},
  \bibinfo{author}{H.~Tuckwell}, \bibinfo{title}{Transient termination of
  spiking by noise in coupled neurons}, \bibinfo{journal}{EPL (Europhysics
  Letters)} \bibinfo{volume}{81}~(\bibinfo{number}{2}) (\bibinfo{year}{2007})
  \bibinfo{pages}{20005}.

\bibitem[{Krauss et~al.(2016)Krauss, Tziridis, Metzner, Schilling, Hoppe, and
  Schulze}]{krauss2016stochastic}
\bibinfo{author}{P.~Krauss}, \bibinfo{author}{K.~Tziridis},
  \bibinfo{author}{C.~Metzner}, \bibinfo{author}{A.~Schilling},
  \bibinfo{author}{U.~Hoppe}, \bibinfo{author}{H.~Schulze},
  \bibinfo{title}{Stochastic resonance controlled upregulation of internal
  noise after hearing loss as a putative cause of tinnitus-related neuronal
  hyperactivity}, \bibinfo{journal}{Frontiers in neuroscience}
  \bibinfo{volume}{10} (\bibinfo{year}{2016}) \bibinfo{pages}{597}.

\bibitem[{Krauss et~al.(2017)Krauss, Metzner, Schilling, Sch{\"u}tz, Tziridis,
  Fabry, and Schulze}]{krauss2017adaptive}
\bibinfo{author}{P.~Krauss}, \bibinfo{author}{C.~Metzner},
  \bibinfo{author}{A.~Schilling}, \bibinfo{author}{C.~Sch{\"u}tz},
  \bibinfo{author}{K.~Tziridis}, \bibinfo{author}{B.~Fabry},
  \bibinfo{author}{H.~Schulze}, \bibinfo{title}{Adaptive stochastic resonance
  for unknown and variable input signals}, \bibinfo{journal}{Scientific
  reports} \bibinfo{volume}{7}~(\bibinfo{number}{1}) (\bibinfo{year}{2017})
  \bibinfo{pages}{1--8}.

\bibitem[{Krauss et~al.(2018)Krauss, Tziridis, Schilling, and
  Schulze}]{krauss2018cross}
\bibinfo{author}{P.~Krauss}, \bibinfo{author}{K.~Tziridis},
  \bibinfo{author}{A.~Schilling}, \bibinfo{author}{H.~Schulze},
  \bibinfo{title}{Cross-modal stochastic resonance as a universal principle to
  enhance sensory processing}, \bibinfo{journal}{Frontiers in neuroscience}
  \bibinfo{volume}{12} (\bibinfo{year}{2018}) \bibinfo{pages}{578}.

\bibitem[{Schilling et~al.(2020)Schilling, Gerum, Zankl, Schulze, Metzner, and
  Krauss}]{schilling2020intrinsic}
\bibinfo{author}{A.~Schilling}, \bibinfo{author}{R.~Gerum},
  \bibinfo{author}{A.~Zankl}, \bibinfo{author}{H.~Schulze},
  \bibinfo{author}{C.~Metzner}, \bibinfo{author}{P.~Krauss},
  \bibinfo{title}{Intrinsic noise improves speech recognition in a
  computational model of the auditory pathway}, \bibinfo{journal}{bioRxiv} .

\bibitem[{Schilling et~al.(2021)Schilling, Tziridis, Schulze, and
  Krauss}]{SCHILLING2021139}
\bibinfo{author}{A.~Schilling}, \bibinfo{author}{K.~Tziridis},
  \bibinfo{author}{H.~Schulze}, \bibinfo{author}{P.~Krauss},
  \bibinfo{title}{Chapter 6 - The stochastic resonance model of auditory
  perception: A unified explanation of tinnitus development, Zwicker tone
  illusion, and residual inhibition}, in: \bibinfo{editor}{B.~Langguth},
  \bibinfo{editor}{T.~Kleinjung}, \bibinfo{editor}{D.~{De Ridder}},
  \bibinfo{editor}{W.~Schlee}, \bibinfo{editor}{S.~Vanneste} (Eds.),
  \bibinfo{booktitle}{Tinnitus - An Interdisciplinary Approach Towards
  Individualized Treatment: Towards understanding the complexity of tinnitus},
  vol. \bibinfo{volume}{262} of \emph{\bibinfo{series}{Progress in Brain
  Research}}, \bibinfo{publisher}{Elsevier}, \bibinfo{pages}{139--157},
  \bibinfo{year}{2021}.

\bibitem[{Yamakou and Jost(2017)}]{yamakou2017simple}
\bibinfo{author}{M.~E. Yamakou}, \bibinfo{author}{J.~Jost}, \bibinfo{title}{A
  simple parameter can switch between different weak-noise--induced phenomena
  in a simple neuron model}, \bibinfo{journal}{EPL (Europhysics Letters)}
  \bibinfo{volume}{120}~(\bibinfo{number}{1}) (\bibinfo{year}{2017})
  \bibinfo{pages}{18002}.

\bibitem[{Yamakou and Jost(2019)}]{yamakou2019control}
\bibinfo{author}{M.~E. Yamakou}, \bibinfo{author}{J.~Jost},
  \bibinfo{title}{Control of coherence resonance by self-induced stochastic
  resonance in a multiplex neural network}, \bibinfo{journal}{Physical Review
  E} \bibinfo{volume}{100}~(\bibinfo{number}{2}) (\bibinfo{year}{2019})
  \bibinfo{pages}{022313}.

\bibitem[{DeVille et~al.(2005)DeVille, Vanden-Eijnden, and
  Muratov}]{deville2005two}
\bibinfo{author}{R.~L. DeVille}, \bibinfo{author}{E.~Vanden-Eijnden},
  \bibinfo{author}{C.~B. Muratov}, \bibinfo{title}{Two distinct mechanisms of
  coherence in randomly perturbed dynamical systems},
  \bibinfo{journal}{Physical Review E}
  \bibinfo{volume}{72}~(\bibinfo{number}{3}) (\bibinfo{year}{2005})
  \bibinfo{pages}{031105}.

\bibitem[{Zamani et~al.(2020)Zamani, Novikov, and
  Gutkin}]{zamani2020concomitance}
\bibinfo{author}{A.~Zamani}, \bibinfo{author}{N.~Novikov},
  \bibinfo{author}{B.~Gutkin}, \bibinfo{title}{Concomitance of inverse
  stochastic resonance and stochastic resonance in a minimal bistable spiking
  neural circuit}, \bibinfo{journal}{Communications in Nonlinear Science and
  Numerical Simulation} \bibinfo{volume}{82} (\bibinfo{year}{2020})
  \bibinfo{pages}{105024}.

\bibitem[{Wiesenfeld and Moss(1995)}]{wiesenfeld1995stochastic}
\bibinfo{author}{K.~Wiesenfeld}, \bibinfo{author}{F.~Moss},
  \bibinfo{title}{Stochastic resonance and the benefits of noise: from ice ages
  to crayfish and SQUIDs}, \bibinfo{journal}{Nature}
  \bibinfo{volume}{373}~(\bibinfo{number}{6509}) (\bibinfo{year}{1995})
  \bibinfo{pages}{33--36}.

\bibitem[{Lindner et~al.(2004)Lindner, Garc{\i}a-Ojalvo, Neiman, and
  Schimansky-Geier}]{lindner2004effects}
\bibinfo{author}{B.~Lindner}, \bibinfo{author}{J.~Garc{\i}a-Ojalvo},
  \bibinfo{author}{A.~Neiman}, \bibinfo{author}{L.~Schimansky-Geier},
  \bibinfo{title}{Effects of noise in excitable systems},
  \bibinfo{journal}{Physics reports}
  \bibinfo{volume}{392}~(\bibinfo{number}{6}) (\bibinfo{year}{2004})
  \bibinfo{pages}{321--424}.

\bibitem[{Guo et~al.(2017)Guo, Perc, Zhang, Xu, and Yao}]{guo2017frequency}
\bibinfo{author}{D.~Guo}, \bibinfo{author}{M.~Perc},
  \bibinfo{author}{Y.~Zhang}, \bibinfo{author}{P.~Xu},
  \bibinfo{author}{D.~Yao}, \bibinfo{title}{Frequency-difference-dependent
  stochastic resonance in neural systems}, \bibinfo{journal}{Physical Review E}
  \bibinfo{volume}{96}~(\bibinfo{number}{2}) (\bibinfo{year}{2017})
  \bibinfo{pages}{022415}.

\bibitem[{Benzi et~al.(1981)Benzi, Sutera, and Vulpiani}]{benzi1981mechanism}
\bibinfo{author}{R.~Benzi}, \bibinfo{author}{A.~Sutera},
  \bibinfo{author}{A.~Vulpiani}, \bibinfo{title}{The mechanism of stochastic
  resonance}, \bibinfo{journal}{Journal of Physics A: mathematical and general}
  \bibinfo{volume}{14}~(\bibinfo{number}{11}) (\bibinfo{year}{1981})
  \bibinfo{pages}{L453}.

\bibitem[{Xu et~al.(2019)Xu, Lu, Ge, and Jia}]{xu2019effects}
\bibinfo{author}{Y.~Xu}, \bibinfo{author}{L.~Lu}, \bibinfo{author}{M.~Ge},
  \bibinfo{author}{Y.~Jia}, \bibinfo{title}{Effects of temporally correlated
  noise on coherence resonance chimeras in FitzHugh-Nagumo neurons},
  \bibinfo{journal}{The European Physical Journal B}
  \bibinfo{volume}{92}~(\bibinfo{number}{11}) (\bibinfo{year}{2019})
  \bibinfo{pages}{1--10}.

\bibitem[{Pikovsky and Kurths(1997)}]{pikovsky1997coherence}
\bibinfo{author}{A.~S. Pikovsky}, \bibinfo{author}{J.~Kurths},
  \bibinfo{title}{Coherence Resonance in a Noise-Driven Excitable System},
  \bibinfo{journal}{Phys. Rev. Lett.} \bibinfo{volume}{78}
  (\bibinfo{year}{1997}) \bibinfo{pages}{775--778}.

\bibitem[{Lindner and Schimansky-Geier(1999)}]{PhysRevE.60.7270}
\bibinfo{author}{B.~Lindner}, \bibinfo{author}{L.~Schimansky-Geier},
  \bibinfo{title}{Analytical approach to the stochastic FitzHugh-Nagumo system
  and coherence resonance}, \bibinfo{journal}{Phys. Rev. E}
  \bibinfo{volume}{60} (\bibinfo{year}{1999}) \bibinfo{pages}{7270--7276}.

\bibitem[{Gammaitoni et~al.(1998)Gammaitoni, H{\"a}nggi, Jung, and
  Marchesoni}]{gammaitoni1998stochastic}
\bibinfo{author}{L.~Gammaitoni}, \bibinfo{author}{P.~H{\"a}nggi},
  \bibinfo{author}{P.~Jung}, \bibinfo{author}{F.~Marchesoni},
  \bibinfo{title}{Stochastic resonance}, \bibinfo{journal}{Reviews of modern
  physics} \bibinfo{volume}{70}~(\bibinfo{number}{1}) (\bibinfo{year}{1998})
  \bibinfo{pages}{223}.

\bibitem[{Zhou et~al.(2001)Zhou, Kurths, and Hu}]{zhou2001array}
\bibinfo{author}{C.~Zhou}, \bibinfo{author}{J.~Kurths},
  \bibinfo{author}{B.~Hu}, \bibinfo{title}{Array-enhanced coherence resonance:
  nontrivial effects of heterogeneity and spatial independence of noise},
  \bibinfo{journal}{Physical review letters}
  \bibinfo{volume}{87}~(\bibinfo{number}{9}) (\bibinfo{year}{2001})
  \bibinfo{pages}{098101}.

\bibitem[{Neiman et~al.(1997)Neiman, Saparin, and Stone}]{neiman1997coherence}
\bibinfo{author}{A.~Neiman}, \bibinfo{author}{P.~I. Saparin},
  \bibinfo{author}{L.~Stone}, \bibinfo{title}{Coherence resonance at noisy
  precursors of bifurcations in nonlinear dynamical systems},
  \bibinfo{journal}{Physical Review E}
  \bibinfo{volume}{56}~(\bibinfo{number}{1}) (\bibinfo{year}{1997})
  \bibinfo{pages}{270}.

\bibitem[{Zhu(2020)}]{zhu2020phase}
\bibinfo{author}{J.~Zhu}, \bibinfo{title}{Phase sensitivity for coherence
  resonance oscillators}, \bibinfo{journal}{Nonlinear Dynamics}
  \bibinfo{volume}{102}~(\bibinfo{number}{4}) (\bibinfo{year}{2020})
  \bibinfo{pages}{2281--2293}.

\bibitem[{Gutkin et~al.(2009)Gutkin, Jost, and Tuckwell}]{gutkin2009inhibition}
\bibinfo{author}{B.~S. Gutkin}, \bibinfo{author}{J.~Jost},
  \bibinfo{author}{H.~C. Tuckwell}, \bibinfo{title}{Inhibition of rhythmic
  neural spiking by noise: the occurrence of a minimum in activity with
  increasing noise}, \bibinfo{journal}{Naturwissenschaften}
  \bibinfo{volume}{96}~(\bibinfo{number}{9}) (\bibinfo{year}{2009})
  \bibinfo{pages}{1091--1097}.

\bibitem[{Uzuntarla et~al.(2013)Uzuntarla, Cressman, Ozer, and
  Barreto}]{uzuntarla2013dynamical}
\bibinfo{author}{M.~Uzuntarla}, \bibinfo{author}{J.~R. Cressman},
  \bibinfo{author}{M.~Ozer}, \bibinfo{author}{E.~Barreto},
  \bibinfo{title}{Dynamical structure underlying inverse stochastic resonance
  and its implications}, \bibinfo{journal}{Physical Review E}
  \bibinfo{volume}{88}~(\bibinfo{number}{4}) (\bibinfo{year}{2013})
  \bibinfo{pages}{042712}.

\bibitem[{Yamakou and Jost(2018{\natexlab{a}})}]{yamakou2018weak}
\bibinfo{author}{M.~E. Yamakou}, \bibinfo{author}{J.~Jost},
  \bibinfo{title}{Weak-noise-induced transitions with inhibition and modulation
  of neural oscillations}, \bibinfo{journal}{Biological cybernetics}
  \bibinfo{volume}{112}~(\bibinfo{number}{5})
  (\bibinfo{year}{2018}{\natexlab{a}}) \bibinfo{pages}{445--463}.

\bibitem[{Krau{\ss} et~al.(2019)Krau{\ss}, Prebeck, Schilling, and
  Metzner}]{krauss2019recurrence}
\bibinfo{author}{P.~Krau{\ss}}, \bibinfo{author}{K.~Prebeck},
  \bibinfo{author}{A.~Schilling}, \bibinfo{author}{C.~Metzner},
  \bibinfo{title}{Recurrence resonance” in three-neuron motifs},
  \bibinfo{journal}{Frontiers in computational neuroscience}
  \bibinfo{volume}{13} (\bibinfo{year}{2019}) \bibinfo{pages}{64}.

\bibitem[{Freidlin(2001{\natexlab{a}})}]{freidlin2001stable}
\bibinfo{author}{M.~I. Freidlin}, \bibinfo{title}{On stable oscillations and
  equilibriums induced by small noise}, \bibinfo{journal}{Journal of
  Statistical Physics} \bibinfo{volume}{103}~(\bibinfo{number}{1})
  (\bibinfo{year}{2001}{\natexlab{a}}) \bibinfo{pages}{283--300}.

\bibitem[{Freidlin(2001{\natexlab{b}})}]{freidlin2001stochastic}
\bibinfo{author}{M.~Freidlin}, \bibinfo{title}{On stochastic perturbations of
  dynamical systems with fast and slow components},
  \bibinfo{journal}{Stochastics and Dynamics}
  \bibinfo{volume}{1}~(\bibinfo{number}{02})
  (\bibinfo{year}{2001}{\natexlab{b}}) \bibinfo{pages}{261--281}.

\bibitem[{Muratov et~al.(2005)Muratov, Vanden-Eijnden, and
  Weinan}]{muratov2005self}
\bibinfo{author}{C.~B. Muratov}, \bibinfo{author}{E.~Vanden-Eijnden},
  \bibinfo{author}{E.~Weinan}, \bibinfo{title}{Self-induced stochastic
  resonance in excitable systems}, \bibinfo{journal}{Physica D: Nonlinear
  Phenomena} \bibinfo{volume}{210}~(\bibinfo{number}{3-4})
  (\bibinfo{year}{2005}) \bibinfo{pages}{227--240}.

\bibitem[{Muratov and Vanden-Eijnden(2008)}]{muratov2008noise}
\bibinfo{author}{C.~B. Muratov}, \bibinfo{author}{E.~Vanden-Eijnden},
  \bibinfo{title}{Noise-induced mixed-mode oscillations in a relaxation
  oscillator near the onset of a limit cycle}, \bibinfo{journal}{Chaos: An
  Interdisciplinary Journal of Nonlinear Science}
  \bibinfo{volume}{18}~(\bibinfo{number}{1}) (\bibinfo{year}{2008})
  \bibinfo{pages}{015111}.

\bibitem[{DeVille and Vanden-Eijnden(2007)}]{deville2007nontrivial}
\bibinfo{author}{R.~L. DeVille}, \bibinfo{author}{E.~Vanden-Eijnden},
  \bibinfo{title}{A nontrivial scaling limit for multiscale Markov chains},
  \bibinfo{journal}{Journal of Statistical Physics}
  \bibinfo{volume}{126}~(\bibinfo{number}{1}) (\bibinfo{year}{2007})
  \bibinfo{pages}{75--94}.

\bibitem[{DeVille et~al.(2007)DeVille, Vanden-Eijnden et~al.}]{deville2007self}
\bibinfo{author}{R.~L. DeVille}, \bibinfo{author}{E.~Vanden-Eijnden}, et~al.,
  \bibinfo{title}{Self-induced stochastic resonance for Brownian ratchets under
  load}, \bibinfo{journal}{Communications in Mathematical Sciences}
  \bibinfo{volume}{5}~(\bibinfo{number}{2}) (\bibinfo{year}{2007})
  \bibinfo{pages}{431--466}.

\bibitem[{Yamakou and Jost(2018{\natexlab{b}})}]{yamakou2018coherent}
\bibinfo{author}{M.~E. Yamakou}, \bibinfo{author}{J.~Jost},
  \bibinfo{title}{Coherent neural oscillations induced by weak synaptic noise},
  \bibinfo{journal}{Nonlinear Dynamics}
  \bibinfo{volume}{93}~(\bibinfo{number}{4})
  (\bibinfo{year}{2018}{\natexlab{b}}) \bibinfo{pages}{2121--2144}.

\bibitem[{Yamakou et~al.(2020)Yamakou, Hjorth, and
  Martens}]{yamakou2020optimal}
\bibinfo{author}{M.~E. Yamakou}, \bibinfo{author}{P.~G. Hjorth},
  \bibinfo{author}{E.~A. Martens}, \bibinfo{title}{Optimal self-induced
  stochastic resonance in multiplex neural networks: electrical versus chemical
  synapses}, \bibinfo{journal}{Frontiers in Computational Neuroscience}
  \bibinfo{volume}{14} (\bibinfo{year}{2020}) \bibinfo{pages}{62}.

\bibitem[{Shen et~al.(2010)Shen, Chen, and Aihara}]{shen2010self}
\bibinfo{author}{J.~Shen}, \bibinfo{author}{L.~Chen},
  \bibinfo{author}{K.~Aihara}, \bibinfo{title}{Self-induced stochastic
  resonance in MicroRNA regulation of a cancer network}, in:
  \bibinfo{booktitle}{The fourth international conference on computational
  systems biology}, \bibinfo{organization}{Citeseer},
  \bibinfo{pages}{251--257}, \bibinfo{year}{2010}.

\bibitem[{Zhang et~al.(2021{\natexlab{a}})Zhang, Yang, Wang, Liu, and
  Yang}]{zhang2021stochastic}
\bibinfo{author}{S.~Zhang}, \bibinfo{author}{J.~Yang},
  \bibinfo{author}{C.~Wang}, \bibinfo{author}{H.~Liu},
  \bibinfo{author}{C.~Yang}, \bibinfo{title}{Stochastic Resonance and
  Self-Induced Stochastic Resonance in Bearing Fault Diagnosis},
  \bibinfo{journal}{Fluctuation and Noise Letters}
  (\bibinfo{year}{2021}{\natexlab{a}}) \bibinfo{pages}{2150047}.

\bibitem[{Yamakou and Tran(2020)}]{yamakou2020levy}
\bibinfo{author}{M.~E. Yamakou}, \bibinfo{author}{T.~D. Tran},
  \bibinfo{title}{Levy noise-induced self-induced stochastic resonance in a
  memristive neuron}, \bibinfo{journal}{arXiv preprint arXiv:2012.03032} .

\bibitem[{Zhu and Nakao(2021)}]{zhu2021stochastic}
\bibinfo{author}{J.~Zhu}, \bibinfo{author}{H.~Nakao},
  \bibinfo{title}{Stochastic Periodic Orbits in Fast-Slow Systems with
  Self-Induced Stochastic Resonance}, \bibinfo{journal}{arXiv preprint
  arXiv:2104.04210} .

\bibitem[{Morris and Lecar(1981)}]{morris1981voltage}
\bibinfo{author}{C.~Morris}, \bibinfo{author}{H.~Lecar},
  \bibinfo{title}{Voltage oscillations in the barnacle giant muscle fiber},
  \bibinfo{journal}{Biophysical journal}
  \bibinfo{volume}{35}~(\bibinfo{number}{1}) (\bibinfo{year}{1981})
  \bibinfo{pages}{193--213}.

\bibitem[{Markram(2012)}]{markram2012human}
\bibinfo{author}{H.~Markram}, \bibinfo{title}{The human brain project},
  \bibinfo{journal}{Scientific American}
  \bibinfo{volume}{306}~(\bibinfo{number}{6}) (\bibinfo{year}{2012})
  \bibinfo{pages}{50--55}.

\bibitem[{Van~Essen et~al.(2013)Van~Essen, Smith, Barch, Behrens, Yacoub,
  Ugurbil, Consortium et~al.}]{van2013wu}
\bibinfo{author}{D.~C. Van~Essen}, \bibinfo{author}{S.~M. Smith},
  \bibinfo{author}{D.~M. Barch}, \bibinfo{author}{T.~E. Behrens},
  \bibinfo{author}{E.~Yacoub}, \bibinfo{author}{K.~Ugurbil},
  \bibinfo{author}{W.-M.~H. Consortium}, et~al., \bibinfo{title}{The WU-Minn
  human connectome project: an overview}, \bibinfo{journal}{Neuroimage}
  \bibinfo{volume}{80} (\bibinfo{year}{2013}) \bibinfo{pages}{62--79}.

\bibitem[{Krauss et~al.(2019{\natexlab{a}})Krauss, Schuster, Dietrich,
  Schilling, Schulze, and Metzner}]{krauss2019weight}
\bibinfo{author}{P.~Krauss}, \bibinfo{author}{M.~Schuster},
  \bibinfo{author}{V.~Dietrich}, \bibinfo{author}{A.~Schilling},
  \bibinfo{author}{H.~Schulze}, \bibinfo{author}{C.~Metzner},
  \bibinfo{title}{Weight statistics controls dynamics in recurrent neural
  networks}, \bibinfo{journal}{PloS one}
  \bibinfo{volume}{14}~(\bibinfo{number}{4})
  (\bibinfo{year}{2019}{\natexlab{a}}) \bibinfo{pages}{e0214541}.

\bibitem[{Milo et~al.(2002)Milo, Shen-Orr, Itzkovitz, Kashtan, Chklovskii, and
  Alon}]{milo2002network}
\bibinfo{author}{R.~Milo}, \bibinfo{author}{S.~Shen-Orr},
  \bibinfo{author}{S.~Itzkovitz}, \bibinfo{author}{N.~Kashtan},
  \bibinfo{author}{D.~Chklovskii}, \bibinfo{author}{U.~Alon},
  \bibinfo{title}{Network motifs: simple building blocks of complex networks},
  \bibinfo{journal}{Science} \bibinfo{volume}{298}~(\bibinfo{number}{5594})
  (\bibinfo{year}{2002}) \bibinfo{pages}{824--827}.

\bibitem[{Krauss et~al.(2019{\natexlab{b}})Krauss, Zankl, Schilling, Schulze,
  and Metzner}]{krauss2019analysis}
\bibinfo{author}{P.~Krauss}, \bibinfo{author}{A.~Zankl},
  \bibinfo{author}{A.~Schilling}, \bibinfo{author}{H.~Schulze},
  \bibinfo{author}{C.~Metzner}, \bibinfo{title}{Analysis of structure and
  dynamics in three-neuron motifs}, \bibinfo{journal}{Frontiers in
  computational neuroscience} \bibinfo{volume}{13}
  (\bibinfo{year}{2019}{\natexlab{b}}) \bibinfo{pages}{5}.

\bibitem[{Li(2008)}]{li2008functions}
\bibinfo{author}{C.~Li}, \bibinfo{title}{Functions of neuronal network motifs},
  \bibinfo{journal}{Physical Review E}
  \bibinfo{volume}{78}~(\bibinfo{number}{3}) (\bibinfo{year}{2008})
  \bibinfo{pages}{037101}.

\bibitem[{Song et~al.(2005)Song, Sj{\"o}str{\"o}m, Reigl, Nelson, and
  Chklovskii}]{song2005highly}
\bibinfo{author}{S.~Song}, \bibinfo{author}{P.~J. Sj{\"o}str{\"o}m},
  \bibinfo{author}{M.~Reigl}, \bibinfo{author}{S.~Nelson},
  \bibinfo{author}{D.~B. Chklovskii}, \bibinfo{title}{Highly nonrandom features
  of synaptic connectivity in local cortical circuits}, \bibinfo{journal}{PLoS
  biology} \bibinfo{volume}{3}~(\bibinfo{number}{3}) (\bibinfo{year}{2005})
  \bibinfo{pages}{e68}.

\bibitem[{Battiston(2017)}]{battiston2017structure}
\bibinfo{author}{F.~Battiston}, \bibinfo{title}{The structure and dynamics of
  multiplex networks}, Ph.D. thesis, \bibinfo{school}{Queen Mary University of
  London}, \bibinfo{year}{2017}.

\bibitem[{Maksimenko et~al.(2016)Maksimenko, Makarov, Bera, Ghosh, Dana,
  Goremyko, Frolov, Koronovskii, and Hramov}]{PhysRevE.94.052205}
\bibinfo{author}{V.~A. Maksimenko}, \bibinfo{author}{V.~V. Makarov},
  \bibinfo{author}{B.~K. Bera}, \bibinfo{author}{D.~Ghosh},
  \bibinfo{author}{S.~K. Dana}, \bibinfo{author}{M.~V. Goremyko},
  \bibinfo{author}{N.~S. Frolov}, \bibinfo{author}{A.~A. Koronovskii},
  \bibinfo{author}{A.~E. Hramov}, \bibinfo{title}{Excitation and suppression of
  chimera states by multiplexing}, \bibinfo{journal}{Phys. Rev. E}
  \bibinfo{volume}{94} (\bibinfo{year}{2016}) \bibinfo{pages}{052205}.

\bibitem[{Ghosh et~al.(2016)Ghosh, Kumar, Zakharova, and
  Jalan}]{ghosh2016birth}
\bibinfo{author}{S.~Ghosh}, \bibinfo{author}{A.~Kumar},
  \bibinfo{author}{A.~Zakharova}, \bibinfo{author}{S.~Jalan},
  \bibinfo{title}{Birth and death of chimera: Interplay of delay and
  multiplexing}, \bibinfo{journal}{EPL (Europhysics Letters)}
  \bibinfo{volume}{115}~(\bibinfo{number}{6}) (\bibinfo{year}{2016})
  \bibinfo{pages}{60005}.

\bibitem[{Ghosh and Jalan(2016)}]{ghosh2016emergence}
\bibinfo{author}{S.~Ghosh}, \bibinfo{author}{S.~Jalan},
  \bibinfo{title}{Emergence of chimera in multiplex network},
  \bibinfo{journal}{International Journal of Bifurcation and Chaos}
  \bibinfo{volume}{26}~(\bibinfo{number}{07}) (\bibinfo{year}{2016})
  \bibinfo{pages}{1650120}.

\bibitem[{Sawicki et~al.(2018)Sawicki, Omelchenko, Zakharova, and
  Sch{\"o}ll}]{sawicki2018synchronization}
\bibinfo{author}{J.~Sawicki}, \bibinfo{author}{I.~Omelchenko},
  \bibinfo{author}{A.~Zakharova}, \bibinfo{author}{E.~Sch{\"o}ll},
  \bibinfo{title}{Synchronization scenarios of chimeras in multiplex networks},
  \bibinfo{journal}{The European Physical Journal Special Topics}
  \bibinfo{volume}{227}~(\bibinfo{number}{10-11}) (\bibinfo{year}{2018})
  \bibinfo{pages}{1161--1171}.

\bibitem[{Goremyko et~al.(2017)Goremyko, Kirsanov, Nedaivozov, Makarov, and
  Hramov}]{goremyko2017pattern}
\bibinfo{author}{M.~V. Goremyko}, \bibinfo{author}{D.~V. Kirsanov},
  \bibinfo{author}{V.~O. Nedaivozov}, \bibinfo{author}{V.~V. Makarov},
  \bibinfo{author}{A.~E. Hramov}, \bibinfo{title}{Pattern formation in adaptive
  multiplex network in application to analysis of the complex structure of
  neuronal network of the brain}, in: \bibinfo{booktitle}{Dynamics and
  Fluctuations in Biomedical Photonics XIV}, vol. \bibinfo{volume}{10063},
  \bibinfo{organization}{International Society for Optics and Photonics},
  \bibinfo{pages}{100631C}, \bibinfo{year}{2017}.

\bibitem[{Singh et~al.(2015)Singh, Ghosh, Jalan, and
  Kurths}]{singh2015synchronization}
\bibinfo{author}{A.~Singh}, \bibinfo{author}{S.~Ghosh},
  \bibinfo{author}{S.~Jalan}, \bibinfo{author}{J.~Kurths},
  \bibinfo{title}{Synchronization in delayed multiplex networks},
  \bibinfo{journal}{EPL (Europhysics Letters)}
  \bibinfo{volume}{111}~(\bibinfo{number}{3}) (\bibinfo{year}{2015})
  \bibinfo{pages}{30010}.

\bibitem[{Semenova and Zakharova(2018)}]{semenova2018weak}
\bibinfo{author}{N.~Semenova}, \bibinfo{author}{A.~Zakharova},
  \bibinfo{title}{Weak multiplexing induces coherence resonance},
  \bibinfo{journal}{Chaos: An Interdisciplinary Journal of Nonlinear Science}
  \bibinfo{volume}{28}~(\bibinfo{number}{5}) (\bibinfo{year}{2018})
  \bibinfo{pages}{051104}.

\bibitem[{Kandel et~al.(2000)Kandel, Schwartz, Jessell, Siegelbaum, Hudspeth,
  and Mack}]{kandel2000principles}
\bibinfo{author}{E.~R. Kandel}, \bibinfo{author}{J.~H. Schwartz},
  \bibinfo{author}{T.~M. Jessell}, \bibinfo{author}{S.~Siegelbaum},
  \bibinfo{author}{A.~J. Hudspeth}, \bibinfo{author}{S.~Mack},
  \bibinfo{title}{Principles of neural science}, vol.~\bibinfo{volume}{4},
  \bibinfo{publisher}{McGraw-hill New York}, \bibinfo{year}{2000}.

\bibitem[{Aboitiz et~al.(1992)Aboitiz, Scheibel, Fisher, and
  Zaidel}]{aboitiz1992individual}
\bibinfo{author}{F.~Aboitiz}, \bibinfo{author}{A.~B. Scheibel},
  \bibinfo{author}{R.~S. Fisher}, \bibinfo{author}{E.~Zaidel},
  \bibinfo{title}{Individual differences in brain asymmetries and fiber
  composition in the human corpus callosum}, \bibinfo{journal}{Brain research}
  \bibinfo{volume}{598}~(\bibinfo{number}{1-2}) (\bibinfo{year}{1992})
  \bibinfo{pages}{154--161}.

\bibitem[{Sch{\"u}z and Prei$\beta$l(1996)}]{schuz1996basic}
\bibinfo{author}{A.~Sch{\"u}z}, \bibinfo{author}{H.~Prei$\beta$l},
  \bibinfo{title}{Basic connectivity of the cerebral cortex and some
  considerations on the corpus callosum}, \bibinfo{journal}{Neuroscience \&
  Biobehavioral Reviews} \bibinfo{volume}{20}~(\bibinfo{number}{4})
  (\bibinfo{year}{1996}) \bibinfo{pages}{567--570}.

\bibitem[{Pereda(2014)}]{pereda2014electrical}
\bibinfo{author}{A.~E. Pereda}, \bibinfo{title}{Electrical synapses and their
  functional interactions with chemical synapses}, \bibinfo{journal}{Nature
  Reviews Neuroscience} \bibinfo{volume}{15}~(\bibinfo{number}{4})
  (\bibinfo{year}{2014}) \bibinfo{pages}{250--263}.

\bibitem[{Van Der~Loos and Glaser(1972)}]{van1972autapses}
\bibinfo{author}{H.~Van Der~Loos}, \bibinfo{author}{E.~M. Glaser},
  \bibinfo{title}{Autapses in neocortex cerebri: synapses between a pyramidal
  cell's axon and its own dendrites}, \bibinfo{journal}{Brain research}
  \bibinfo{volume}{48} (\bibinfo{year}{1972}) \bibinfo{pages}{355--360}.

\bibitem[{L{\"u}bke et~al.(1996)L{\"u}bke, Markram, Frotscher, and
  Sakmann}]{lubke1996frequency}
\bibinfo{author}{J.~L{\"u}bke}, \bibinfo{author}{H.~Markram},
  \bibinfo{author}{M.~Frotscher}, \bibinfo{author}{B.~Sakmann},
  \bibinfo{title}{Frequency and dendritic distribution of autapses established
  by layer 5 pyramidal neurons in the developing rat neocortex: comparison with
  synaptic innervation of adjacent neurons of the same class},
  \bibinfo{journal}{Journal of Neuroscience}
  \bibinfo{volume}{16}~(\bibinfo{number}{10}) (\bibinfo{year}{1996})
  \bibinfo{pages}{3209--3218}.

\bibitem[{Yilmaz et~al.(2016)Yilmaz, Ozer, Baysal, and
  Perc}]{yilmaz2016autapse}
\bibinfo{author}{E.~Yilmaz}, \bibinfo{author}{M.~Ozer},
  \bibinfo{author}{V.~Baysal}, \bibinfo{author}{M.~Perc},
  \bibinfo{title}{Autapse-induced multiple coherence resonance in single
  neurons and neuronal networks}, \bibinfo{journal}{Scientific Reports}
  \bibinfo{volume}{6}~(\bibinfo{number}{1}) (\bibinfo{year}{2016})
  \bibinfo{pages}{1--14}.

\bibitem[{Herrmann and Klaus(2004)}]{herrmann2004autapse}
\bibinfo{author}{C.~S. Herrmann}, \bibinfo{author}{A.~Klaus},
  \bibinfo{title}{Autapse turns neuron into oscillator},
  \bibinfo{journal}{International Journal of Bifurcation and Chaos}
  \bibinfo{volume}{14}~(\bibinfo{number}{02}) (\bibinfo{year}{2004})
  \bibinfo{pages}{623--633}.

\bibitem[{Guo et~al.(2016)Guo, Wu, Chen, Perc, Zhang, Ma, Cui, Xu, Xia, and
  Yao}]{guo2016regulation}
\bibinfo{author}{D.~Guo}, \bibinfo{author}{S.~Wu}, \bibinfo{author}{M.~Chen},
  \bibinfo{author}{M.~Perc}, \bibinfo{author}{Y.~Zhang},
  \bibinfo{author}{J.~Ma}, \bibinfo{author}{Y.~Cui}, \bibinfo{author}{P.~Xu},
  \bibinfo{author}{Y.~Xia}, \bibinfo{author}{D.~Yao},
  \bibinfo{title}{Regulation of irregular neuronal firing by autaptic
  transmission}, \bibinfo{journal}{Scientific reports}
  \bibinfo{volume}{6}~(\bibinfo{number}{1}) (\bibinfo{year}{2016})
  \bibinfo{pages}{1--14}.

\bibitem[{Yin et~al.(2018)Yin, Zheng, Ke, He, Zhang, Li, Wang, Mi, Long, Rasch
  et~al.}]{yin2018autapses}
\bibinfo{author}{L.~Yin}, \bibinfo{author}{R.~Zheng}, \bibinfo{author}{W.~Ke},
  \bibinfo{author}{Q.~He}, \bibinfo{author}{Y.~Zhang}, \bibinfo{author}{J.~Li},
  \bibinfo{author}{B.~Wang}, \bibinfo{author}{Z.~Mi}, \bibinfo{author}{Y.-s.
  Long}, \bibinfo{author}{M.~J. Rasch}, et~al., \bibinfo{title}{Autapses
  enhance bursting and coincidence detection in neocortical pyramidal cells},
  \bibinfo{journal}{Nature communications}
  \bibinfo{volume}{9}~(\bibinfo{number}{1}) (\bibinfo{year}{2018})
  \bibinfo{pages}{1--12}.

\bibitem[{Bacci and Huguenard(2006)}]{bacci2006enhancement}
\bibinfo{author}{A.~Bacci}, \bibinfo{author}{J.~R. Huguenard},
  \bibinfo{title}{Enhancement of spike-timing precision by autaptic
  transmission in neocortical inhibitory interneurons},
  \bibinfo{journal}{Neuron} \bibinfo{volume}{49}~(\bibinfo{number}{1})
  (\bibinfo{year}{2006}) \bibinfo{pages}{119--130}.

\bibitem[{Wang et~al.(2017)Wang, Guo, Xu, Ma, Tang, Alzahrani, and
  Hobiny}]{wang2017formation}
\bibinfo{author}{C.~Wang}, \bibinfo{author}{S.~Guo}, \bibinfo{author}{Y.~Xu},
  \bibinfo{author}{J.~Ma}, \bibinfo{author}{J.~Tang},
  \bibinfo{author}{F.~Alzahrani}, \bibinfo{author}{A.~Hobiny},
  \bibinfo{title}{Formation of autapse connected to neuron and its biological
  function}, \bibinfo{journal}{Complexity} \bibinfo{volume}{2017}.

\bibitem[{Seung et~al.(2000)Seung, Lee, Reis, and Tank}]{seung2000autapse}
\bibinfo{author}{H.~S. Seung}, \bibinfo{author}{D.~D. Lee},
  \bibinfo{author}{B.~Y. Reis}, \bibinfo{author}{D.~W. Tank},
  \bibinfo{title}{The autapse: a simple illustration of short-term analog
  memory storage by tuned synaptic feedback}, \bibinfo{journal}{Journal of
  computational neuroscience} \bibinfo{volume}{9}~(\bibinfo{number}{2})
  (\bibinfo{year}{2000}) \bibinfo{pages}{171--185}.

\bibitem[{Wang et~al.(2014)Wang, Wang, Chen, and Chen}]{wang2014effect}
\bibinfo{author}{H.~Wang}, \bibinfo{author}{L.~Wang},
  \bibinfo{author}{Y.~Chen}, \bibinfo{author}{Y.~Chen}, \bibinfo{title}{Effect
  of autaptic activity on the response of a Hodgkin-Huxley neuron},
  \bibinfo{journal}{Chaos: An Interdisciplinary Journal of Nonlinear Science}
  \bibinfo{volume}{24}~(\bibinfo{number}{3}) (\bibinfo{year}{2014})
  \bibinfo{pages}{033122}.

\bibitem[{Liu and Yang(2018)}]{liu2018coherence}
\bibinfo{author}{X.~Liu}, \bibinfo{author}{X.~Yang}, \bibinfo{title}{Coherence
  resonance in a modified FHN neuron with autapse and phase noise},
  \bibinfo{journal}{International Journal of Modern Physics B}
  \bibinfo{volume}{32}~(\bibinfo{number}{30}) (\bibinfo{year}{2018})
  \bibinfo{pages}{1850332}.

\bibitem[{Protachevicz et~al.(2020)Protachevicz, Iarosz, Caldas, Antonopoulos,
  Batista, and Kurths}]{protachevicz2020influence}
\bibinfo{author}{P.~R. Protachevicz}, \bibinfo{author}{K.~C. Iarosz},
  \bibinfo{author}{I.~L. Caldas}, \bibinfo{author}{C.~G. Antonopoulos},
  \bibinfo{author}{A.~M. Batista}, \bibinfo{author}{J.~Kurths},
  \bibinfo{title}{Influence of autapses on synchronisation in neural networks
  with chemical synapses}, \bibinfo{journal}{Frontiers in Systems Neuroscience}
  \bibinfo{volume}{14} (\bibinfo{year}{2020}) \bibinfo{pages}{91}.

\bibitem[{Fan et~al.(2018)Fan, Wang, Wang, Lai, and Wang}]{fan2018autapses}
\bibinfo{author}{H.~Fan}, \bibinfo{author}{Y.~Wang}, \bibinfo{author}{H.~Wang},
  \bibinfo{author}{Y.-C. Lai}, \bibinfo{author}{X.~Wang},
  \bibinfo{title}{Autapses promote synchronization in neuronal networks},
  \bibinfo{journal}{Scientific reports}
  \bibinfo{volume}{8}~(\bibinfo{number}{1}) (\bibinfo{year}{2018})
  \bibinfo{pages}{1--13}.

\bibitem[{Yang et~al.(2017)Yang, Yu, and Sun}]{yang2017autapse}
\bibinfo{author}{X.~Yang}, \bibinfo{author}{Y.~Yu}, \bibinfo{author}{Z.~Sun},
  \bibinfo{title}{Autapse-induced multiple stochastic resonances in a modular
  neuronal network}, \bibinfo{journal}{Chaos: An Interdisciplinary Journal of
  Nonlinear Science} \bibinfo{volume}{27}~(\bibinfo{number}{8})
  (\bibinfo{year}{2017}) \bibinfo{pages}{083117}.

\bibitem[{Song et~al.(2018)Song, Wang, and Chen}]{song2018coherence}
\bibinfo{author}{X.~Song}, \bibinfo{author}{H.~Wang},
  \bibinfo{author}{Y.~Chen}, \bibinfo{title}{Coherence resonance in an autaptic
  Hodgkin--Huxley neuron with time delay}, \bibinfo{journal}{Nonlinear
  Dynamics} \bibinfo{volume}{94}~(\bibinfo{number}{1}) (\bibinfo{year}{2018})
  \bibinfo{pages}{141--150}.

\bibitem[{Jia et~al.(2021)Jia, Gu, Li, and Ding}]{jia2021inhibitory}
\bibinfo{author}{Y.~Jia}, \bibinfo{author}{H.~Gu}, \bibinfo{author}{Y.~Li},
  \bibinfo{author}{X.~Ding}, \bibinfo{title}{Inhibitory autapses enhance
  coherence resonance of a neuronal network}, \bibinfo{journal}{Communications
  in Nonlinear Science and Numerical Simulation} \bibinfo{volume}{95}
  (\bibinfo{year}{2021}) \bibinfo{pages}{105643}.

\bibitem[{Zhang et~al.(2021{\natexlab{b}})Zhang, Li, and
  Xing}]{zhang2021autapse}
\bibinfo{author}{N.~Zhang}, \bibinfo{author}{D.~Li}, \bibinfo{author}{Y.~Xing},
  \bibinfo{title}{Autapse-induced multiple inverse stochastic resonance in a
  neural system}, \bibinfo{journal}{The European Physical Journal B}
  \bibinfo{volume}{94}~(\bibinfo{number}{1})
  (\bibinfo{year}{2021}{\natexlab{b}}) \bibinfo{pages}{1--11}.

\bibitem[{Fries(2005)}]{fries2005mechanism}
\bibinfo{author}{P.~Fries}, \bibinfo{title}{A mechanism for cognitive dynamics:
  neuronal communication through neuronal coherence}, \bibinfo{journal}{Trends
  in cognitive sciences} \bibinfo{volume}{9}~(\bibinfo{number}{10})
  (\bibinfo{year}{2005}) \bibinfo{pages}{474--480}.

\bibitem[{Benchenane et~al.(2010)Benchenane, Peyrache, Khamassi, Tierney,
  Gioanni, Battaglia, and Wiener}]{benchenane2010coherent}
\bibinfo{author}{K.~Benchenane}, \bibinfo{author}{A.~Peyrache},
  \bibinfo{author}{M.~Khamassi}, \bibinfo{author}{P.~L. Tierney},
  \bibinfo{author}{Y.~Gioanni}, \bibinfo{author}{F.~P. Battaglia},
  \bibinfo{author}{S.~I. Wiener}, \bibinfo{title}{Coherent theta oscillations
  and reorganization of spike timing in the hippocampal-prefrontal network upon
  learning}, \bibinfo{journal}{Neuron}
  \bibinfo{volume}{66}~(\bibinfo{number}{6}) (\bibinfo{year}{2010})
  \bibinfo{pages}{921--936}.

\bibitem[{Fries(2015)}]{fries2015rhythms}
\bibinfo{author}{P.~Fries}, \bibinfo{title}{Rhythms for cognition:
  communication through coherence}, \bibinfo{journal}{Neuron}
  \bibinfo{volume}{88}~(\bibinfo{number}{1}) (\bibinfo{year}{2015})
  \bibinfo{pages}{220--235}.

\bibitem[{Wouapi et~al.(2020)Wouapi, Fotsin, Louodop, Feudjio, Njitacke, and
  Djeudjo}]{wouapi2020various}
\bibinfo{author}{K.~M. Wouapi}, \bibinfo{author}{B.~H. Fotsin},
  \bibinfo{author}{F.~P. Louodop}, \bibinfo{author}{K.~F. Feudjio},
  \bibinfo{author}{Z.~T. Njitacke}, \bibinfo{author}{T.~H. Djeudjo},
  \bibinfo{title}{Various firing activities and finite-time synchronization of
  an improved Hindmarsh--Rose neuron model under electric field effect},
  \bibinfo{journal}{Cognitive neurodynamics}
  \bibinfo{volume}{14}~(\bibinfo{number}{3}) (\bibinfo{year}{2020})
  \bibinfo{pages}{375--397}.

\bibitem[{Wouapi et~al.(2021)Wouapi, Fotsin, Ngouonkadi, Kemwoue, and
  Njitacke}]{wouapi2021complex}
\bibinfo{author}{M.~K. Wouapi}, \bibinfo{author}{B.~H. Fotsin},
  \bibinfo{author}{E.~B.~M. Ngouonkadi}, \bibinfo{author}{F.~F. Kemwoue},
  \bibinfo{author}{Z.~T. Njitacke}, \bibinfo{title}{Complex bifurcation
  analysis and synchronization optimal control for Hindmarsh--Rose neuron model
  under magnetic flow effect}, \bibinfo{journal}{Cognitive Neurodynamics}
  \bibinfo{volume}{15}~(\bibinfo{number}{2}) (\bibinfo{year}{2021})
  \bibinfo{pages}{315--347}.

\bibitem[{Boaretto et~al.(2021)Boaretto, Manchein, Prado, and
  Lopes}]{boaretto2021role}
\bibinfo{author}{B.~R. Boaretto}, \bibinfo{author}{C.~Manchein},
  \bibinfo{author}{T.~L. Prado}, \bibinfo{author}{S.~R. Lopes},
  \bibinfo{title}{The role of individual neuron ion conductances in the
  synchronization processes of neuron networks}, \bibinfo{journal}{Neural
  Networks} \bibinfo{volume}{137} (\bibinfo{year}{2021})
  \bibinfo{pages}{97--105}.

\bibitem[{Yu et~al.(2021)Yu, Lu, Wang, Yang, and Jia}]{yu2021synchronization}
\bibinfo{author}{D.~Yu}, \bibinfo{author}{L.~Lu}, \bibinfo{author}{G.~Wang},
  \bibinfo{author}{L.~Yang}, \bibinfo{author}{Y.~Jia},
  \bibinfo{title}{Synchronization mode transition induced by bounded noise in
  multiple time-delays coupled FitzHugh--Nagumo model},
  \bibinfo{journal}{Chaos, Solitons \& Fractals} \bibinfo{volume}{147}
  (\bibinfo{year}{2021}) \bibinfo{pages}{111000}.

\bibitem[{Masoliver et~al.(2017)Masoliver, Malik, Sch{\"o}ll, and
  Zakharova}]{masoliver2017coherence}
\bibinfo{author}{M.~Masoliver}, \bibinfo{author}{N.~Malik},
  \bibinfo{author}{E.~Sch{\"o}ll}, \bibinfo{author}{A.~Zakharova},
  \bibinfo{title}{Coherence resonance in a network of FitzHugh-Nagumo systems:
  Interplay of noise, time-delay, and topology}, \bibinfo{journal}{Chaos: An
  Interdisciplinary Journal of Nonlinear Science}
  \bibinfo{volume}{27}~(\bibinfo{number}{10}) (\bibinfo{year}{2017})
  \bibinfo{pages}{101102}.

\bibitem[{Lu et~al.(2020)Lu, Jia, Ge, Xu, and Li}]{lu2020inverse}
\bibinfo{author}{L.~Lu}, \bibinfo{author}{Y.~Jia}, \bibinfo{author}{M.~Ge},
  \bibinfo{author}{Y.~Xu}, \bibinfo{author}{A.~Li}, \bibinfo{title}{Inverse
  stochastic resonance in Hodgkin--Huxley neural system driven by Gaussian and
  non-Gaussian colored noises}, \bibinfo{journal}{Nonlinear Dynamics}
  \bibinfo{volume}{100}~(\bibinfo{number}{1}) (\bibinfo{year}{2020})
  \bibinfo{pages}{877--889}.

\bibitem[{Wang et~al.(2021)Wang, Yu, Ding, Li, and Jia}]{wang2021effects}
\bibinfo{author}{G.~Wang}, \bibinfo{author}{D.~Yu}, \bibinfo{author}{Q.~Ding},
  \bibinfo{author}{T.~Li}, \bibinfo{author}{Y.~Jia}, \bibinfo{title}{Effects of
  electric field on multiple vibrational resonances in Hindmarsh-Rose neuronal
  systems}, \bibinfo{journal}{Chaos, Solitons \& Fractals}
  \bibinfo{volume}{150} (\bibinfo{year}{2021}) \bibinfo{pages}{111210}.

\bibitem[{Liu et~al.(2014)Liu, Liu, and Liu}]{liu2014bifurcation}
\bibinfo{author}{C.~Liu}, \bibinfo{author}{X.~Liu}, \bibinfo{author}{S.~Liu},
  \bibinfo{title}{Bifurcation analysis of a Morris--Lecar neuron model},
  \bibinfo{journal}{Biological cybernetics}
  \bibinfo{volume}{108}~(\bibinfo{number}{1}) (\bibinfo{year}{2014})
  \bibinfo{pages}{75--84}.

\bibitem[{Iqbal et~al.(2017)Iqbal, Rehan, and Hong}]{iqbal2017modeling}
\bibinfo{author}{M.~Iqbal}, \bibinfo{author}{M.~Rehan}, \bibinfo{author}{K.-S.
  Hong}, \bibinfo{title}{Modeling of inter-neuronal coupling medium and its
  impact on neuronal synchronization}, \bibinfo{journal}{PloS one}
  \bibinfo{volume}{12}~(\bibinfo{number}{5}) (\bibinfo{year}{2017})
  \bibinfo{pages}{e0176986}.

\bibitem[{Wang et~al.(2006)Wang, Lu, Chen, and Guo}]{wang2006chaos}
\bibinfo{author}{Q.~Y. Wang}, \bibinfo{author}{Q.~S. Lu},
  \bibinfo{author}{G.~R. Chen}, \bibinfo{author}{D.~H. Guo},
  \bibinfo{title}{Chaos synchronization of coupled neurons with gap junctions},
  \bibinfo{journal}{Physics Letters A}
  \bibinfo{volume}{356}~(\bibinfo{number}{1}) (\bibinfo{year}{2006})
  \bibinfo{pages}{17--25}.

\bibitem[{Xu et~al.(2017)Xu, Ying, Jia, and Hayat}]{autaptic_regulation_ying}
\bibinfo{author}{Y.~Xu}, \bibinfo{author}{H.~Ying}, \bibinfo{author}{Y.~Jia},
  \bibinfo{author}{T.~Hayat}, \bibinfo{title}{Autaptic regulation of electrical
  activities in neuron under electromagnetic induction},
  \bibinfo{journal}{Scientific Reports} \bibinfo{volume}{7}
  (\bibinfo{year}{2017}) \bibinfo{pages}{43452}.

\bibitem[{Destexhe et~al.(1994)Destexhe, Mainen, and
  Sejnowski}]{destexhe1994efficient}
\bibinfo{author}{A.~Destexhe}, \bibinfo{author}{Z.~F. Mainen},
  \bibinfo{author}{T.~J. Sejnowski}, \bibinfo{title}{An efficient method for
  computing synaptic conductances based on a kinetic model of receptor
  binding}, \bibinfo{journal}{Neural computation}
  \bibinfo{volume}{6}~(\bibinfo{number}{1}) (\bibinfo{year}{1994})
  \bibinfo{pages}{14--18}.

\bibitem[{Destexhe et~al.(1998)Destexhe, Mainen, and
  Sejnowski}]{destexhe1998kinetic}
\bibinfo{author}{A.~Destexhe}, \bibinfo{author}{Z.~F. Mainen},
  \bibinfo{author}{T.~J. Sejnowski}, \bibinfo{title}{Kinetic models of synaptic
  transmission}, \bibinfo{journal}{Methods in neuronal modeling}
  \bibinfo{volume}{2} (\bibinfo{year}{1998}) \bibinfo{pages}{1--25}.

\bibitem[{Greengard(2001)}]{greengard2001neurobiology}
\bibinfo{author}{P.~Greengard}, \bibinfo{title}{The neurobiology of slow
  synaptic transmission}, \bibinfo{journal}{Science}
  \bibinfo{volume}{294}~(\bibinfo{number}{5544}) (\bibinfo{year}{2001})
  \bibinfo{pages}{1024--1030}.

\bibitem[{Markram(1997)}]{markram1997network}
\bibinfo{author}{H.~Markram}, \bibinfo{title}{A network of tufted layer 5
  pyramidal neurons.}, \bibinfo{journal}{Cerebral cortex (New York, NY: 1991)}
  \bibinfo{volume}{7}~(\bibinfo{number}{6}) (\bibinfo{year}{1997})
  \bibinfo{pages}{523--533}.

\bibitem[{Pitk{\"a}nen et~al.(2000)Pitk{\"a}nen, Pikkarainen, Nurminen, and
  Ylinen}]{pitkanen2000reciprocal}
\bibinfo{author}{A.~Pitk{\"a}nen}, \bibinfo{author}{M.~Pikkarainen},
  \bibinfo{author}{N.~Nurminen}, \bibinfo{author}{A.~Ylinen},
  \bibinfo{title}{Reciprocal connections between the amygdala and the
  hippocampal formation, perirhinal cortex, and postrhinal cortex in rat: a
  review}, \bibinfo{journal}{Annals of the new York Academy of Sciences}
  \bibinfo{volume}{911}~(\bibinfo{number}{1}) (\bibinfo{year}{2000})
  \bibinfo{pages}{369--391}.

\bibitem[{Zupanc and Corr{\^e}a(2005)}]{zupanc2005reciprocal}
\bibinfo{author}{G.~K. Zupanc}, \bibinfo{author}{S.~A. Corr{\^e}a},
  \bibinfo{title}{Reciprocal neural connections between the central
  posterior/prepacemaker nucleus and nucleus G in the gymnotiform fish,
  Apteronotus leptorhynchus}, \bibinfo{journal}{Brain, behavior and evolution}
  \bibinfo{volume}{65}~(\bibinfo{number}{1}) (\bibinfo{year}{2005})
  \bibinfo{pages}{14--25}.

\bibitem[{Perin et~al.(2011)Perin, Berger, and Markram}]{perin2011synaptic}
\bibinfo{author}{R.~Perin}, \bibinfo{author}{T.~K. Berger},
  \bibinfo{author}{H.~Markram}, \bibinfo{title}{A synaptic organizing principle
  for cortical neuronal groups}, \bibinfo{journal}{Proceedings of the National
  Academy of Sciences} \bibinfo{volume}{108}~(\bibinfo{number}{13})
  (\bibinfo{year}{2011}) \bibinfo{pages}{5419--5424}.

\bibitem[{Bastos et~al.(2012)Bastos, Usrey, Adams, Mangun, Fries, and
  Friston}]{bastos2012canonical}
\bibinfo{author}{A.~M. Bastos}, \bibinfo{author}{W.~M. Usrey},
  \bibinfo{author}{R.~A. Adams}, \bibinfo{author}{G.~R. Mangun},
  \bibinfo{author}{P.~Fries}, \bibinfo{author}{K.~J. Friston},
  \bibinfo{title}{Canonical microcircuits for predictive coding},
  \bibinfo{journal}{Neuron} \bibinfo{volume}{76}~(\bibinfo{number}{4})
  (\bibinfo{year}{2012}) \bibinfo{pages}{695--711}.

\bibitem[{Izhikevich(2000)}]{izhikevich2000neural}
\bibinfo{author}{E.~M. Izhikevich}, \bibinfo{title}{Neural excitability,
  spiking and bursting}, \bibinfo{journal}{International journal of bifurcation
  and chaos} \bibinfo{volume}{10}~(\bibinfo{number}{06}) (\bibinfo{year}{2000})
  \bibinfo{pages}{1171--1266}.

\bibitem[{Sch{\"o}ll et~al.(2009)Sch{\"o}ll, Hiller, H{\"o}vel, and
  Dahlem}]{scholl2009time}
\bibinfo{author}{E.~Sch{\"o}ll}, \bibinfo{author}{G.~Hiller},
  \bibinfo{author}{P.~H{\"o}vel}, \bibinfo{author}{M.~A. Dahlem},
  \bibinfo{title}{Time-delayed feedback in neurosystems},
  \bibinfo{journal}{Philosophical Transactions of the Royal Society A:
  Mathematical, Physical and Engineering Sciences}
  \bibinfo{volume}{367}~(\bibinfo{number}{1891}) (\bibinfo{year}{2009})
  \bibinfo{pages}{1079--1096}.

\bibitem[{Kramers(1940)}]{kramers1940brownian}
\bibinfo{author}{H.~A. Kramers}, \bibinfo{title}{Brownian motion in a field of
  force and the diffusion model of chemical reactions},
  \bibinfo{journal}{Physica} \bibinfo{volume}{7}~(\bibinfo{number}{4})
  (\bibinfo{year}{1940}) \bibinfo{pages}{284--304}.

\bibitem[{Kuehn(2015)}]{kuehn2015multiple}
\bibinfo{author}{C.~Kuehn}, \bibinfo{title}{Multiple time scale dynamics}, vol.
  \bibinfo{volume}{191}, \bibinfo{publisher}{Springer}, \bibinfo{year}{2015}.

\bibitem[{Lee~DeVille et~al.(2005)Lee~DeVille, Vanden-Eijnden, and
  Muratov}]{SISRvsCR}
\bibinfo{author}{R.~E. Lee~DeVille}, \bibinfo{author}{E.~Vanden-Eijnden},
  \bibinfo{author}{C.~B. Muratov}, \bibinfo{title}{Two distinct mechanisms of
  coherence in randomly perturbed dynamical systems}, \bibinfo{journal}{Phys.
  Rev. E} \bibinfo{volume}{72} (\bibinfo{year}{2005}) \bibinfo{pages}{031105}.

\bibitem[{Pei et~al.(1996)Pei, Wilkens, and Moss}]{pei1996noise}
\bibinfo{author}{X.~Pei}, \bibinfo{author}{L.~Wilkens},
  \bibinfo{author}{F.~Moss}, \bibinfo{title}{Noise-mediated spike timing
  precision from aperiodic stimuli in an array of Hodgekin-Huxley-type
  neurons}, \bibinfo{journal}{Physical review letters}
  \bibinfo{volume}{77}~(\bibinfo{number}{22}) (\bibinfo{year}{1996})
  \bibinfo{pages}{4679}.

\bibitem[{R{\"o}{\ss}ler(2009)}]{rossler2009second}
\bibinfo{author}{A.~R{\"o}{\ss}ler}, \bibinfo{title}{Second order Runge--Kutta
  methods for It{\^o} stochastic differential equations},
  \bibinfo{journal}{SIAM Journal on Numerical Analysis}
  \bibinfo{volume}{47}~(\bibinfo{number}{3}) (\bibinfo{year}{2009})
  \bibinfo{pages}{1713--1738}.

\bibitem[{Rumelhart and Zipser(1985)}]{rumelhart1985feature}
\bibinfo{author}{D.~E. Rumelhart}, \bibinfo{author}{D.~Zipser},
  \bibinfo{title}{Feature discovery by competitive learning},
  \bibinfo{journal}{Cognitive science}
  \bibinfo{volume}{9}~(\bibinfo{number}{1}) (\bibinfo{year}{1985})
  \bibinfo{pages}{75--112}.

\bibitem[{Grossberg(1987)}]{grossberg1987competitive}
\bibinfo{author}{S.~Grossberg}, \bibinfo{title}{Competitive learning: From
  interactive activation to adaptive resonance}, \bibinfo{journal}{Cognitive
  science} \bibinfo{volume}{11}~(\bibinfo{number}{1}) (\bibinfo{year}{1987})
  \bibinfo{pages}{23--63}.

\bibitem[{Rabinovich et~al.(2001)Rabinovich, Volkovskii, Lecanda, Huerta,
  Abarbanel, and Laurent}]{rabinovich2001dynamical}
\bibinfo{author}{M.~Rabinovich}, \bibinfo{author}{A.~Volkovskii},
  \bibinfo{author}{P.~Lecanda}, \bibinfo{author}{R.~Huerta},
  \bibinfo{author}{H.~Abarbanel}, \bibinfo{author}{G.~Laurent},
  \bibinfo{title}{Dynamical encoding by networks of competing neuron groups:
  winnerless competition}, \bibinfo{journal}{Physical review letters}
  \bibinfo{volume}{87}~(\bibinfo{number}{6}) (\bibinfo{year}{2001})
  \bibinfo{pages}{068102}.

\bibitem[{Savin et~al.(2010)Savin, Joshi, and Triesch}]{savin2010independent}
\bibinfo{author}{C.~Savin}, \bibinfo{author}{P.~Joshi},
  \bibinfo{author}{J.~Triesch}, \bibinfo{title}{Independent component analysis
  in spiking neurons}, \bibinfo{journal}{PLoS Comput Biol}
  \bibinfo{volume}{6}~(\bibinfo{number}{4}) (\bibinfo{year}{2010})
  \bibinfo{pages}{e1000757}.

\bibitem[{Krotov and Hopfield(2019)}]{krotov2019unsupervised}
\bibinfo{author}{D.~Krotov}, \bibinfo{author}{J.~J. Hopfield},
  \bibinfo{title}{Unsupervised learning by competing hidden units},
  \bibinfo{journal}{Proceedings of the National Academy of Sciences}
  \bibinfo{volume}{116}~(\bibinfo{number}{16}) (\bibinfo{year}{2019})
  \bibinfo{pages}{7723--7731}.

\bibitem[{Wu et~al.(2017)Wu, Xu, and Ma}]{wu2017levy}
\bibinfo{author}{J.~Wu}, \bibinfo{author}{Y.~Xu}, \bibinfo{author}{J.~Ma},
  \bibinfo{title}{L{\'e}vy noise improves the electrical activity in a neuron
  under electromagnetic radiation}, \bibinfo{journal}{PLoS One}
  \bibinfo{volume}{12}~(\bibinfo{number}{3}) (\bibinfo{year}{2017})
  \bibinfo{pages}{e0174330}.

\bibitem[{Segev et~al.(2002)Segev, Benveniste, Hulata, Cohen, Palevski, Kapon,
  Shapira, and Ben-Jacob}]{segev2002long}
\bibinfo{author}{R.~Segev}, \bibinfo{author}{M.~Benveniste},
  \bibinfo{author}{E.~Hulata}, \bibinfo{author}{N.~Cohen},
  \bibinfo{author}{A.~Palevski}, \bibinfo{author}{E.~Kapon},
  \bibinfo{author}{Y.~Shapira}, \bibinfo{author}{E.~Ben-Jacob},
  \bibinfo{title}{Long term behavior of lithographically prepared in vitro
  neuronal networks}, \bibinfo{journal}{Physical review letters}
  \bibinfo{volume}{88}~(\bibinfo{number}{11}) (\bibinfo{year}{2002})
  \bibinfo{pages}{118102}.

\end{thebibliography}
%\begin{thebibliography}{00}
%\addcontentsline{toc}{chapter}{\numberline{}Bibliography}
%%%%%%%%%%%%%%%%%%%%%%%%%%%%%%%%%%%%%%%%%%%%%%%%%%%%%%%%%%%%%%%%%%%%%%%%%%%%%%%%%%%%%%%%%%%%%%%%%%%%%%%%%%%%%%%%%%
%\end{thebibliography}

\end{document}